\documentclass[]{emulateapj}
\usepackage{natbib}
\usepackage{amssymb}
\usepackage{natbib}
\usepackage{amssymb}
\usepackage{color}
\usepackage{amsmath,mathtools}
\usepackage{epsfig}
\usepackage[FIGTOPCAP]{subfigure}
\usepackage{afterpage}
\usepackage{enumerate}
\usepackage{multirow}
\usepackage{verbatim}
\usepackage{tikz}
\usetikzlibrary{shapes.geometric, arrows}
\usepackage{morefloats}
\usepackage{wasysym}

\newcommand{\mockalph}[1]{}
\newcommand{\noop}[1]{}

\newcommand{\unit}[1]{\ensuremath{\, \mathrm{#1}}}
\newcommand{\exuv}{\ensuremath{\epsilon_\mathrm{XUV}}}
\newcommand{\mearth}{\unit{M_\oplus}}
\newcommand{\rearth}{\unit{R_\oplus}}
\newcommand{\msun}{\unit{M_\odot}}
\newcommand{\lsun}{\unit{L_\odot}}
\newcommand{\mstar}{\unit{M_\star}}

\newcommand{\avg}[1]{\langle #1 \rangle}

\shorttitle{Habitable Evaporated Cores}
\shortauthors{Luger et al. 2015}

\begin{document}

\title{Habitable Evaporated Cores: \\Transforming Mini-Neptunes into Super-Earths \\in the Habitable Zones of M Dwarfs}

\author{R. Luger\altaffilmark{1,2}, R. Barnes\altaffilmark{1,2}, E. Lopez\altaffilmark{3}, J. Fortney\altaffilmark{3}, B. Jackson\altaffilmark{4}, and V. Meadows\altaffilmark{1,2}}
\affil{\altaffilmark{1}Astronomy Department, University of Washington, Box 351580, Seattle, WA 98195, USA; rodluger@uw.edu\\
\altaffilmark{2}Virtual Planetary Laboratory, Seattle, WA 98195, USA\\
\altaffilmark{3}Department of Astronomy and Astrophysics, University of California, Santa Cruz, California\\
\altaffilmark{4}Carnegie Institute of Washington, Washington, DC}

\begin{abstract}
We show that photoevaporation of small gaseous exoplanets (``mini-Neptunes'') in the habitable zones of M dwarfs can remove several Earth masses of hydrogen and helium from these planets and transform them into potentially habitable worlds. We couple X-ray/extreme ultraviolet (XUV)-driven escape, thermal evolution, tidal evolution and orbital migration to explore the types of systems that may harbor such ``habitable evaporated cores'' (HECs). We find that HECs are most likely to form from planets with $\sim 1\mearth$ solid cores with up to about 50\% H/He by mass, though whether or not a given mini-Neptune forms a HEC is highly dependent on the early XUV evolution of the host star. As terrestrial planet formation around M dwarfs by accumulation of local material is likely to form planets that are small and dry, evaporation of small migrating mini-Neptunes could be one of the dominant formation mechanisms for volatile-rich Earths around these stars.
\end{abstract}

\section{Introduction}
\label{sec:intro}
Due to their small radii and low luminosities, M dwarfs currently offer the best opportunity for the detection of terrestrial planets in the habitable zone (HZ), the region around a star where liquid water can exist on the surface of a planet \citep{KAS93,KOP13}. Upcoming missions such as the Transiting Exoplanet Survey Satellite (TESS) and the repurposed Kepler spacecraft (K2) will be capable of detecting potentially habitable Earths and super-Earths around M dwarfs \citep{RIC10,HOW14}. In particular, the detection of potentially habitable planets around M dwarfs is easier because the habitable zones of these stars can be as close in as $\sim$ 0.02 AU \citep{KOP13}. However, such proximity implies that terrestrial planets forming within the HZs of low mass stars are likely to be small ($\lesssim 0.3 \mathrm{M_\oplus}$) and form dry \citep{RAY07,LIS07}. Moreover, M dwarfs are extremely active when young, bombarding their planets with high energy radiation and bursts of relativistic particles during flaring events, which can erode their atmospheres and potentially sterilize the surface \citep{SCA07,SEG10}. Strong tidal heating and orbital evolution may further impact the habitability of planets around these stars \citep{BAR13}. Many planets formed \emph{in situ} in the HZ of M dwarfs may therefore be uninhabitable.

However, planets need not form and remain in place. It is now commonly accepted that both disk-driven migration and planet-planet interactions can lead to substantial orbital changes, potentially bringing planets from outside the snow line (the region of the disk beyond which water and other volatiles condense into ices, facilitating the formation of massive planetary cores) to within the HZ \citep{HAY85,IL08a, IL08b, OI09}. Disk-driven migration relies on the exchange of angular momentum between a planet and the surrounding gaseous disk, which induces rapid inward migration for planets in the terrestrial mass range \citep{WAR97}.  Around solar-type stars, disk dispersal typically occurs on the order of a few Myr \citep{WAL88,STR89}. While there is evidence that disk lifetimes may exceed $\sim$ 5 Myr for low mass stars \citep{CAR06,PAS09}, disks are no longer present after $\sim$ 10-20 Myr around all stellar types \citep{RIB13}. Planets migrating via interactions with the disk will thus settle into their new orbits relatively early. Planet-planet scattering, on the other hand, is by nature a stochastic process and may take place at any point during a system's lifetime. However, such interactions are far more frequent during the final stages of planet assembly (up to a few tens of Myr), after which planets relax into their long-term quasi-stable orbits \citep{FR08}.

Given the abundance of ices beyond the snow line, planets that migrate into the HZ from the outer regions of the disk should have abundant water, therefore satisfying that important criterion for habitability within the HZ. However, the situation is complicated by the fact that these planets may have accumulated thick gaseous envelopes, which could render them uninhabitable. Investigating whether these so-called ``mini-Neptunes'' can lose their envelopes and form planets with solid surfaces is therefore critical to understanding the habitability of planets around low mass stars. 

In this study we focus on mini-Neptunes with initial masses in the range $1\mearth \leq M_p \leq 10\mearth$ with up to 50\% hydrogen/helium by mass that have migrated into the HZs of mid- to late M dwarf stars. We investigate whether it is possible for atmospheric escape processes to remove the thick H/He envelopes of mini-Neptunes in the HZ, effectively turning them into volatile-rich Earths and super-Earths (terrestrial planets more massive than Earth),which we refer to as ``habitable evaporated cores'' (HECs). We consider two atmospheric loss processes: XUV-driven escape, in which stellar X-ray/extreme ultraviolet (XUV) photons heat the atmosphere and drive a hydrodynamic wind away from the planet, and a simple model of Roche lobe overflow (RLO), in which the atmosphere is so extended that part of it lies exterior to the planet's Roche lobe; that gas is therefore no longer gravitationally bound to the planet. We further couple the effects of atmospheric mass loss to the thermal and tidal evolution of these planets. Planets cool as they age, undergoing changes of up to an order of magnitude in radius, which can greatly affect the mass loss rate. Tidal forces arising from the differential strength of gravity distort both the planet and the star away from sphericity, introducing torques that lead to the evolution of the orbital and spin parameters of both bodies.

While many studies have considered the separate effects of atmospheric escape \citep{ERK07,MUR09,TIA09,OJ12,LAM13}, Roche lobe overflow \citep{TRI98, GLB03}, thermal evolution \citep{LOP12}, and tidal evolution \citep{JGB08,FER08,CL11} on exoplanets, none have considered the coupling of these effects in the HZ. For some systems, in particular those that may harbor HECs, modeling the coupling of these processes is essential to accurately determine the evolution, since several feedbacks can ensue. Tidal forces in the HZ typically act to decrease a planet's semi-major axis, leading to higher stellar fluxes and faster mass loss. The mass loss, in turn, affects the rate of tidal evolution primarily via the changing planet radius, which is also governed by the cooling rate of the envelope. Changes to the star's radius and luminosity lead to further couplings that need to be treated with care.

\cite{JAC10} considered the effect of the coupling between mass loss and tidal evolution on the hot super-Earth CoRoT-7 b, finding that the two effects are strongly linked and must be considered simultaneously to obtain an accurate understanding of the planet's evolution. However, an analogous study has not been performed in the HZ, in great part because both tidal effects and atmospheric mass loss are generally orders of magnitude weaker at such distances from the star. This is not necessarily true around M dwarfs, for two reasons: (a) their low luminosities result in a HZ that is much closer in, exposing planets to strong tidal effects and possible RLO; and (b) M dwarfs are extremely active early on, so that the XUV flux in the HZ can be orders of magnitude higher than that around a solar-type star \citep[see, for instance,][]{SCA07}.

In this paper we present the results of the first model to couple tides, thermal evolution, and atmospheric mass loss in the habitable zone, showing that for certain systems the coupling is key in determining the long-term evolution of the planet. We demonstrate that it is possible to turn mini-Neptunes into HECs within the habitable zone, providing an important pathway to the formation of potentially habitable, volatile-rich planets around M dwarfs.

In \S\ref{sec:background} we provide a detailed description of the relevant physics. In \S\ref{sec:model} we describe our model, followed by our results in \S\ref{sec:results}. We then discuss our main findings and the corresponding caveats in \S\ref{sec:discuss}, followed by a summary in \S\ref{sec:conclusions}. We present auxiliary derivations and calculations in the Appendix.

\section{Stellar and Planetary Evolution}
\label{sec:background}
In the following sections we review the luminosity evolution of low mass stars (\S\ref{sec:lumevol}), the habitable zone and its evolution (\S\ref{sec:hz}), atmospheric escape processes from planets (\S\ref{sec:massloss}), and tidal evolution of star-planet systems (\S\ref{sec:tides}).

\subsection{Stellar Luminosity}
\label{sec:lumevol}
Following their formation from a giant molecular cloud, stars contract under their own gravity until they reach the \emph{main sequence}, at which point the core temperature is high enough to ignite hydrogen fusion. While the Sun is thought to have spent $\lesssim$ 50 Myr in this pre-main sequence (PMS) phase \citep{BAR98}, M dwarfs take hundreds of Myr to fully contract; the lowest mass M dwarfs reach the main sequence only after $\sim$ 1 Gyr \citep[e.g.,][]{RH05}. During their contraction, M dwarfs remain at a roughly constant effective temperature \citep{HAY61}, so that their luminosity is primarily a function of their surface area, which is significantly larger than when they arrive on the main sequence. M dwarfs therefore remain super-luminous for several hundred Myr, with total (bolometric) luminosities higher than the main sequence value by up to two orders of magnitude. As we discuss below, this will significantly affect the atmospheric evolution of any planets these stars may host.

XUV emissions ($1 \mathrm{\AA} \lesssim \lambda \lesssim 1000 \mathrm{\AA}$) from M dwarfs also vary strongly with time. This is because the XUV luminosity of M dwarfs is rooted in the vigorous magnetic fields generated in their large convection zones \citep{SCA07}. Stellar magnetic fields are largely controlled by rotation \citep{PAR55}, which slows down with time due to angular momentum loss \citep{SKU72}, leading to an XUV activity that declines with stellar age. However, due to the difficulty of accurately determining both the XUV luminosities (usually inferred from line proxies) and the ages (often determined kinematically) of M dwarfs, the exact functional form of the evolution is still very uncertain \citep[for a review, see][]{SCA07}. Further complications arise due to the fact that the process(es) that generate magnetic fields in late M dwarfs may be quite different from those in earlier type stars. The rotational dynamo of \cite{PAR55} involves the amplification of toroidal fields generated at the radiative-convective boundary; late M dwarfs, however, are fully convective, and have no such boundary. Instead, their magnetic fields may be formed by turbulent dynamos \citep{DUR93}, which may evolve differently in time from those around higher mass stars \citep{RH05}.

In a comprehensive study of the XUV emissions of solar-type stars of different ages, \cite{RIB05} found that the XUV evolution is well modeled by a simple power law with an initial short-lived ``saturation'' phase:
\begin{align}
\label{eq:lxuvevol}
\frac{L_\mathrm{XUV}}{L_\mathrm{bol}} = \left\{
				\begin{array}{lcr}
					\left(\frac{L_\mathrm{XUV}}{L_\mathrm{bol}}\right)_\mathrm{sat} &\ & t \leq t_\mathrm{sat} \\
					\left(\frac{L_\mathrm{XUV}}{L_\mathrm{bol}}\right)_\mathrm{sat}\left(\frac{t}{t_\mathrm{sat}}\right)^{-\beta} &\ & t > t_\mathrm{sat},
				\end{array}
				\right.
\end{align}
where $L_\mathrm{XUV}$ and $L_\mathrm{bol}$ are the XUV and bolometric luminosities, respectively, and $\beta = -1.23$. Prior to $t = t_\mathrm{sat}$, the XUV luminosity is said to be ``saturated,'' as observations show that the ratio $L_\mathrm{XUV}/L_\mathrm{bol}$ remains relatively constant at early times.
 
\cite{JDW12} find that $t_\mathrm{sat} \approx 100 \unit{Myr}$ and $(L_\mathrm{XUV}/L_{\mathrm{bol}})_\mathrm{sat} \approx 10^{-3}$ for K dwarfs. Similar studies for M dwarfs, however, are still being developed \citep[e.g.,][]{EG11}, but it is likely that the saturation timescale is much longer for late-type M dwarfs. \cite{WRI11} show that X-ray emission in low mass stars is saturated for $P_\mathrm{rot}/\tau \lesssim 0.1$, where $P_\mathrm{rot}$ is the stellar rotation period and $\tau$ is the convective turnover time. The extent of the convective zone increases with decreasing stellar mass; below 0.35M$_\odot$, M dwarfs are fully convective \citep{CB97}, resulting in larger values of $\tau$ \citep[see, e.g.,][]{PIZ00}. Low mass stars also have longer spin-down times \citep{STA94}; together, these effects should lead to significantly longer saturation times compared to solar-type stars. This is consistent with observational studies; \cite{WES08} find that the magnetic activity lifetime increases from $\lesssim 1 \unit{Gyr}$ for early (i.e., most massive) M dwarfs to $\gtrsim 7 \unit{Gyr}$ for late (least massive) M dwarfs, possibly due to the onset of full convection around spectral type M3. Finally, \cite{STE13} find that for M dwarfs, $\beta = -1.10 \pm 0.02$ in the X-ray and $\beta = -0.79 \pm 0.05$ in the FUV (far ultraviolet), suggesting a slightly shallower slope in the XUV compared to the value from \cite{RIB05}.

\subsection{The Habitable Zone}
\label{sec:hz}
The habitable zone (HZ) is classically defined as the region around a star where an Earth-like planet can sustain liquid water on its surface \citep{HAR79,KAS93}. Interior to the HZ, high surface temperatures lead to the runaway evaporation of a planet's oceans, which increases the atmospheric infrared opacity and reduces the ability of the surface to cool in a process known as the \emph{runaway greenhouse}. Exterior to the HZ, greenhouse gases---gases, like water vapor, that absorb strongly in the infrared---are unable to maintain surface temperatures above the freezing point, and the oceans freeze globally. Recently, \cite{KOP13} re-calculated the location of the HZ boundaries as a function of stellar luminosity and effective temperature using an updated one-dimensional, radiative-convective, cloud-free climate model. Their five boundaries are the (1) Recent Venus, (2) Runaway Greenhouse, (3) Moist Greenhouse, (4) Maximum Greenhouse, and (5) Early Mars habitable zones.

The first and last limits can be considered ``optimistic'' empirical limits, since prior to $\sim 1$ and $\sim 3.8$ Gyr ago, respectively, Venus and Mars may have had liquid surface water. The ability of a planet to maintain liquid water and to sustain life at these limits is still unclear and probably depends on a host of properties of its climate. Conversely, the other three limits are the ``pessimistic'' theoretical limits, corresponding to where a cloud-free, Earth-like planet would lose its entire water inventory due to the greenhouse effect (2 and 3) and to where the addition of CO$_2$ to the atmosphere would be incapable of sustaining surface temperatures above freezing (4). 

\begin{figure}[h]
  \begin{center}
      \psfig{file=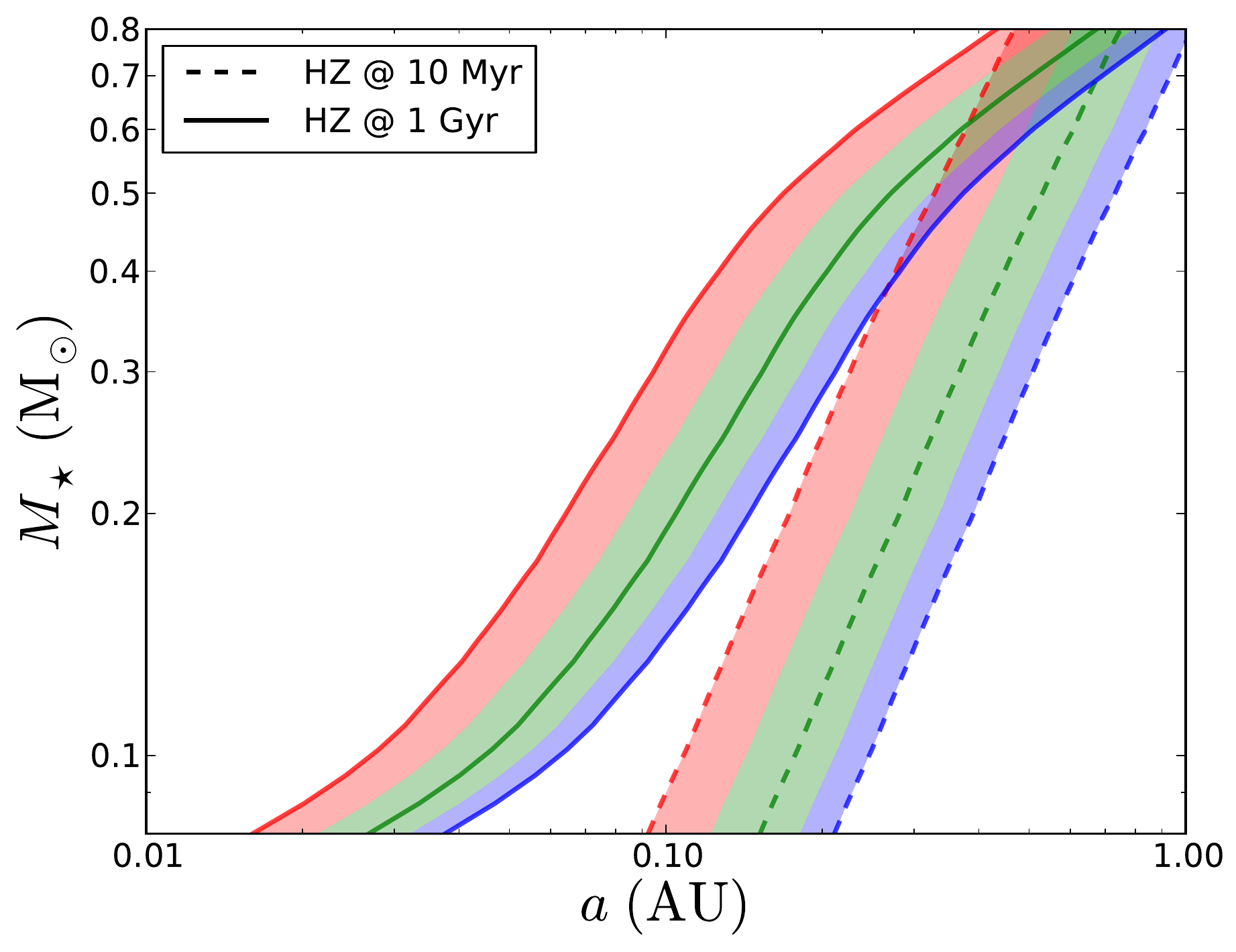,width=\columnwidth}
       \caption{Location of the inner habitable zone (red), central habitable zone (green) and outer habitable zone (blue) as a function of stellar mass at 10 Myr (dashed) and 1 Gyr (solid). After 1 Gyr, the evolution of the HZ is negligible for M dwarfs.}
     \label{fig:hz}
  \end{center}
\end{figure}

Because stellar luminosities vary with time, the location of the HZ is not fixed. In Figure~\ref{fig:hz} we plot the HZ at 10 Myr (dashed lines) and at 1 Gyr (solid lines), calculated from the HZ model of \cite{KOP13} and the stellar evolution models of \cite{BAR98}. While for K and G dwarfs the change in the HZ is negligible, the HZ of M dwarfs moves in substantially, changing by nearly an order of magnitude for the least massive stars. Due to this evolution, planets observed in the HZ of M dwarfs today likely spent a significant amount of time interior to the inner edge of the HZ, provided they either formed \emph{in situ} or migrated to their current positions relatively early. \cite{LUG14} explore in detail the effects of the evolution of the HZ on terrestrial planets.

Finally, we note that the location of the HZ is also a function of the eccentricity $e$. This is due to the fact that at a fixed semi-major axis $a$, the orbit-averaged flux $\avg{F}$ is higher for more eccentric orbits \citep{WP02}:

\begin{align}
\label{eq:savg}
\avg{F} = \frac{L_\mathrm{bol}}{4\pi a^2\sqrt{1-e^2}}.
\end{align}

\subsection{Atmospheric Mass Loss}
\label{sec:massloss}
Planetary atmospheres constantly evolve. Several mechanisms exist through which planets can lose significant quantities of their atmospheres to space, leading to dramatic changes in composition and in some cases complete atmospheric erosion. The early Earth itself could have been rich in hydrogen, with mixing ratios as high as 30\% in the prebiotic atmosphere \citep{TIA05,HAS07}. A variety of processes subsequently led to the loss of most of this hydrogen; \cite{WAT81} argue that on the order of $10^{23}$ g of hydrogen could have escaped in the first billion years. Similarly, \cite{KP83} calculated that early Venus could have lost an Earth ocean equivalent of water in the same amount of time. Currently, observational evidence for atmospheric escape exists for two ``hot Jupiters,'' HD 209458b \citep{VID03} and HD 189733b \citep{LEC10}, and one ``hot Neptune,'' GJ 436b \citep{KUL14}, whose proximity to their parent stars leads to the rapid hydrodynamic loss of hydrogen.

Atmospheric escape mechanisms fall into two major categories: thermal escape, in which the heating of the upper atmosphere accelerates the gas to velocities exceeding the escape velocity, and nonthermal escape, which encompasses a variety of mechanisms and may involve energy exchange among ions or erosion due to stellar winds. While nonthermal processes certainly do play a role in the evaporation of super-Earth and mini-Neptune atmospheres, the high exospheric temperatures resulting from strong XUV irradiation probably make thermal escape the dominant mechanism for planets around M dwarfs at early times. However, the escape can be greatly enhanced by flares and coronal mass ejections, which can completely erode the atmosphere of a planet lacking a strong magnetic field \citep{LAM07, SCA07}. For a review of the nonthermal mechanisms of escape, the reader is referred to \cite{HUN82}.

\subsubsection{Jeans Escape}
\label{sec:massloss:jeans}
In the low temperature limit, atmospheric mass loss occurs via the ballistic escape of individual atoms from the collisionless exosphere, where the low densities ensure that atoms with outward velocities exceeding the escape velocity will escape the planet. Originally developed by \cite{JEA25}, who assumed a hydrostatic, isothermal atmosphere, the mass loss rate of a pure hydrogen atmosphere is given by \citep{OPI63}
\begin{align}
\frac{dM_p}{dt} = 4\pi R_\mathrm{exo}^2 nm_H v_t \frac{(1+\lambda_J)e^{-\lambda_J}}{2\sqrt{\pi}}
\label{eq:dmdtjeans}
\end{align}
where $R_\mathrm{exo}$ is the radius of the exobase, $n$ is the number density of hydrogen atoms at the exobase, $m_H$ is the mass of a hydrogen atom, $v_t$ is the thermal velocity of the gas, and $\lambda_J$ is the Jeans escape parameter, defined as the ratio of the potential energy to the kinetic energy of the gas and given by
\begin{align}
\lambda_J \equiv \frac{GM_pm_H}{kT_\mathrm{exo}R_\mathrm{exo}},
\label{eq:jeans}
\end{align}
where $G$ is the gravitational constant, $M_p$ is the mass of the planet, and $T_\mathrm{exo}$ is the temperature in the (isothermal) exosphere. Since in the Jeans regime the thermal velocity of the gas is less than the escape velocity, the bulk of the gas remains bound to the planet, and only atoms in the tail of the Maxwell-Boltzmann distribution are able to escape. Jeans escape is thus slow. As an example, the present Jeans escape flux for hydrogen on Venus is on the order of $10\ \unit{cm^{-2}s^{-1}}$ \citep{LAM06}, corresponding to the feeble rate of $\sim 10^{-4}\ \unit{g/s}$, which would remove only one part in $10^{11}$ of the atmosphere per billion years.

For higher exospheric temperatures and/or larger values of $R_\mathrm{exo}$, however, corresponding to low values of $\lambda_J$, the atmosphere enters a regime in which the velocity of the average atom in the exosphere approaches the escape velocity of the planet. In this regime, the bulk of the upper atmosphere ceases to be hydrostatic (and isothermal), (\ref{eq:jeans}) is no longer valid, and the escape rate must be calculated from hydrodynamic models. 

\subsubsection{Hydrodynamic Escape}
\label{sec:massloss:hydro}
One of the primary mechanisms for heating the exosphere and decreasing $\lambda_J$ is via strong XUV irradiation. XUV photons are absorbed close to the base of the thermosphere, where they deposit energy and heat the gas via the ionization of atomic hydrogen. This heating is balanced by the adiabatic expansion of the upper atmosphere, downward heat conduction, and cooling by recombination radiation. For sufficiently high XUV fluxes, the expansion of the atmosphere accelerates the gas to supersonic speeds, at which point a hydrodynamic wind is established similar to the solar Parker wind \citep{PAR65}. Once the gas reaches the exosphere, it will escape the planet provided its kinetic energy exceeds the energy required to lift it out of the planet's gravitational well.

Since the kinetic energy of a hydrogen atom at the exobase is $\frac{3}{2}kT_\mathrm{exo}$, the condition $\lambda_J < 1.5$ implies that the kinetic energy of the gas is greater than the absolute value of its gravitational binding energy, and it should therefore begin to escape in bulk in a process commonly referred to as ``blow-off.'' Unlike in the Jeans escape regime, where the mass loss occurs on a per-particle basis, blow-off leads to the rapid loss of large portions of the upper atmosphere, irrespective of particle species, as atoms and molecules heavier than hydrogen are carried along by the hydrodynamic wind. However, contrary to what \cite{OPI63} suggests, the mass loss in this stage is not \emph{arbitrarily} high, since once blow-off begins the upper atmosphere can no longer be treated as isothermal. As the exosphere expands it also cools, so that in the absence of an energy source the value of $\lambda_J$ will tend to increase, thereby moderating the blow-off. The mass loss is, in this sense, ``energy-limited,'' and may be calculated by equating the energy input to the energy required to drive the escape.

Originally proposed by \cite{WAT81}, the energy-limited model assumes that the XUV flux is absorbed in a thin layer at radius $R_\mathrm{XUV}$ where the optical depth to stellar XUV photons is unity. Recently updated to include tidal effects by \cite{ERK07}, this model approximates the mass loss as
\begin{align}
\frac{dM_p}{dt} \approx \frac{\epsilon_\mathrm{XUV}\pi F_\mathrm{XUV}R_p R_\mathrm{XUV}^2}{GM_pK_\mathrm{tide}(\xi)}
\label{eq:dmdt}
\end{align}
where \exuv\ is the heating efficiency parameter (see below), $F_\mathrm{XUV}$ is the incident XUV flux, $R_p$ is the planetary radius, and $K_\mathrm{tide}$ is a tidal enhancement factor, accounting for the fact that for sufficiently close-in planets, the stellar gravity reduces the gravitational binding energy of the gas such that it need only reach the Roche radius to escape the planet. \cite{ERK07} show that
\begin{align}
\label{eq:ktide}
K_\mathrm{tide}(\xi) = \left(1 - \frac{3}{2\xi} + \frac{1}{2\xi^3} \right),
\end{align}
where the parameter $\xi$ is defined as
\begin{align}
\label{eq:xi}
\xi &\equiv \frac{R_\mathrm{Roche}}{R_\mathrm{XUV}}
\end{align}
with
\begin{align}
\label{eq:roche}
R_\mathrm{Roche} &\equiv \left(\frac{M_p}{3M_\star}\right)^{1/3}a,
\end{align}
where M$_\star$ is the mass of the star and $a$ is the semi-major axis. For simplicity, as in \cite{LOP12}, we replace $R_p$ with $R_\mathrm{XUV}$ in (\ref{eq:dmdt}), which is approximately valid given that $R_\mathrm{XUV}$ is typically only 10-20\% larger than $R_p$ \citep{MUR09,LOP12}.

Since the input XUV energy is balanced in part by cooling radiation (via Lyman $\alpha$ emission in the case of hydrogen) and heat conduction, only a fraction of it goes into the adiabatic expansion that drives escape. Rather than running complex hydrodynamic and radiative transfer models to determine the net heating rate, many authors \citep[]{JAC10,LEI11,LOP12,KOS12,LAM13} choose to fold the balance between XUV heating and cooling into an efficiency parameter, \exuv, defined as the fraction of the incoming XUV energy that is converted into $PdV$ work. Because of the complex dependence of \exuv\ on the atmospheric composition and structure, its value is still poorly constrained. \cite{LOP12} estimate $\exuv = 0.2 \pm 0.1$ for super-Earths and mini-Neptunes based on values found throughout the literature. Earlier work by \cite{CHA96} estimates $\exuv \lesssim 0.30$ for Venus-like planets but the author argues that the actual value may be closer to $0.15$. Recently, \cite{OJ12} found X-ray efficiencies $\lesssim 0.1$ for planets more massive than Neptune, but higher efficiencies ($\sim 0.15$) for terrestrial planets. Moreover, \cite{SHE14} argue that studies that assume efficiencies higher than about 0.2 probably lead to overestimates in the escape rate. On the other hand, some studies suggest higher heating efficiencies: \cite{KOS12} use hydrodynamic and photochemical models of the hot Jupiter HD209458b to calculate $\exuv = 0.44$.

As we have already implied, unlike Jeans escape, hydrodynamic blow-off is fast. \cite{CHA96} calculates the maximum hydrodynamic escape rate from the early Venusian atmosphere to be $\sim 10^6$ g/s, ten orders of magnitude higher than the present Jeans escape flux (see \S~\ref{sec:massloss:jeans}). Although there has been debate over the validity of the blow-off assumption \citep[see, for instance, the discussion in][]{TIA08}, recently \cite{LAM13} showed that for super-Earths exposed to high levels of XUV irradiation, the energy-limited approximation yields mass loss rates that are consistent with hydrodynamic models to within a factor of about two, which is within the uncertainties of the problem.

\subsubsection{Controlled Hydrodynamic Escape}
It is also worth noting that there may be an intermediate regime between Jeans escape and blow-off known as ``modified Jeans escape'' or ``controlled hydrodynamic escape'' \citep{ERK13}. In this regime, which occurs for intermediate XUV fluxes and/or higher planetary surface gravity, blow-off conditions are not met but the atmosphere still expands, so that the hydrostatic Jeans formalism is not valid. In order to calculate the escape rate, one must replace the classical Maxwellian velocity distribution with one that includes the bulk expansion velocity of the atmosphere. This yields escape rates lower than those due to a hydrodynamic flow, but significantly higher than those predicted by the hydrostatic Jeans equation (\ref{eq:dmdtjeans}).

\subsubsection{Jeans Escape or Hydrodynamic Escape?}
\label{sec:massloss:blowoff}
Since the location of the habitable zone is governed primarily by the total (bolometric) flux incident on a planet, the higher ratio of $L_\mathrm{XUV}$ to $L_\mathrm{bol}$ of M dwarfs implies a much larger XUV flux in the HZ compared to solar-type stars. The present-day solar XUV luminosity is $L_\mathrm{XUV}/L_\mathrm{bol} \approx 3.4\times10^{-6}$ \cite[see Table 4 in][]{RIB05}, while for active M dwarfs this ratio is $\sim 10^{-3}$ (e.g., \citeauthor{SCA07} \citeyear{SCA07}). Therefore we should expect planets in the HZ of M dwarfs to experience XUV fluxes several orders of magnitude greater than the present Earth level ($F_\mathrm{XUV_\oplus}\approx 4.64 \unit{erg/s/cm^2}$). Recent papers \citep{LAM07,LAM13,ERK13} show that terrestrial planets experiencing XUV fluxes corresponding to 10$\times$ and 100$\times$ $F_\mathrm{XUV_\oplus}$ are in the hydrodynamic flow regime, and we may thus expect the same for super-Earths/mini-Neptunes in the HZ of active M dwarfs. 

\begin{figure*}[t]
  \begin{center}
      \psfig{file=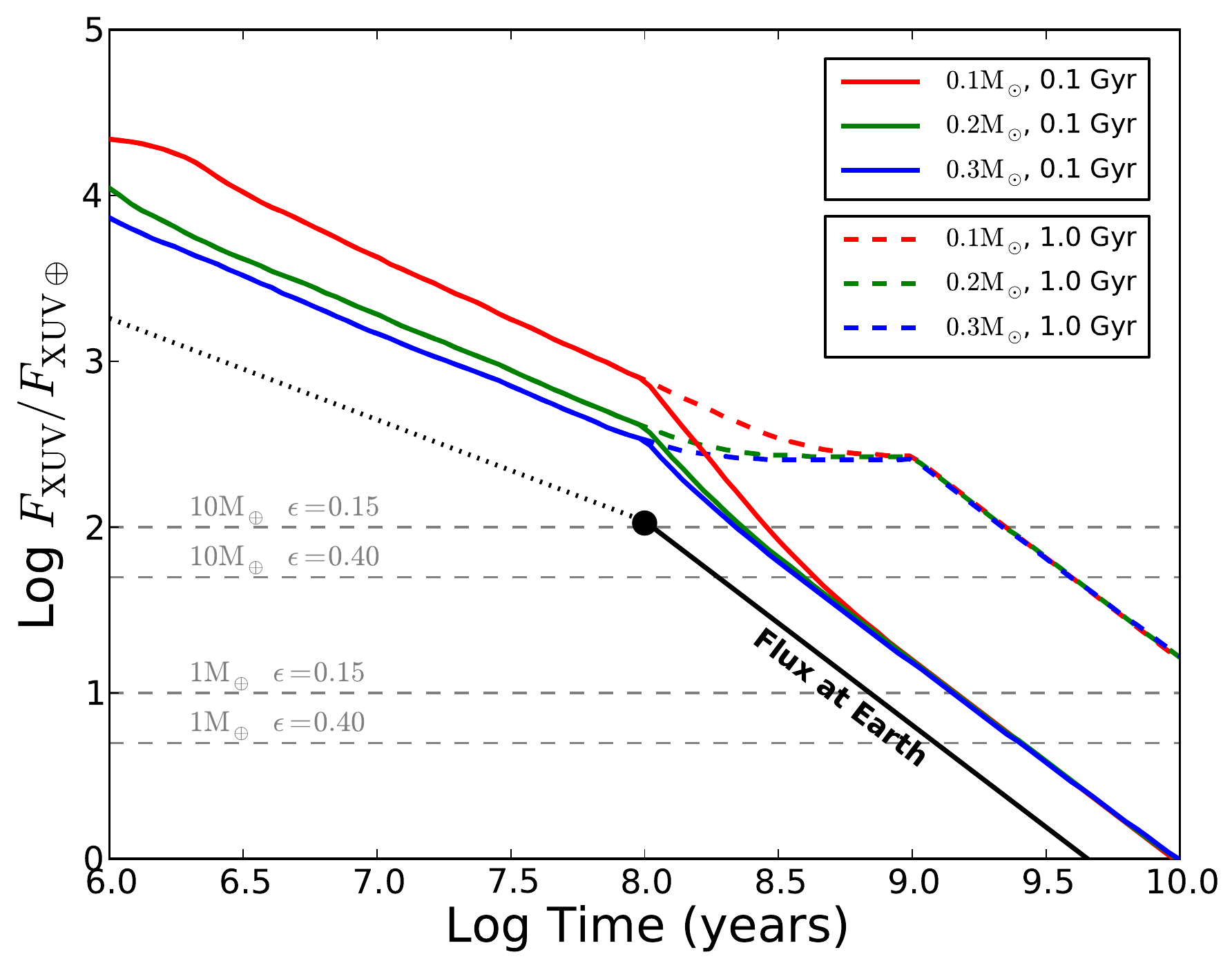,width=5in}
       \caption{Evolution of the XUV flux received by planets close to the inner edge of the HZ (at 1 Gyr) for stars of mass 0.1, 0.2, and 0.3 M$_\odot$. Solid lines correspond to an XUV saturation time of 0.1 Gyr; dashed lines correspond to 1 Gyr. The flux at Earth is indicated by the black line. The dot corresponds to the earliest time for which \cite{RIB05} has data for solar-type stars; this is also roughly the time at which Earth formed. An XUV luminosity saturated at $10^{-3}L_\mathrm{bol}$ is roughly indicated by the dotted line. Finally, the dashed gray lines indicate the minimum XUV fluxes required to sustain blowoff according to the study of \cite{ERK13}. Super-Earths in the HZs of M dwarfs remain in the blowoff regime for at least a few 100 Myr; Earths undergo blowoff for much longer. For an XUV saturation time of 1 Gyr, blowoff occurs for several Gyr for all planets.}
     \label{fig:xuvevol}
  \end{center}
\end{figure*}

In Figure~\ref{fig:xuvevol} we plot the evolution of the XUV flux received by a planet located close to the inner edge of the HZ (defined at 5 Gyr), for three different M dwarf masses and two different XUV saturation times (see \S\ref{sec:lumevol}). The dashed lines correspond to the critical fluxes in \cite{ERK13} above which hydrodynamic escape occurs, for 1 and 10 M$_\oplus$ and two values of $\epsilon_\mathrm{XUV}$. Earth-mass planets remain in the hydrodynamic escape regime for at least 1 Gyr in all cases and in excess of 10 Gyr for active M dwarfs. The duration of this regime is shorter for 10 M$_\oplus$ planets, but still on the order of several 100 Myr.

We note also that tidal effects can significantly increase the critical value of $\lambda_J$ below which hydrodynamic escape occurs \citep[see discussion in][]{ERK07}. Hydrodynamic escape ensues when the thermal energy of the gas exceeds its potential energy, which occurs when
\begin{align}
1.5 \geq \frac{GM_pK_\mathrm{tide}m_H}{kT_\mathrm{exo}R_\mathrm{exo}} \nonumber
\end{align}
or
\begin{align}
\lambda_J &\leq \frac{1.5}{K_\mathrm{tide}} \nonumber \\
		&\equiv \lambda_\mathrm{crit}
\end{align}
provided we maintain the original definition of the Jeans parameter (\ref{eq:jeans}). Due to the strong tidal forces acting on the planets we consider here, this effect should greatly increase the critical value of the escape parameter, effectively reducing the value of $F_\mathrm{XUV}$ for which hydrodynamic escape occurs.

\subsubsection{Energy-Limited or Radiation/Recombination-Limited?}
\label{sec:massloss:rrlim}
Hydrodynamic escape from planetary atmospheres need not be energy-limited. In the limit of high extreme ultraviolet (EUV) flux (low-energy XUV photons with $100 \mathrm{\AA} \lesssim \lambda \lesssim 1000 \mathrm{\AA}$), \cite{MUR09} showed that the escape is ``radiation/recombination-limited,'' scaling roughly as $\dot{M} \propto \left(F_\mathrm{EUV}\right)^{1/2}$. In this regime, the upper atmosphere thermostats to $T \sim 10^4\ \mathrm{K}$, photoionization is balanced by radiative recombination (as opposed to $PdV$ work), and a large fraction of the gas energy budget is lost to Lyman $\alpha$ cooling. The mass loss rate is found by solving the mass continuity equation, yielding
\begin{align}
\label{eq:dmdtgeneral}
\dot{M} = 4\pi r_s^2\rho_s c_s
\end{align}
where $r_s$ is the radius of the sonic point (where the wind velocity equals the local sound speed $c_s$) and $\rho_s$ is the density at $r_s$. Since the bulk of the flow is ionized, the density is fixed by ionization-recombination balance, scaling roughly as $\left(F_\mathrm{EUV}r_s\right)^{1/2}$. For a $0.7M_\mathrm{J}$ hot Jupiter, the radiation/recombination limited mass loss rate is \citep{MUR09}
\begin{align}
\label{eq:dmdtmur09}
\dot{M}_\mathrm{RR} \approx 4\times 10^{12}\ \mathrm{g/s}\left( \frac{F_\mathrm{EUV}}{5\times 10^5\ \mathrm{erg/cm^2/s}} \right)^{1/2}.
\end{align}
\cite{OJ12} re-derive this expression with explicit scalings on several planet properties:
\begin{align}
\label{eq:dmdtoj12}
\dot{M}&_\mathrm{RR} \approx 2.4\times 10^{11}\ \mathrm{g/s}\ \left( 1+x \right)^{3/2}f_\mathrm{parker}\left( \frac{\Phi_\star}{10^{40}\ \mathrm{s}^{-1}} \right)^{1/2} \nonumber \\
&\times \left( \frac{a}{0.1\ \mathrm{AU}} \right) \left( \frac{R_p}{10\rearth} \right)^{3/2}\left( \frac{A}{1/3} \right)^{1/2}\left( \frac{c_\mathrm{EUV}}{10\ \mathrm{km/s}} \right),
\end{align}
where the quantity $(1+x)$ is the radius of the ionization front in units of $R_p$, $f_\mathrm{parker}$ is the Mach number of the flow, $\Phi_\star$ is the stellar EUV luminosity in photons per second, $A$ is a geometrical factor and $c_\mathrm{EUV}$ is the isothermal sound speed of the gas. Taking $x = 0$, $f_\mathrm{parker} = 1$, $A = 1/3$, $c_\mathrm{EUV} = 10\ \mathrm{km/s}$, and an average EUV photon energy $h\nu = 20\ \mathrm{eV}$, this becomes
\begin{align}
\label{eq:dmdtrr}
\dot{M}_\mathrm{RR} \approx 7.11\times 10^{7}\ \mathrm{g/s}\left( \frac{F_\mathrm{EUV}}{\mathrm{erg/cm^2/s}} \right)^{1/2}\left( \frac{R_p}{\rearth} \right)^{3/2}.
\end{align}
The transition from energy-limited to radiation\slash recombination-limited escape is found by equating the two expressions, namely (\ref{eq:dmdt}) and (\ref{eq:dmdtrr}), and solving for the critical EUV flux. For hot Jupiters and mini-Neptunes alike, the transition occurs at roughly $F_\mathrm{EUV} \sim 10^4 \unit{erg/cm^2/s}$. Below this flux, the escape is energy-limited; above it, the escape is radiation\slash recombination-limited and thus increases more slowly with the flux. 

Mini-Neptunes that migrate early into the HZ of M dwarfs are exposed to EUV fluxes up to an order of magnitude larger than this critical value. During this period, which lasts on the order of a few hundred Myr, the mass loss rate may be radiation\slash recombination-limited.

However, whether the high-flux mass loss rate is more accurately described by (\ref{eq:dmdt}) or (\ref{eq:dmdtrr}) will depend on whether the flux is dominated by X-ray or EUV radiation. \cite{OJ12} show that the mass loss rate for X-ray driven hydrodynamic winds scales linearly with the X-ray flux; this is because the sonic point for X-ray flows tends to occur \emph{below} the ionization front. Even though recombination radiation still removes energy from the flow, it does so once the gas is already supersonic and thus causally decoupled from the planet, such that it cannot bottleneck the escape. Although the authors caution that the dependence of the mass loss rate on planet mass and radius must be determined numerically, this regime is analogous to the energy-limited regime and can be roughly approximated by (\ref{eq:dmdt}).

For X-ray luminosities $L_\mathrm{X} \gtrsim 10^{-6}\lsun$, close-in planets undergo X-ray driven hydrodynamic escape \citep[see Figure 11 in][]{OJ12}. If X-rays contribute significantly to the XUV emissions of young M dwarfs, their X-ray luminosities may exceed this value for as long as 1 Gyr, and close-in mini-Neptunes will undergo energy-limited escape. Unfortunately, given the lack of observational constraints on the X-ray/EUV luminosities of young M dwarfs, it is unclear at this point whether the hydrodynamic escape will be EUV-driven (radiation\slash recombination-limited) or X-ray-driven (energy-limited).

\subsubsection{The Effect of Eccentricity}
\label{sec:massloss:ecc}
Most of the formalism that has been developed to analytically treat hydrodynamic blow-off only considers circular orbits. For planets on sufficiently eccentric orbits, neither the stellar flux nor the tidal effects may be treated as constant over the course of an orbit. Due to this fact, an expression analogous to (\ref{eq:dmdt}) for eccentric orbits seems to be lacking in the literature. In this section we derive such an expression.

There are two separate effects that enhance the mass loss for planets on eccentric orbits. Most papers account for the first effect, which is the increase of the orbit-averaged stellar flux by a factor of $1/\sqrt{1-e^2}$
\citep[see, for instance,][]{KOP13}. However, for $e \lesssim 0.3$, this effect is quite small. The second, more important effect is that the Roche lobe radius is no longer constant over the course of an orbit, and (\ref{eq:roche}) is not valid. Instead, we must replace $a$ with the instantaneous planet-star separation $r(t)$:
\begin{align}
\label{eq:rocheoft}
R_\mathrm{Roche}(t) = \left(\frac{M_p}{3M_\star}\right)^{1/3}r(t).
\end{align}
One might wonder whether this replacement is valid. Specifically, if $R_\mathrm{Roche}(t)$ changes faster than the atmosphere is able to respond to the changes in the gravitational potential, we would expect that the time dependence of the mass loss rate would be a complicated function of the tides generated in the atmosphere. On the other hand, if the orbital period is very large compared to the dynamical timescale of the planet, the atmosphere will have sufficient time to assume the equilibrium shape dictated by the new potential. This limit is known as the \emph{quasi-static approximation} \citep{SEP07}. In the Appendix, we show that all the planets in our runs with eccentricities $e \lesssim 0.4$ are in the quasi-static regime and that (\ref{eq:rocheoft}) is therefore valid. In some runs, we allow the eccentricity to increase beyond 0.4. We discuss the implications of this in \S\ref{sec:discuss:caveats}.

Since $R_\mathrm{Roche} = R_\mathrm{Roche}(t)$, $\xi$, $K_\mathrm{tide}$, and $dM/dt$ (Equations~\ref{eq:dmdt}, \ref{eq:ktide} and \ref{eq:xi}) are now also functions of $t$, varying significantly over a single orbit. To account for this, we may calculate the time-averaged mass loss rate over the course of one orbit, $\avg{\dot{M}}_t$, such that the total amount of mass lost in time $\Delta t$ is $\Delta M = \avg{\dot{M}}_t \Delta t$. To this end, in the Appendix we derive the eccentric version of $K_\mathrm{tide}$:

\begin{align}
\label{eq:kecc}
\frac{1}{K_\mathrm{ecc}} \equiv \frac{1}{2\pi}\int_0^{2\pi} \left[(1-e\cos E) - \frac{3}{2\xi} + \frac{1}{2\xi^3(1-e\cos E)^2}\right]^{-1}dE
\end{align}
where $\xi$ is the time-independent parameter given by (\ref{eq:xi}) and (\ref{eq:roche}) and $E$ is the eccentric anomaly. The average mass loss rate is then simply
\begin{align}
\label{eq:dmdtavg}
\avg{\dot{M}}_t = \frac{\dot{M}_0}{K_\mathrm{ecc}},
\end{align}
where 
\begin{align}
\dot{M}_0 &\equiv \frac{R_\mathrm{XUV}^3\exuv L_\mathrm{XUV}}{4GM_pa^2}
\end{align}
is the zero-eccentricity mass loss rate in the absence of tidal enhancement. Note that the flux-enhancement factor $1/\sqrt{1-e^2}$ is already folded into $K_\mathrm{ecc}$, since it must be incorporated when integrating (\ref{ap:eq:dmdt}). 

For $\xi \gtrsim 10$, the integral may be approximated by the analytic expression
\begin{align}
\label{eq:keccapp}
K_\mathrm{ecc} \approx \sqrt{1- \frac{3}{\xi_0} - \frac{9}{4\xi_0^2} -e^2},
\end{align}
which greatly reduces computing time.

As we show in the Appendix, the decreased Roche lobe distance for eccentric orbits has a large effect on the amount of mass lost, particularly for low values of $\xi$ and for high $e$ (see Figure~\ref{fig:kecc}). Moreover, higher eccentricities result in Roche lobe overflow at larger values of $a$ compared to the circular case, since the planet may overflow near pericenter, leading to mass loss rates potentially orders of magnitude higher. We discuss RLO in detail in the Model Description section (\S\ref{sec:model:escape}).
 
\subsection{Tidal Theory}
\label{sec:tides}
The final aspect of planetary evolution we discuss is the effect of tidal interactions with the host star. Below we review two different approaches to analytically calculate the orbital evolution of the system.

\subsubsection{Constant Phase Lag}
\label{sec:tides:cpl}
Classical tidal theory predicts that torques arising from interactions between tidal deformations on a planet and its host star lead to the secular evolution of the orbit and the spin of both bodies. In this paper we employ the ``equilibrium tide'' model of \cite{DAR80}, which approximates the tidal potential as a superposition of Legendre polynomials representing waves propagating along the surfaces of the bodies; these add up to give the familiar tidal ``bulge.'' Because of viscous forces in the bodies' interiors, the tidal bulges do not instantaneously align with the line connecting the two bodies. Instead, the $N^{th}$ wave on the $i^{th}$ body will lag or lead by an angle $\varepsilon_{N,i}$, assumed to be constant in the \emph{constant phase lag} (CPL) model. In general, different waves may have different lag angles, and it is unclear how the $\varepsilon_{N,i}$ vary as a function of frequency. A common approach \citep[see][]{FER08} is to assume that the magnitudes of the lag angles are equal \citep[see][]{GS66}, while their signs may change depending on the orbital and rotational frequencies involved. This allows us to introduce the \emph{tidal quality factor}
\begin{align}
\label{eq:Q}
Q_i &\equiv \frac{1}{\varepsilon_{0,i}},
\end{align}
which in turn allows us to express the lags (in radians) as
\begin{align}
\varepsilon_{N,i} &= \pm\frac{1}{Q_i}.
\end{align}
The parameter $Q_i$ is a measure of the dissipation within the $i^{th}$ body; it is inversely proportional to the amount of orbital and rotational energy lost to heat per cycle, in analogy with a damped-driven harmonic oscillator. The merit of this approach is that the tidal response of a body can be captured in a single parameter. Planets with high values of $Q_p$ have smaller phase lags, dissipate less energy and undergo slower orbital evolution; planets with low values of $Q_p$ have larger phase lags, higher dissipation rates, and therefore faster evolution. Measurements in the solar system constrain the value of $Q_p$ for terrestrial bodies in the range 10-500, while gas giants are consistent with $Q_p \sim 10^4-10^5$ \citep{GS66}. Values of $Q_\star$ for the Sun and other main sequence stars are poorly constrained but are likely to be $\gtrsim 10^5-10^6$ \citep{SCH10,PEN12}. Intuitively, this makes sense, given that the dissipation due to internal friction in rocky bodies should be much higher than that in bodies dominated by gaseous atmospheres. One should bear in mind, however, that the exact dependence of $Q_i$ on the properties of a body's interior is likely to be extremely complicated. Given the dearth of data on the composition and internal structure of exoplanets, it is at this point impossible to infer precise values of $Q_p$ for these planets. 

By calculating the forces and torques due to the tides raised on both the planet and the star, one can arrive at the secular expressions for the evolution of the planet's orbital parameters, which are given by a set of coupled nonlinear differential equations; these are reproduced in the Appendix. 

\subsubsection{Constant Time Lag}
\label{sec:tides:ctl}
Unlike the CPL model, which assumes the phase lag of the tidal bulge is constant, the \emph{constant time lag} (CTL) model assumes that it is the time interval between the bulge and the passage of the perturbing body that is constant. Originally proposed by \cite{ALE73} and updated by \cite{LECON10}, this model allows for a continuum of tidal wave frequencies and therefore avoids unphysical discontinuities present in the CPL model. However, implicit in the CTL theory is the assumption that the lag angles are directly proportional to the driving frequency \citep{GRE09}, which is likely also an oversimplification. We note, however, that in the low eccentricity limit, both the CPL and the CTL models arrive at qualitatively similar results. At higher eccentricities, the CTL model is probably better suited, given that it is derived to eighth order in $e$ (versus second order in the CPL model).

The tidal quality factors $Q_i$ do not enter the CTL calculations at any point; instead, the dissipation is characterized by the time lags $\tau_i$. Although there is no general conversion between $Q_i$ and $\tau_i$,
\cite{LECON10} show that provided annual tides dominate the evolution,
\begin{align}
\tau_i \approx \frac{1}{nQ_i},
\end{align}
where $n$ is the mean motion (or the orbital frequency) of the secondary body (in this case, the planet).

For a planet with $Q_p = 10^4$ in the center of the HZ of a late M dwarf, $\tau_p \approx 10$ s; rocky planets with lower $Q_p$ may have values on the order of hundreds of seconds. Since $\tau \propto n^{-1}$, close-in planets should have much lower time lags. For reference, \cite{LECON10} argue that hot Jupiters should have $2\times 10^{-3} \unit{s} \lesssim \tau_p \lesssim 2\times 10^{-2} \unit{s}$.

The tidal evolution expressions are reproduced in the Appendix. For a more detailed review of tidal theory, the reader is referred to \cite{FER08}, \cite{HLB11}, and the Appendices in \cite{BAR13}.

\section{Model Description}
\label{sec:model}
Our model evolves planet-star systems forward in time in order to determine whether HECs can form from mini-Neptunes that have migrated into the HZs of M dwarfs. We perform our calculations on a grid of varying planetary, orbital, and stellar properties in order to determine the types of systems that may harbor HECs. The complete list is provided in Table~\ref{tab:params}, where we indicate the ranges of values we consider as well as the default values adopted in the plots in \S\ref{sec:results} (unless otherwise indicated).

\begin{deluxetable}{lccc}
\tabletypesize{\small}
\tablecaption{Free Parameters and Their Ranges \label{tab:params}}
\tablewidth{0pt}
\tablehead{
\colhead{Parameter} & \colhead{Range} & \colhead{Default} & \colhead{Notes}
}
\startdata
$\mstar (\msun)$ & $0.08-0.4$ & - & Late-mid MD\\
$M_p (\mearth)$ & $1 - 10$ & - & -\\
$R_\mathrm{XUV} (R_p)$ & $1.0-1.2$ & 1.2 & See \S\ref{sec:massloss:hydro}\\
$a$ & IHZ - OHZ & - & See \S\ref{sec:model:stellar}\\
$e$ & 0.0 - 0.95 & - & -\\
$P_{0,\star}$ (days) & $1.0 - 100$ & 30.0 & Initial rot. per.\\
$f_H$ & $10^{-6} - 0.5$ & - & H mass fraction\\
$\exuv$ & $0.1 - 0.4$ & 0.3 & -\\
$\xi_\mathrm{min}$ & $1+10^{-5} - 3$ & 3 & See \S\ref{sec:model:escape}\\
Atmos. esc. & R/R-Lim / E-Lim & - & See \S\ref{sec:model:escape} \\
Tidal model & CPL/CTL & CTL & -\\
$Q_\star$ & $10^5 - 10^6$ & $10^5$ & CPL only\\
$Q_p$ & $10^1 - 10^5$ & $10^4$ & CPL only\\
$\tau_\star$ (s) & $10^{-2} - 10^{-1}$ & $10^{-1}$ & CTL only\\
$\tau_p$ (s) & $10^{-3} - 10^3$ & $10^{-1}$ & CTL only\\
$\beta$ & $0.7 - 1.23$ & 1.23 & See Eq. (\ref{eq:lxuvevol})\\
$t_\mathrm{sat}$ (Gyr) & $0.1 - 1.0$ & 1.0 & XUV sat. time\\
$t_0$ (Myr) & $10.0 - 100.0$ & 10.0 & Integration start\\
$t_\mathrm{stop}$ (Gyr) & $0.01 - 5.0$ & 5.0 & Integration end
\enddata \vspace*{0.1in}
\end{deluxetable}
Integrations are performed from $t = t_0$ (the time at which the planet is assumed to have migrated into the HZ) to $t = t_\mathrm{stop}$ (the current age of the system) using the adaptive timestepping method described in Appendix E of \cite{BAR13}.

\subsection{Stellar Model}
\label{sec:model:stellar}
We use the evolutionary tracks of \cite{BAR98} for solar metallicity to calculate $L_\mathrm{bol}$ and $T_\mathrm{eff}$ as a function of time. We then use (\ref{eq:lxuvevol}) to calculate $L_\mathrm{XUV}$, given $(L_\mathrm{XUV}/L_{\mathrm{bol}})_\mathrm{sat} = 10^{-3}$ and values of $t_\mathrm{sat}$ and $\beta$ given in Table~\ref{tab:params}.

Using $L_\mathrm{bol}$ and $T_\mathrm{eff}$, we calculate the location of the HZ from the equations given in \cite{KOP13}, adding the eccentricity correction (\ref{eq:savg}). Given the uncertainty in the actual HZ boundaries and their dependence on a host of properties of a planet's climate, we choose our inner edge (IHZ) to be the average of the Recent Venus and the Runaway Greenhouse limits and our outer edge (OHZ) to be the average of the Maximum Greenhouse and the Early Mars limits. Throughout this paper we will also refer to the center of the HZ (CHZ), which we take to be the average of the IHZ and OHZ. Since we are concerned with the formation of ultimately habitable planets, we take the locations of the IHZ, CHZ, and OHZ to be their values at 1 Gyr, at which point the stellar luminosity becomes roughly constant.

\subsection{Planet Radius Model}
\label{sec:model:rad}
\begin{figure}[h]
  \begin{center}
    \leavevmode
      \psfig{file=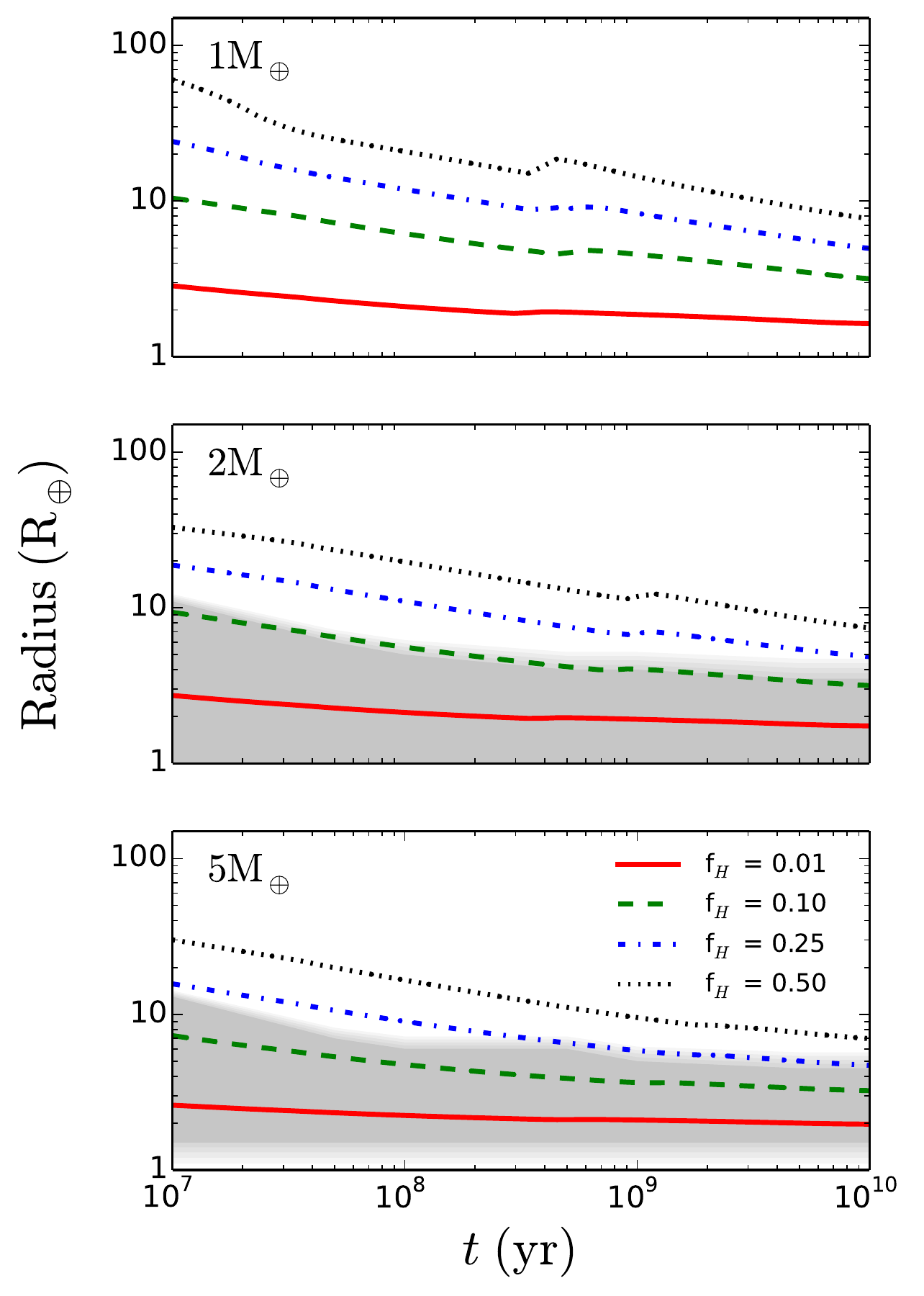,width=\columnwidth}
       \caption{Evolution of the radius as a function of time due to thermal contraction of the envelope, in the absence of tidal effects and atmospheric mass loss. From top to bottom, the plots correspond to planets with initial total masses (core + envelope) of 1, 2, and 5 \mearth. Line styles correspond to different initial hydrogen mass fractions: 1$\%$ (red, solid), 10$\%$ (green, dashed), 25$\%$ (blue, dot-dashed), and 50$\%$ (black, dotted). For comparison, the grey shaded regions in the bottom two plots are the spread in radii calculated by \cite{MOR12b} for $f_H \lesssim 0.20$. See text for a discussion.}
     \label{fig:radius}
  \end{center}
\end{figure}

To determine the planetary radius $R_p$ as a function of the core mass $M_c$, the envelope mass fraction $f_H \equiv M_e/M_p$, and the planet age, we use the planet structure model described in \cite{LOP12} and \cite{LF14}, which is an extension of the model of \cite{FMB07} to low-mass low-density (LMLD) planets. These models perform full thermal evolution calculations of the interior as a function of time. In our runs, the core is taken to be Earth-like, with a mixture of 2/3 silicate rock and 1/3 iron, and the envelope is modeled as a H/He adiabat. A grid of values of $R_p$ is then computed in the range $1 \mearth \leq M_c \leq 10 \mearth$, $10^{-6} \leq f_H \leq 0.5$, and $10^7 \unit{years} \leq t \leq 10^{10} \unit{years}$.  For values between grid points, we perform a simple trilinear interpolation. For gas-rich planets, $R_p$ is the 20 mbar radius; for gas-free planets, it corresponds to the surface radius. The evolution of $R_p$ with age due solely to thermal contraction is plotted in Figure~\ref{fig:radius} for a few different core masses and values of $f_H$. 

We note that the models of \cite{FMB07} and \cite{LOP12} are in general agreement with those of \cite{MOR12a,MOR12b} and, by extension, \cite{ROG11}. \cite{MOR12a} presented a validation of their model against that of \cite{FMB07}, showing that for planets spanning 0.1 to 10 Jupiter masses, the two models predict the same radius to within a few percent. At the lower masses relevant to our study, the two models are also in agreement. To demonstrate this, in Figure \ref{fig:radius} we shade the regions corresponding to the spread in radii at a given mass and age in Figure 9 of \cite{MOR12b}. Since those authors used a coupled formation/evolution code, at low planet mass the maximum envelope mass fraction $f_H$ is small; for a total mass of 4$\mathrm{M}_\oplus$, \cite{MOR12b} find that all planets have $f_H < 0.2$. At 2$\mathrm{M}_\oplus$, most planets have $f_H \lesssim 0.1$. We can see from Figure \ref{fig:radius} that at these values of $f_H$, the two models predict very similar radius evolution. Note that \cite{MOR12b} did not consider planets less massive than 2$\mathrm{M}_\oplus$.

The maximum envelope fraction merits further discussion. Since we do not model the formation of mini-Neptunes, we do not place a priori constraints on the value of $f_H$ at a given mass; instead, we allow it to vary in the range $10^{-6} \leq f_H \leq 0.5$ for all planet masses. At masses $\lesssim 5 M_\oplus$, planets accumulate gas slowly and are typically unable to accrete more than $\sim$ 10-20\% of their mass in H/He \citep{ROG11,BH14}; values of $f_H \approx 0.5$ may thus be unphysical. However, as we argue in \S\ref{sec:discuss:howto}, the longer disk lifetimes around M dwarfs \citep{CAR06,PAS09} allow more time for gas accretion, potentially increasing the maximum value of $f_H$. Nevertheless, and more importantly, if a planet with $f_H = 0.5$ loses its entire envelope via atmospheric escape, \emph{any planet with the same core mass and $f_H < 0.5$ will also lose its envelope}. Below, where we present integrations with $f_H = 0.5$, our results are therefore conservative, as planets with $f_H \ll 0.5$ will in general evaporate more quickly.

While our treatment of the radius evolution is an improvement upon past tidal-atmospheric coupling papers \citep[][for instance, calculate $R_p$ for super-Earths by assuming a constant density as mass is lost]{JAC10}, there are still issues with our approach: (1) We do not account for inflation of the radius due to high insolation. Instead, we calculate our radii from grids corresponding to a planet receiving the same flux as Earth. While at late times this is justified, since planets in the HZ by definition receive fluxes similar to Earth, at early times this is probably a poor approximation; recall that planets in the HZ around low mass M dwarfs are exposed to fluxes up to two orders of magnitude higher during the host star's pre-main sequence phase. The primary effect of a higher insolation is to act as a blanket, delaying the planet's cooling and causing it to maintain an inflated radius for longer. This will result in mass loss rates higher than what we calculate here. (2) Since we are determining the radii from pre-computed grids, we also do not model the effect of tidal dissipation on the thermal evolution of the planet. Planets undergoing fast tidal evolution can dissipate large amounts of energy in their interiors, which should lead to significant heating and inflation of their radii. (3) The radius is also likely to depend on the mass loss rate. Setting $R_p$ equal to the tabulated value for a given mass, age, and composition is valid only as long as the timescale on which the planet is able to cool is significantly shorter than the mass loss timescale. Otherwise, the radius will not have enough time to adjust to the rapid loss of mass and the planet will remain somewhat inflated, leading to a regime of runaway mass loss \citep{LOP12}. While the planets considered here are probably not in the runaway regime (\cite{LOP12} found that runaway mass loss occurred only for H/He mass fractions $\gtrsim 90\%$), we might still be significantly underestimating the radii during the early active phase of the parent star.

All points outlined above lead to an \emph{underestimate} of the radius at a given time. Since the mass loss rate  is proportional to $R_p^3$ (\ref{eq:dmdt}) or $R_p^{3/2}$ (\ref{eq:dmdtrr}), calculating the radius in this fashion leads to a \emph{lower bound} on the amount of mass lost and on the strength of the coupling to tidal effects. Because our present goal is to determine whether it is possible to form habitable evaporated cores via this mechanism, this conservative approach is sufficient. Future work will incorporate a self-consistent thermal structure model to better address the radius evolution.

\subsection{Atmospheric Escape Model}
\label{sec:model:escape}
We assume that the escape of H/He from the planet atmosphere is hydrodynamic (blow-off) at all times, which is valid at the XUV fluxes we consider here \citep[see][and Figure~\ref{fig:xuvevol}]{ERK13}. We run two separate sets of integrations: one in which we assume the flow is energy-limited (\ref{eq:dmdt}) for all values of $F_\mathrm{XUV}$, and one in which we switch from energy-limited to radiation\slash recombination-limited (\ref{eq:dmdtrr}) above the critical value of the flux (see \S\ref{sec:massloss:rrlim}). For planets whose orbits are eccentric enough that they switch between the two regimes over the course of one orbit, we make use of the expressions derived in \S\ref{ap:sec:crossing} in the Appendix. These two sets of integrations should roughly bracket the true mass loss rate.

For eccentric orbits, we calculate the mass loss in the energy-limited regime from (\ref{eq:dmdtavg}), with $K_\mathrm{ecc}$ evaluated from (\ref{eq:kecc}). We vary $\epsilon_\mathrm{XUV}$ and $R_\mathrm{XUV}$ in the ranges given in Table~\ref{tab:params}. We choose $\epsilon_\mathrm{XUV} = 0.30$ as our default case. While this is consistent with values cited in the literature (see \S\ref{sec:massloss:hydro}), it could be an overestimate. We discuss the implications of this choice in \S\ref{sec:discuss:constraints}.

Given the large planetary radii at early times, many of the planets we model here are not stable against Roche lobe overflow in the HZ. During RLO, the stellar gravity causes the upper layers of the atmosphere to  suddenly become unbound from the planet; this occurs when $R_\mathrm{p} > R_\mathrm{Roche}$, where $R_\mathrm{Roche}$ is given by (\ref{eq:roche}). For a planet that forms and evolves \emph{in situ}, RLO never occurs, since any gas that would be lifted from the planet in this fashion would have never accreted in the first place. However, an inflated gaseous planet that forms at a large distance from the star may initially be stable against overflow and enter RLO as it migrates inwards (since $R_\mathrm{Roche} \propto a$). This is particularly the case for planets in the HZs of M dwarfs, since $a$ and consequently $R_\mathrm{Roche}$ are small.

Ideally, the tidally-enhanced mass loss rate equation (\ref{eq:dmdt}) should capture this process, but instead it predicts an infinite mass loss rate as $R_\mathrm{XUV} \rightarrow R_\mathrm{Roche}$ (or as $\xi \rightarrow 1$) and unphysically changes sign for $\xi < 1$. This is due to the fact that the energy-limited model implicitly assumes that the bulk of the atmosphere is located at $R_\mathrm{XUV}$ (the single-layer assumption). Realistically, we would expect the planet to quickly lose any mass above the Roche lobe and then return to the stable hydrodynamic escape regime. However, upon loss of the material above $R_\mathrm{Roche}$, the portion of the envelope below the new XUV absorption radius $R_\mathrm{XUV}'$ will not be in hydrostatic equilibrium; instead, an outward flow will attempt to redistribute mass to the evacuated region above, leading to further overflow. 

Several models exist that allow one to calculate the mass loss rate due to RLO \citep[e.g.,][]{RIT88,TRI98,GLB03,SEP07}. These often involve calculating the angular momentum exchange between the outflowing gas and the planet, which can lead to its outward migration, given by
\begin{align}
	\label{eq:dadtrecoil}
	\frac{1}{a}\frac{da}{dt} = -\frac{2}{M_p}\frac{dM_p}{dt},
\end{align}
for a planet on a circular orbit \citep{GLB03,CHA10}. This leads to a corresponding increase in $R_\mathrm{Roche}$ until it reaches $R_\mathrm{XUV}$ and the overflow is halted. By differentiating the stability criterion $R_\mathrm{XUV}(M_p) = R_\mathrm{Roche}(M_p)$, one may then obtain an approximate expression for $dM_p/dt$ in terms of the density profile $dM(<R)/dR$ of the envelope. 

However, for mini-Neptunes that migrate into the HZ early on, RLO should occur \emph{during} the initial migration process, which we do not model in this paper. Instead, we begin our calculations by assuming that our planets are stable to RLO in the HZ. If a planet's radius initially exceeds the Roche lobe radius, we set its envelope mass equal to the \emph{maximum} envelope mass for which it can be stable at its current orbit; the difference between the two envelope masses is the amount of H/He it must have lost prior to its arrival in the HZ. It is important to note that these planets will initially have $R_\mathrm{XUV} = R_\mathrm{Roche}$, which as we mentioned above, leads to an infinite mass loss rate in (\ref{eq:dmdt}). An accurate determination of $\dot{M}_p$ in this case probably requires hydrodynamic simulations. However, the mass loss rate can be approximated by imposing a minimum value $\xi_\mathrm{min}$ in (\ref{eq:dmdt}). For $\xi < \xi_\mathrm{min}$, we set the mass loss rate equal to $\dot{M}_p(\xi = \xi_\mathrm{min})$. This is equivalent to imposing a maximum mass loss enhancement factor $1/K_\mathrm{tide}$, preventing the mass loss rate from reaching unphysical values as $R_\mathrm{XUV} \rightarrow R_\mathrm{Roche}$.

\begin{figure}[h]
  \begin{center}
      \psfig{file=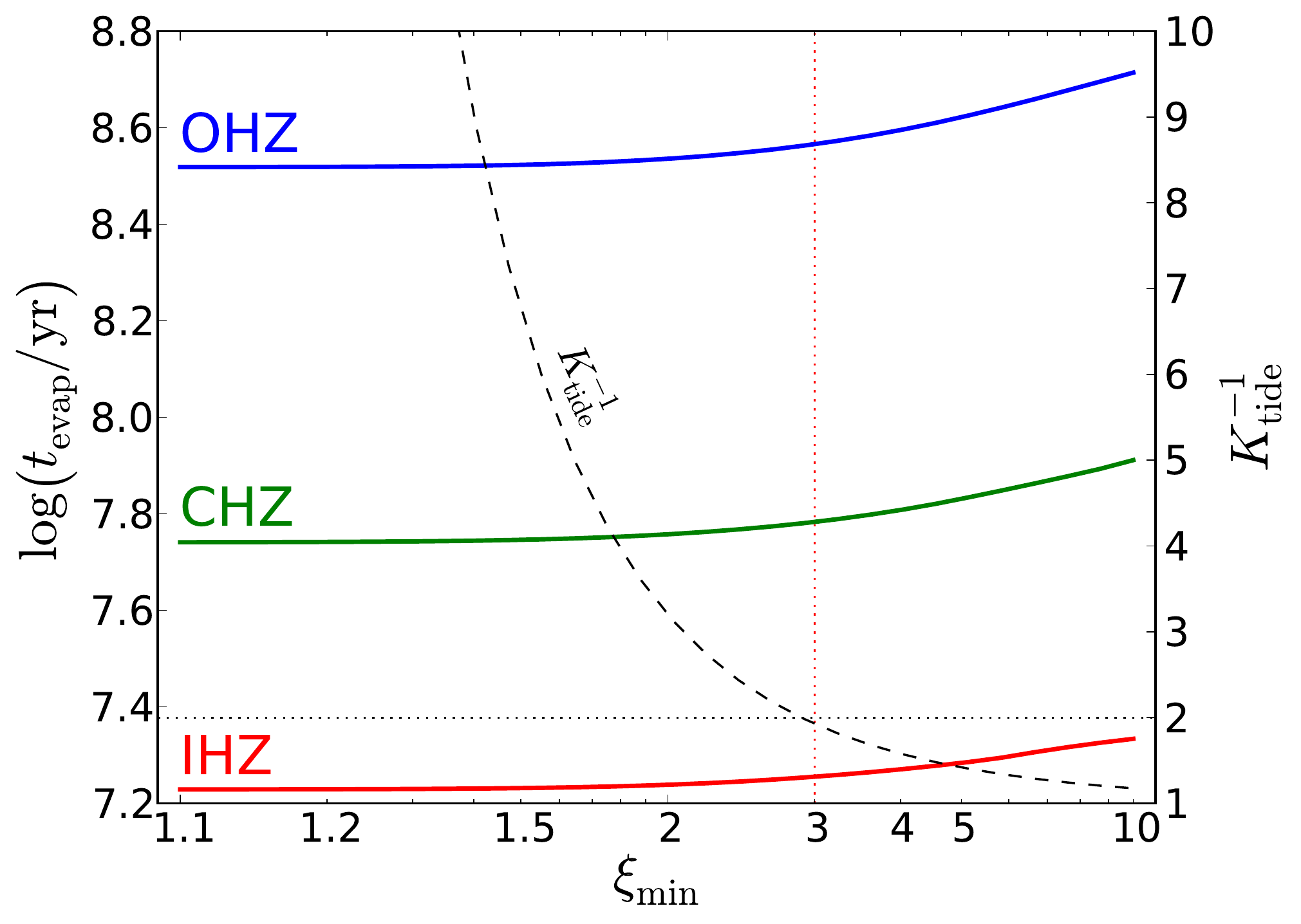,width=\columnwidth}
       \caption{Complete evaporation time $t_\mathrm{evap}$ as a function of the cutoff value $\xi_\mathrm{min}$ for a $2\mearth$ mini-Neptune with $f_{H_0} = 0.5$ on a circular orbit around a $0.08\msun$ star. The red, green, and blue lines correspond to planets in the IHZ, CHZ, and OHZ, respectively. Also plotted is the mass loss enhancement factor $1/K_\mathrm{tide}$ (black dashed line), which approaches infinity as $\xi \rightarrow 1$. Note that for $\xi_\mathrm{min} \lesssim 3$, the evaporation time is relatively insensitive to the exact cutoff value, despite the fact that $1/K_\mathrm{tide}$ blows up. We therefore choose $\xi_\mathrm{min} = 3$ as the default cutoff, corresponding to a maximum enhancement factor $1/K_\mathrm{tide} \approx 2$.}
     \label{fig:minxi}
  \end{center}
\end{figure}

In Figure~\ref{fig:minxi} we show how the time $t_\mathrm{evap}$ at which complete evaporation takes place scales with $\xi_\mathrm{min}$ for a typical mini-Neptune in the IHZ (red), CHZ (green), and OHZ (blue). Also plotted is the mass loss enhancement factor $1/K_\mathrm{tide}$ (Equation~\ref{eq:ktide}, black dashed line) as a function of $\xi = \xi_\mathrm{min}$. Interestingly, despite the fact that the instantaneous mass loss rate ($\dot{M}_p \propto 1/K_\mathrm{tide}$) approaches infinity as $\xi \rightarrow 1$, the evaporation time is relatively constant for $\xi_\mathrm{min} \lesssim 3$. This is because for very large $\dot{M}_p$, the planet loses sufficient mass in an amount of time $\Delta t$ to decrease $R_p$ substantially and terminate the overflow. In other words, cases in which $\xi \approx 1$ are very unstable, and as mass is lost $\xi$ will quickly increase beyond $\sim 3$. Both the net amount of mass lost and the evaporation time are therefore insensitive to the particular value of $\xi_\mathrm{min}$, provided it is less than about 3. We therefore choose $\xi_\mathrm{min} = 3$ as the default value for our runs, noting that this corresponds to a maximum mass loss enhancement of $1/K_\mathrm{tide} \approx 2$.

\subsection{Tidal Model}
We calculate the evolution of the semi-major axis, the eccentricity and the rotation rates from (\ref{eq:dadt})-(\ref{eq:domdt}) and (\ref{eq:dadtctl})-(\ref{eq:domdtctl}) in the Appendix for the CPL and CTL models, respectively. For simplicity, we set the obliquities of all our planets to zero. Since the tidal locking timescale is very short for close-in planets\footnote{Tidal locking refers to the state in which a body's rotation rate is fixed by tidal forces at an equilibrium value. While for circular orbits, this implies $\omega_i = n$, in the general case of an eccentric orbit in the CPL model, the planet's rotation rate assumes a slightly super-synchronous value. See \cite{BAR13} for a discussion.} \citep{GLB03}, we  assume that the planet's rotation rate is given by the equilibrium value (\ref{eq:omeqcpl}) or (\ref{eq:omeqctl}). 

We calculate the stellar spin by assuming different initial periods (see Table~\ref{tab:params}) and evolving it according to the tidal equations, while enforcing conservation of angular momentum as the star contracts during the pre-main sequence phase. We neglect the effects of rotational braking \citep{SKU72}, whereby stars lose angular momentum to winds and spin down over time. This leads to an overestimate of the spin rate at later times, but tidal effects should only be important early on, when the radii and the eccentricity are higher. \cite{BOL12} recently modeled the coupling between stellar spin and tides, following the evolution of a ``slow rotator'' star ($P_0 \approx 10$ days) and a ``fast rotator'' star ($P_0 \approx 1$ day). In both cases, the stars sped up during the first $\sim 300-500$ Myr, after which time angular momentum loss became significant. However, tidal evolution is orders of magnitude weaker at such late times, so we would expect rotational braking to have a minimal effect on the tidal evolution. Moreover, as we show in the Appendix, in general it is the tide raised on the planet that dominates the evolution; as this depends on the \emph{planetary} rotation rate, and not on the stellar rotation rate, our results are relatively insensitive to the details of the spin evolution of the star.

In the CPL model, we adopt typical gas giant values $10^4 \leq Q_p \leq 10^5$ for gas-rich mini-Neptunes and typical terrestrial values $10 \leq Q_p \leq 500$ for planets that have lost their envelopes; we assume stellar values in the range $10^5 \leq Q_\star \leq 10^6$. In the CTL model, we consider time lags in the range $10^{-3} \unit{s} \leq \tau_p \leq 10^{1} \unit{s}$ for gas-rich mini-Neptunes and $10^{-1} \unit{s} \leq \tau_p \leq 10^{3} \unit{s}$ once they lose their envelopes. Following \cite{LECON10}, we consider stellar time lags in the range $10^{-2} \unit{s} \leq \tau_\star \leq 10^{-1} \unit{s}$.

Once a mini-Neptune loses all of its atmosphere, we artificially switch $Q_p$ or $\tau_p$ to the terrestrial value adopted in that run. In reality, as the atmosphere is lost, the transition from high to low $Q_p$ (or low to high $\tau_p$) should be continuous. A detailed treatment of the dependence of $Q_p$ and $\tau_p$ on the envelope mass fraction is deferred to future work.

Finally, we note that the second-order CPL model described above is valid only at low eccentricity. For $e \geq \sqrt{1/19} \approx 0.23$, the phase lag of the dominant tidal wave discontinuously changes from negative to positive, such that the model then predicts outward migration due to the planetary tide. This effect is unphysical, stemming from the fact that the CPL model considers only terms up to second order in the eccentricity \citep[for a detailed discussion of this, see][]{LECON10}. We therefore restrict all our calculations in the CPL framework to values of the eccentricity $e \leq 0.2$. For higher values of $e$, we use the higher-order CTL model.


	\begin{figure*}[t]
        \centering   \subfigure[Energy-Limited]{\label{fig:int0}\includegraphics[width=2.1in]{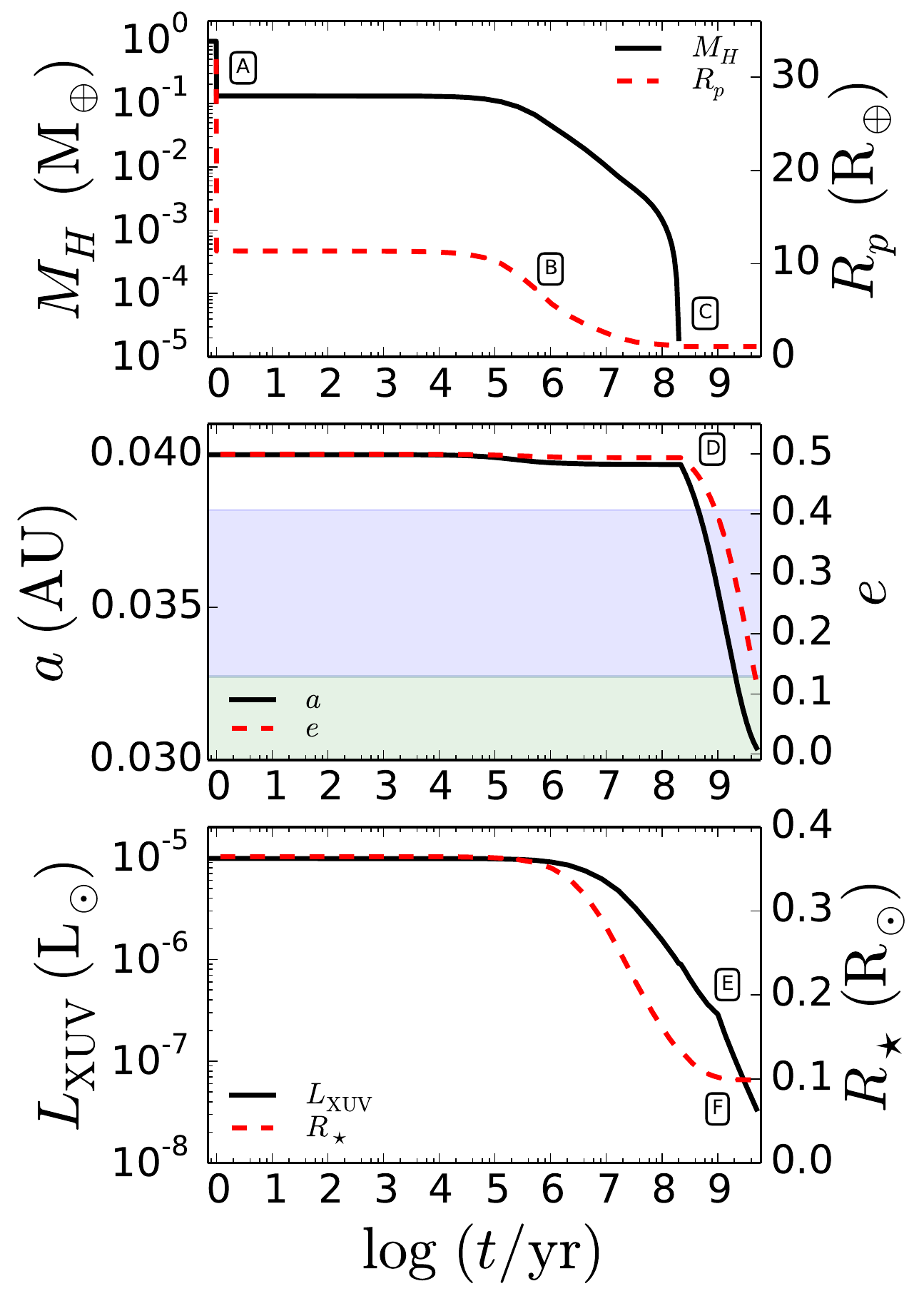}}
        \subfigure[Energy-Limited, $t_0 = 100$ Myr]{\label{fig:int1}\includegraphics[width=2.1in]{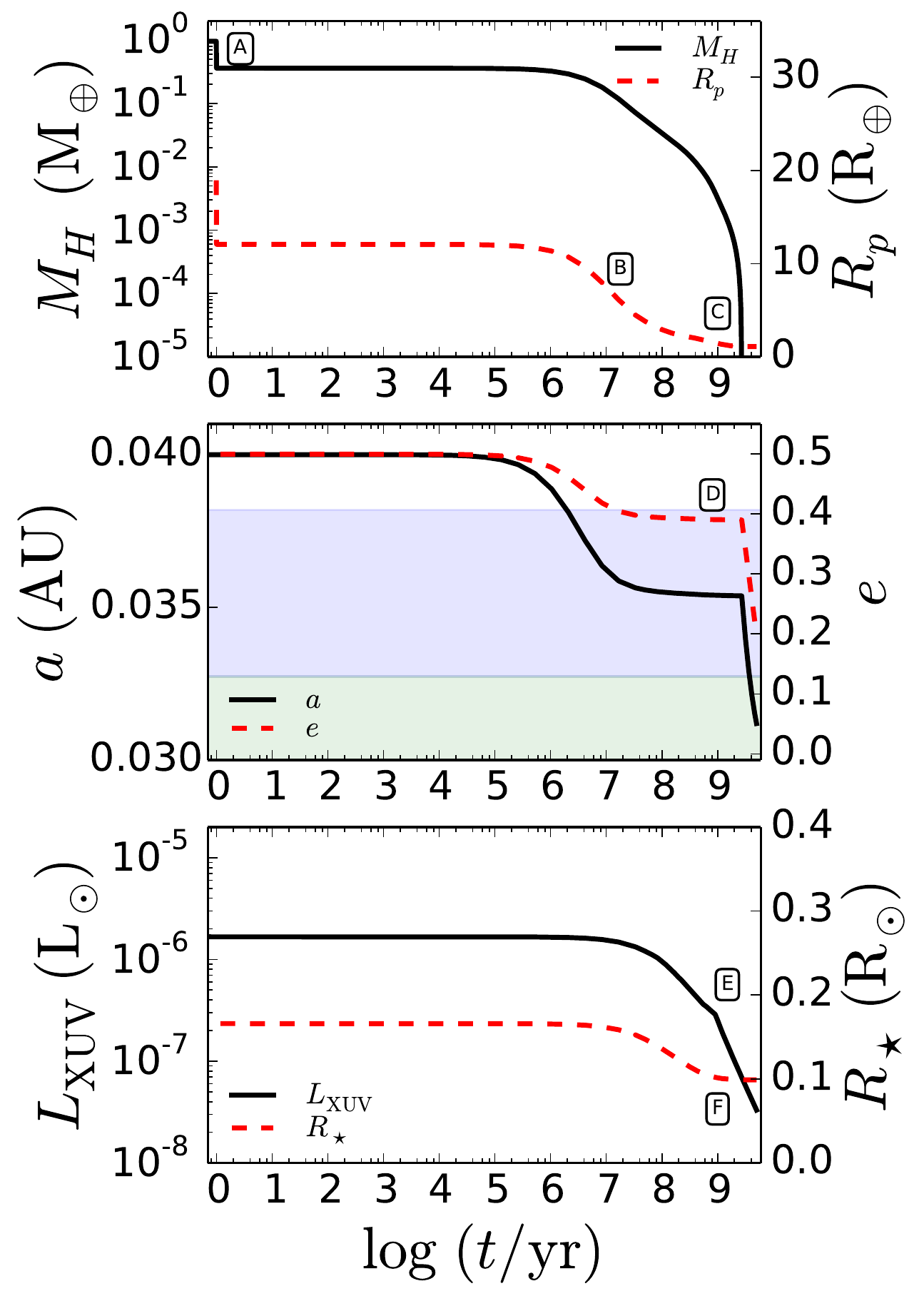}}
        \subfigure[Radiation/Recombination Limited]{\label{fig:int2}\includegraphics[width=2.1in]{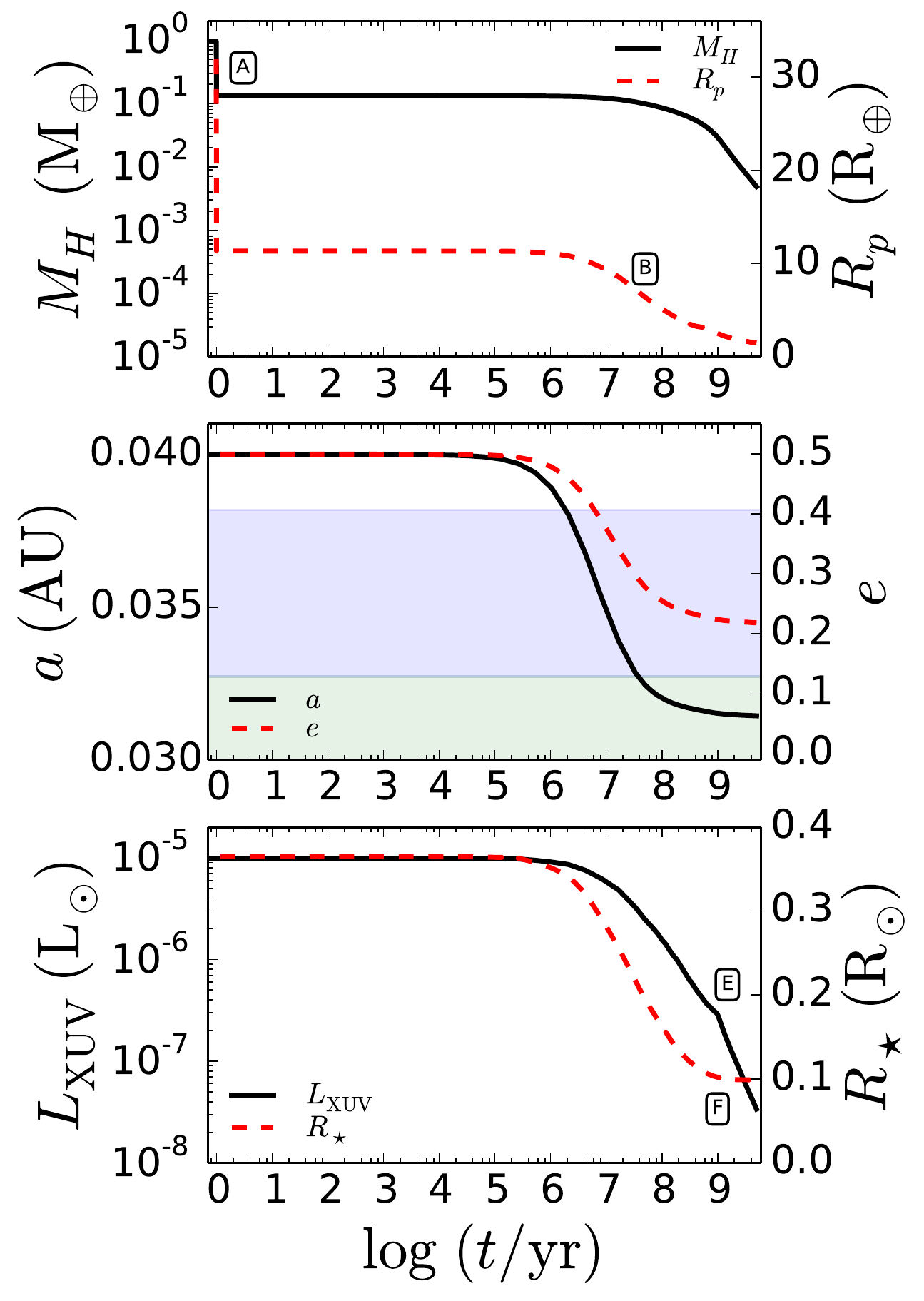}}
        \caption{Three sample integrations of our code. The first row plots show the envelope mass (left axis) and planet radius (right axis) versus time since formation; the second row plots show the semi-major axis (left) and eccentricity (right) versus time; and the third row shows the stellar XUV luminosity (left) and stellar radius (right) versus time. The planet is initially a 1 $\mearth$ core with a 1 $\mearth$ envelope orbiting around a 0.08 $\msun$ M-dwarf with $e = 0.5$ at a semi-major axis of 0.04 AU, just outside the OHZ (light blue shading). Unless otherwise noted, all other parameters are set to their default values (Table~\ref{tab:params}). As it loses mass and tidally evolves, it migrates into the CHZ (light green shading). Note that the evolution of the HZ is not shown; the CHZ and OHZ are taken to be the long-term ($> 1$ Gyr) values. \textbf{(a):} In this run, we force the escape to be energy-limited, as in an X-ray dominated flow. The planet loses its entire envelope at $t \approx 100$ Myr. \textbf{(b):} Same as (a), except the calculation starts at $t_0 = 100$ Myr, corresponding to a planet that undergoes late migration. While the envelope still completely evaporates, this occurs at a much later time, $t \approx 2$ Gyr. \textbf{(c):} Same as (a), except that the escape is radiation/recombination limited above the critical flux. Here, the envelope does not fully evaporate and tidal migration is noticeably weaker. See the text for a discussion of the labels \textbf{A-F}\vspace*{0.1in}.\label{fig:int}}
	\end{figure*}

\section{Results}
\label{sec:results}
\subsection{A Typical Run}
\label{sec:res:typical}
In Figure~\ref{fig:int} we show the time evolution of three mini-Neptunes as a guide to understanding the results presented in the following sections. We plot planet mass and radius (top row), semi-major axis and eccentricity (center row), and stellar XUV luminosity and radius (bottom row), all as a function of time since the planet's initial migration, $t_0$, for 5 Gyr. In the center plots, the post-1 Gyr OHZ and CHZ are shaded blue and green, respectively. At $t = t_0$, the planet is a 1 $\mearth$ core with a 1 $\mearth$ envelope orbiting around a 0.08 $\msun$ M-dwarf with $e = 0.5$ at a semi-major axis of 0.04 AU. We set $\epsilon_\mathrm{XUV} = 0.30$; other parameters are equal to the default values listed in Table~\ref{tab:params}. Because of the high eccentricity, the tidal evolution is calculated according to the CTL model.

In column (a), we force the escape to be energy-limited (\ref{eq:dmdt}), corresponding to an X-ray dominated flow. Prior to the first timestep, nearly 90 percent of the envelope mass is lost to RLO, indicated by the discontinuous drop marked \textbf{A} on the top plot. This is due primarily to the inflated radius shortly after formation, which reaches 30 $\rearth$ for a planet of age $t = t_0 = 10$ Myr. Once this mass is removed, the planet enters the energy-limited escape regime, which operates on a timescale of $\sim 10$ Myr (\textbf{B}). After $\sim 100$ Myr (\textbf{C}), the planet loses its entire envelope and becomes a HEC. 

In the center plot, we see that the planet's orbit steadily decays as it circularizes, with a sharp discontinuity in the slope at $\sim 100$ Myr (\textbf{D}), corresponding to when it transitions from a gaseous (low $\tau_p$) to a rocky (high $\tau_p$) body. The tidal evolution from that point forward is dramatically stronger, and $e$ decreases to $\sim 0.1$ at 5 Gyr. The planet's semi-major axis decays by $25\%$, moving it well into the CHZ. As we noted earlier, the transition from low to high tidal time lags (or, alternatively, from high to low tidal quality factors) is likely to be gradual as the bulk of the energy shifts from being dissipated in the envelope to being dissipated in the core. In this case, the faster inward migration as $\tau_p$ increases is likely to accelerate the rate of mass loss, leading to slightly earlier evaporation times. However, given the large uncertainty in the values of $\tau_p$ and its dependence on planetary and orbital parameters, our current approach should suffice.

In the bottom plot, we see that the bulk of the mass loss occurs when the stellar XUV flux is high. After $t \approx 100$ Myr, the XUV luminosity is low enough that a planet with significant hydrogen left ($f_H \gtrsim 0.01$) is unlikely to completely evaporate. Here, the XUV saturation time is set to 1 Gyr, visible in the kink marked by the label \textbf{E}; prior to that time, the decrease in the XUV luminosity is simply a function of the rate of contraction of the star. After $t \sim 1$ Gyr (\textbf{F}), the stellar radius asymptotes to its main sequence value and the XUV flux decays as a simple power law.

In column (b) we repeat the integration but delay the start time, setting $t_0 = 100$ Myr. This corresponds to a planet that undergoes a late scattering event, bringing it to $a = 0.04$ AU when both its radius and the XUV flux are significantly smaller. In this case, RLO is somewhat less effective, removing only $50\%$ of the envelope initially (\textbf{A}). However, the planet still loses all of its hydrogen at $t \approx 2$ Gyr (\textbf{C}). Interestingly, because of its delayed evaporation, the planet's eccentricity at 5 Gyr is significantly higher than in the previous run. This occurs because the transition from low to high $\tau_p$ (\textbf{D}) occurs much later. In this sense, a planet's current orbital properties can yield useful information about its atmospheric history. However, a more rigorous tidal model that accounts for the gradual change in $Q_p$ and $\tau_p$ as $f_H$ decreases is probably necessary to accurately infer the atmospheric evolution based on a planet's current eccentricity.

Finally, in column (c) we repeat the integration performed in (a), but this time set the flow to be radiation/recombination-limited above the critical flux (\S\ref{sec:massloss:rrlim}). Because of the significantly lower escape rate at early times, the envelope does not completely evaporate, and at 5 Gyr this is a super-Earth with slightly less than 1\% hydrogen by mass. In order for a planet to completely lose its envelope in the radiation/recombination-limited regime, it must either migrate into an orbit closer to the star, have a larger eccentricity, have a smaller core, or be stripped by another process.

\begin{figure}[h]
  \begin{center}
      \psfig{file=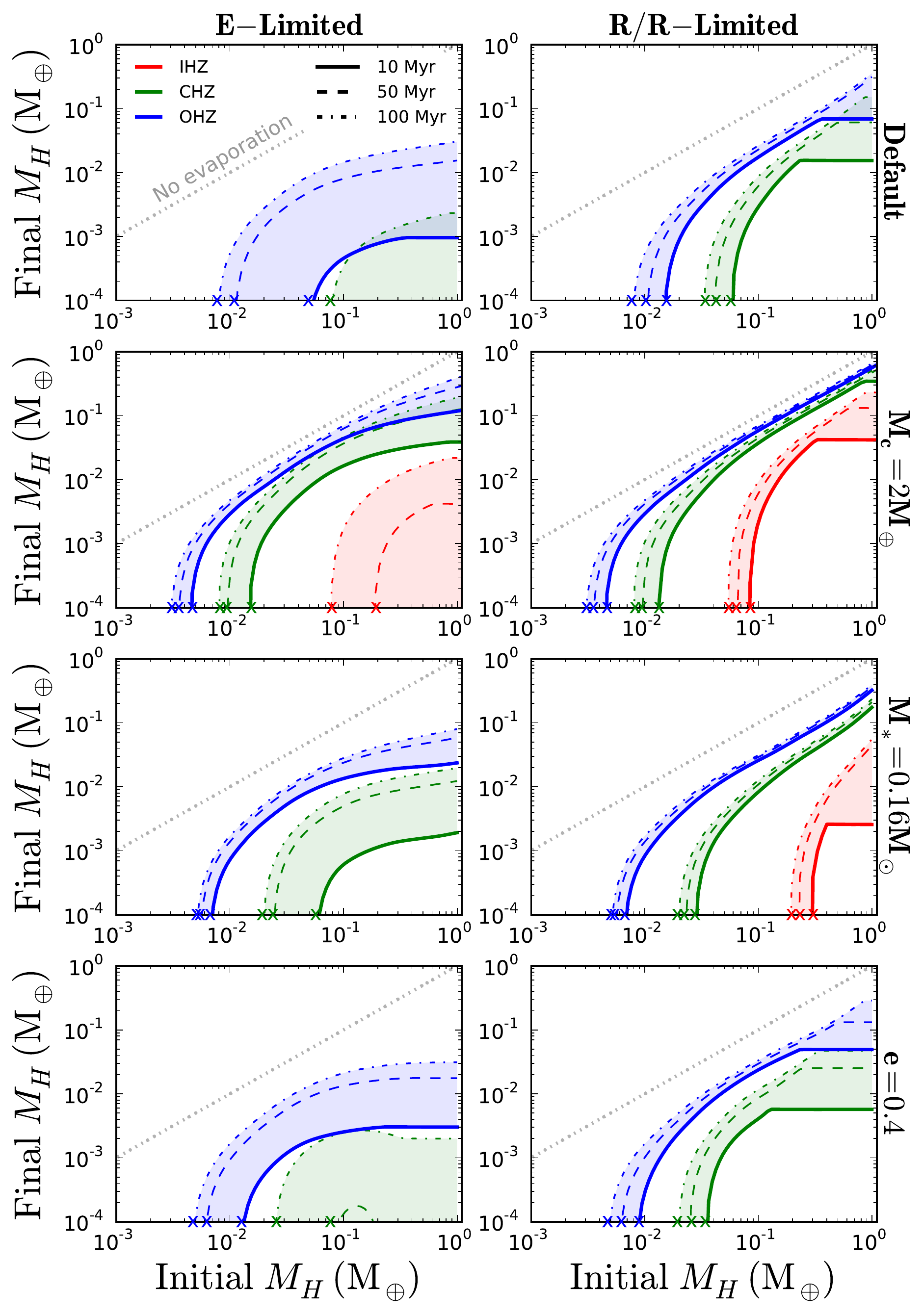,width=\columnwidth}
       \caption{Initial versus final envelope mass ($M_H$) for planets that end up in the IHZ (red), CHZ (green), and OHZ (blue). Line styles correspond to different values of $t_0$ (solid, 10 Myr; dashed, 50 Myr; dash-dotted, 100 Myr). Columns correspond to runs in which the escape mechanism is energy-limited (left) and radiation/recombination-limited (right); rows vary certain parameters as labeled, with all others set to their default values. In the default run, the planet has a 1$\mearth$ core and orbits an M dwarf with $M_\star = 0.08\msun$ in a circular orbit. The dotted gray line corresponds to a planet that undergoes no evaporation. An X marks the critical initial envelope mass below which full evaporation occurs within 5 Gyr. In some plots, curves of a given color/line style are missing; for those runs, the entire envelope was lost for all starting values of $M_H$.}
     \label{fig:mhmh}
  \end{center}
\end{figure}

\subsection{Dependence on $M_c$, $M_H$, $M_\star$, and $t_0$}
\label{sec:res:dep_planet}
In Figure~\ref{fig:mhmh} we plot the initial versus final envelope mass ($M_H$) for planets that end up in the IHZ (red lines), CHZ (green lines), and OHZ (blue lines). Line styles correspond to different values of $t_0$, the time at which the planet migrated into its initial close-in orbit (solid, 10 Myr; dashed, 50 Myr; dash-dotted, 100 Myr). The colored shadings are simply an aid to the eye, highlighting the spread due to different values of $t_0$. For reference, in dotted gray we plot the line corresponding to a planet that undergoes no evaporation. Note that in most plots, the curves approach this line as the initial $M_H$ is increased: at constant core mass, it is in general more difficult to evaporate planets with more massive H/He envelopes. Finally, if a curve of a given color/line style is missing, the final hydrogen mass is zero for all values of the initial $M_H$.

The two columns correspond to runs in which the escape was forced to be energy-limited (left) and radiation/recombination-limited at high XUV flux and energy-limited at low XUV flux (right). Rows correspond to different planet properties, as labeled. In the top (``Default'') row, the planet's core mass is set to $1 \mearth$. The planet orbits an M dwarf with $M_\star = 0.08\msun$ in a circular orbit. All other parameters are set to their default values.

In the second row, we double the core mass. The third row is the same as the first, but for a $0.16\msun$ star; and the fourth row is the same as the first, but for an initial eccentricity $e = 0.4$. Since planets in this run undergo orbital evolution, the different color curves correspond to planets that \emph{end up} in the IHZ, CHZ, and OHZ, respectively (their initial semi-major axes are somewhat larger).

Let us first consider the general trends in the plots. For low enough $M_H$, all curves become nearly vertical. The \emph{critical envelope mass} below which all mass is lost is marked with an X. Note that once planets lose sufficient mass such that $M_H \lesssim 10^{-4}$, the envelope is very unstable to complete erosion under the XUV fluxes considered here (corresponding to a near-vertical slope in this diagram). Many curves also display a flattening towards high initial envelope masses; some have prominent kinks beyond which the final envelope mass is constant. This is due to Roche lobe overflow, which causes any planets with radii larger than the Roche radius to lose mass prior to entering the HZ. Since increasing the envelope mass increases the planet radius, this results in a maximum envelope mass for some planets.

Planets that migrate in early (small $t_0$) lose significantly more mass than planets that migrate in late. This is a consequence of both the decay in the XUV luminosity of the star with time and the quick decrease of the planet radius as the planet cools. Another interesting trend is that the difference in the evaporated amount is much more pronounced in the energy-limited runs than in the radiation/recombination-limited runs. This is due to the $R_p^{3/2}$ scaling of the mass loss rate in the latter regime (versus the steeper $R_p^3$ scaling in the former). The difference in the initial radius across runs with different values of $t_0$ is less significant in the radiation/recombination-limited regime, resulting in more comparable mass loss rates. We also note that the $F_\mathrm{XUV}^{1/2}$ scaling of the mass loss rate in this regime results in a weaker dependence on semi-major axis, as expected: the blue, green, and red curves (for a given $t_0$) are packed more closely together in the right column than in the left column.

Now let us focus on individual rows. For the default run (a 1$\mearth$ core in a circular orbit around a 0.08$\msun$ star), \emph{all} planets in the IHZ lose all of their hydrogen and form HECs, regardless of migration time, envelope mass, or escape mechanism. In the CHZ, only planets that migrate within $\lesssim 50$ Myr and undergo energy-limited escape form HECs for all initial values of $M_H$. However, HECs still form from planets with $M_H \lesssim 0.5-0.9\mearth$ in the CHZ. In the OHZ, this is only possible for planets with less than about 1\% H/He by mass.

At twice the core mass (second row), all curves shift up and to the left, approaching the zero evaporation line for planets in the OHZ. In the IHZ, HECs still form from planets with any initial hydrogen amount for $t_0$ = 10 Myr and in the energy-limited regime. In all other cases, HECs only form from planets with $M_H \lesssim 0.1\mearth$. Of all the parameters we varied in our integrations, changing the core mass has the most dramatic effect on whether or not HECs can form. As we discuss below, habitable evaporated cores with masses greater than about 2$\mearth$ are unlikely.

At higher stellar mass (third row), HECs are again more difficult to form, particularly in the radiation/recombination-limited regime. Due to the more distant HZ, Roche lobe overflow is less effective in removing mass. The shorter super-luminous contraction phase of earlier M dwarfs also results in less total XUV energy deposition in the envelope. However, in the energy-limited regime, HECs still form from planets with up to 50\% H/He envelopes in the IHZ.

Finally, the effect of a higher eccentricity (bottom row) is much more subtle. In general, these planets lose slightly less mass than in the default run, but the plots are qualitatively similar. At high eccentricity, the orbit-integrated mass loss rate is higher (see \S\ref{sec:massloss:ecc}), but because of the orbital evolution, the planet must start out at larger $a$ in order for it to end up in the HZ at 5 Gyr. These effects roughly cancel out: in general, habitable evaporated cores are just as likely on eccentric as on circular orbits. Note, also, that the green curves in the bottom right plot are the only ones that are non-monotonic. The effect is very small, but hints at an interesting coupling between tides and mass loss. At high initial $M_H$, the radius is large enough to drive fast inward migration (see below), exposing the planet to high XUV flux for slightly longer than a planet with smaller $M_H$ (and therefore a smaller radius), resulting in a change in the slope of the curve at initial $M_H \approx 0.15\mearth$.

\subsection{The Role of Tides}
\label{sec:res:dep_ecc}
Next, we consider in detail how tides affect the evolution of HECs. We show in the Appendix that the net effect of tides is to induce inward migration and circularization of planet orbits in the HZ of M dwarfs, an effect that couples strongly to the atmospheric mass loss. For $e \lesssim 0.7$, the flux increases with time as planets tidally evolve, accelerating the rate of mass loss; at higher eccentricities, the flux actually decreases due to the circularization of the orbit (see the Appendix for a derivation). The changing mass and radius of the planet can then act back on the tidal evolution, either accelerating it (in cases where $|dM_p/dt| \gg |dR_p/dt|$) or decelerating it in a negative feedback loop (otherwise).

\begin{figure*}[t]
	\begin{tabular}{lll} 
    \multirow{-3}[-20]{*}{
      \subfigure[Default run]
        {\includegraphics[width=3.1in,height=3.05in]{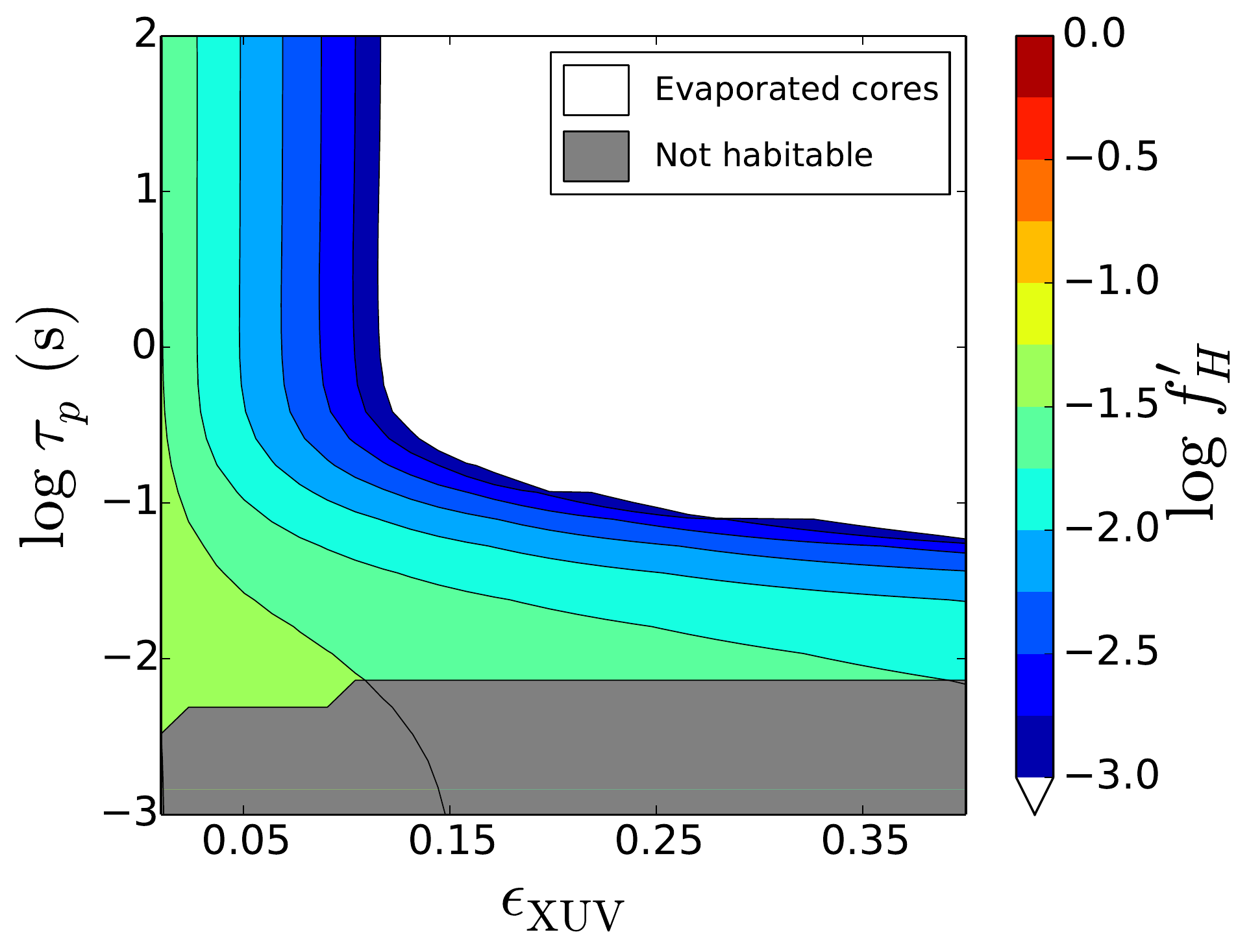}}} & 
        \hspace{-0.25in}
      \subfigure[Higher core mass]
        {\includegraphics[width=2.15in,height=2in]{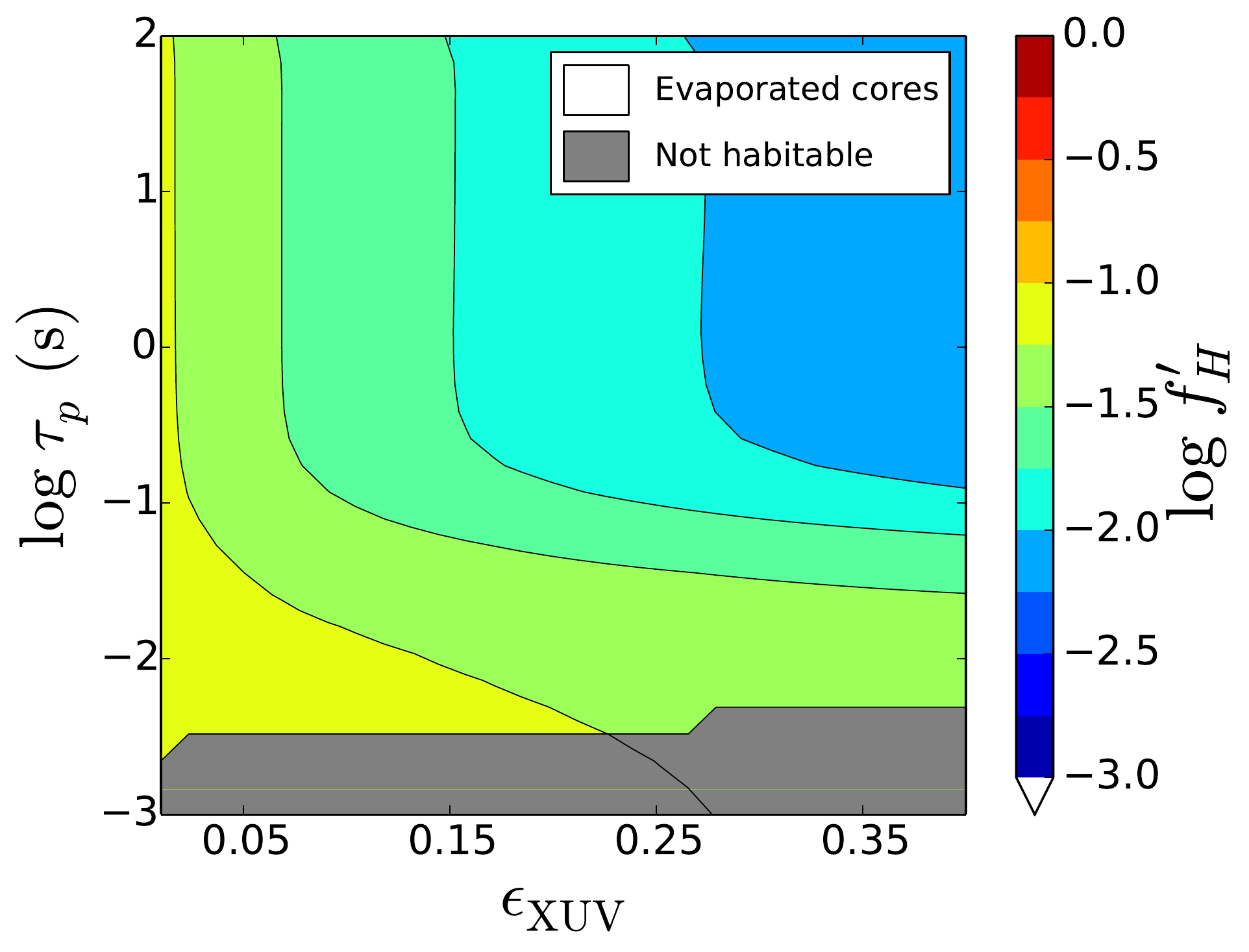}} & 
        \hspace{-0.2in}
      \subfigure[Lower eccentricity]
        {\includegraphics[width=2.15in,height=2in]{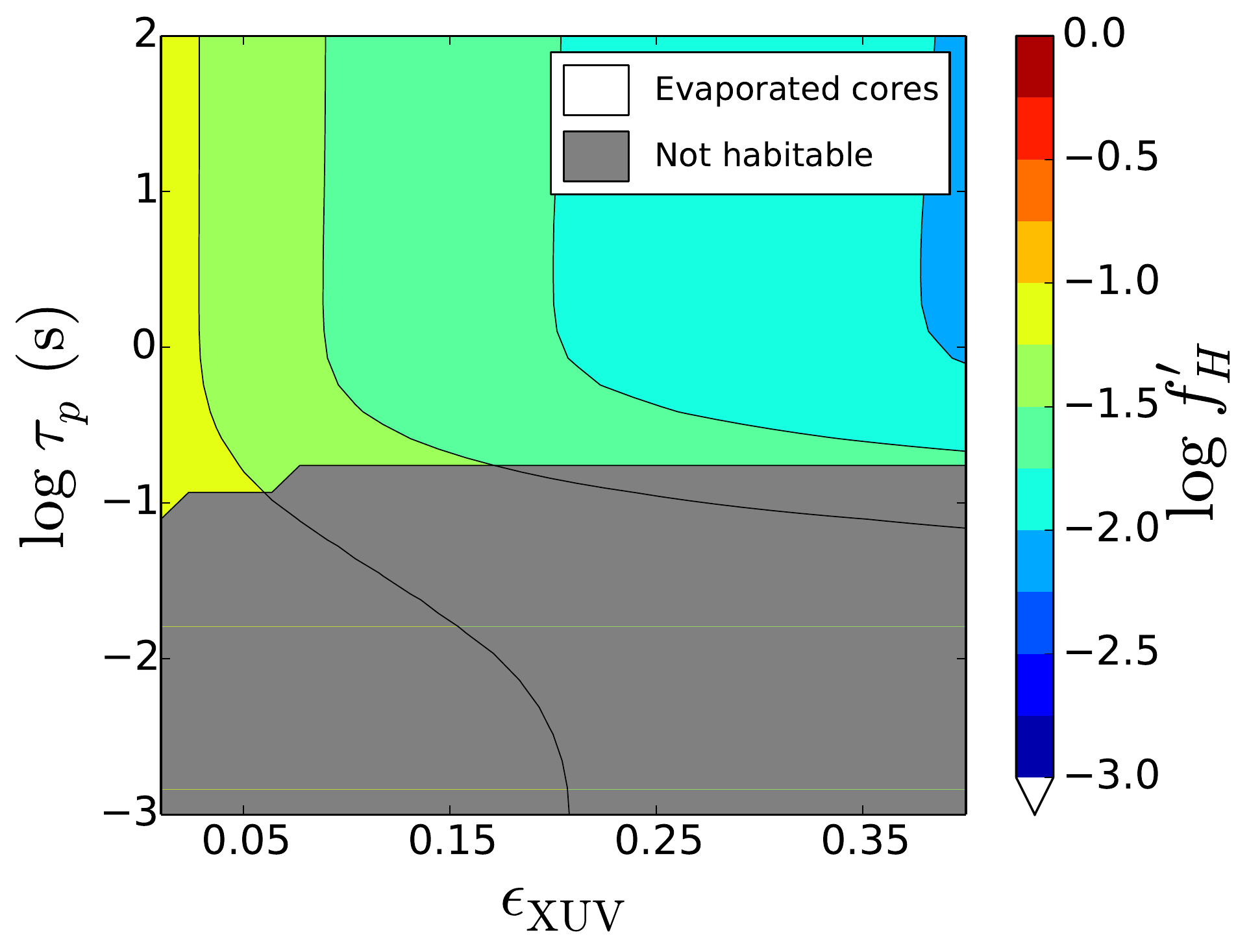}}\\ &
        \hspace{-0.25in} 
      \subfigure[Shorter saturation time]
        {\includegraphics[width=2.15in,height=2in]{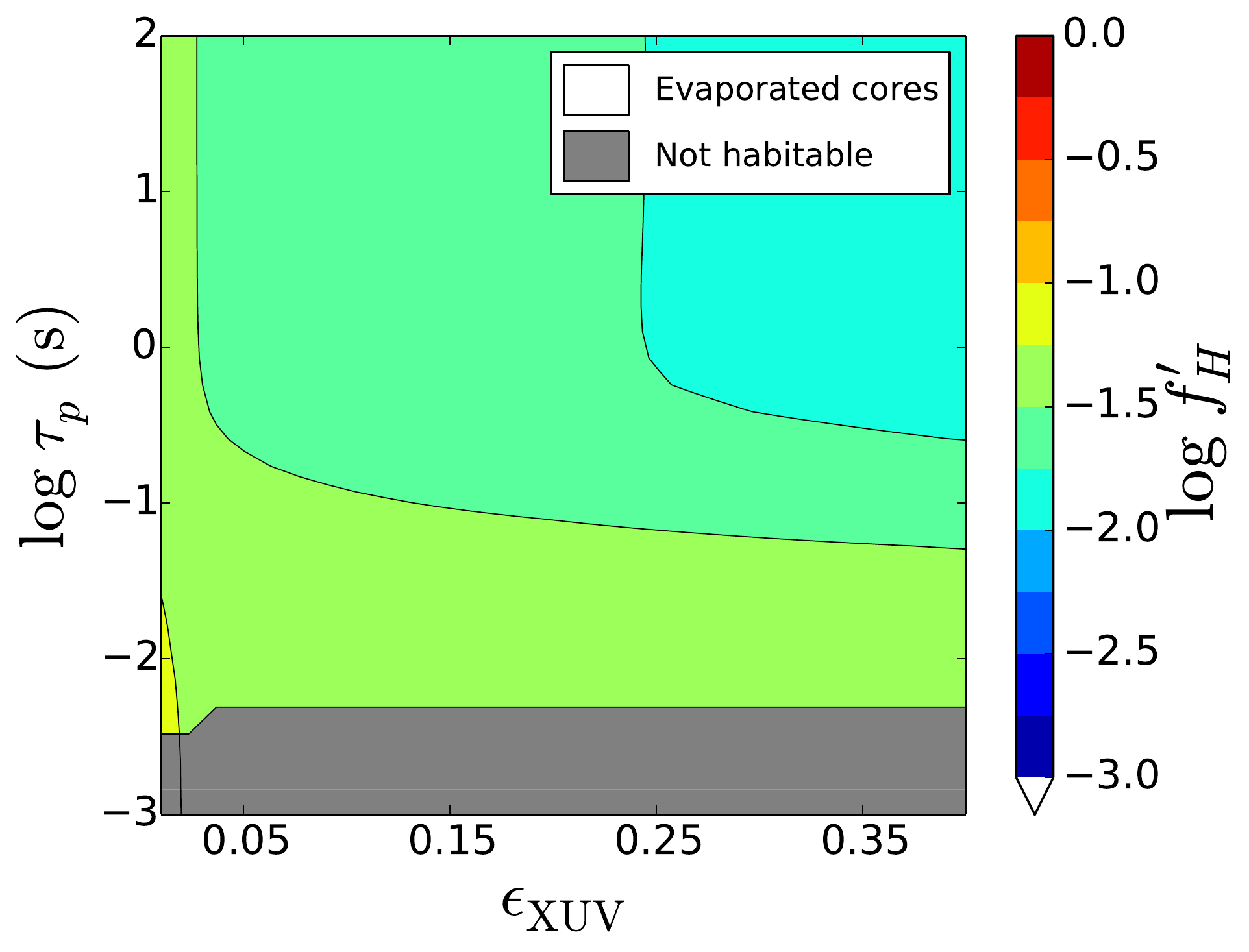}} & 
        \hspace{-0.2in}
      \subfigure[Energy-limited only]
        {\includegraphics[width=2.15in,height=2in]{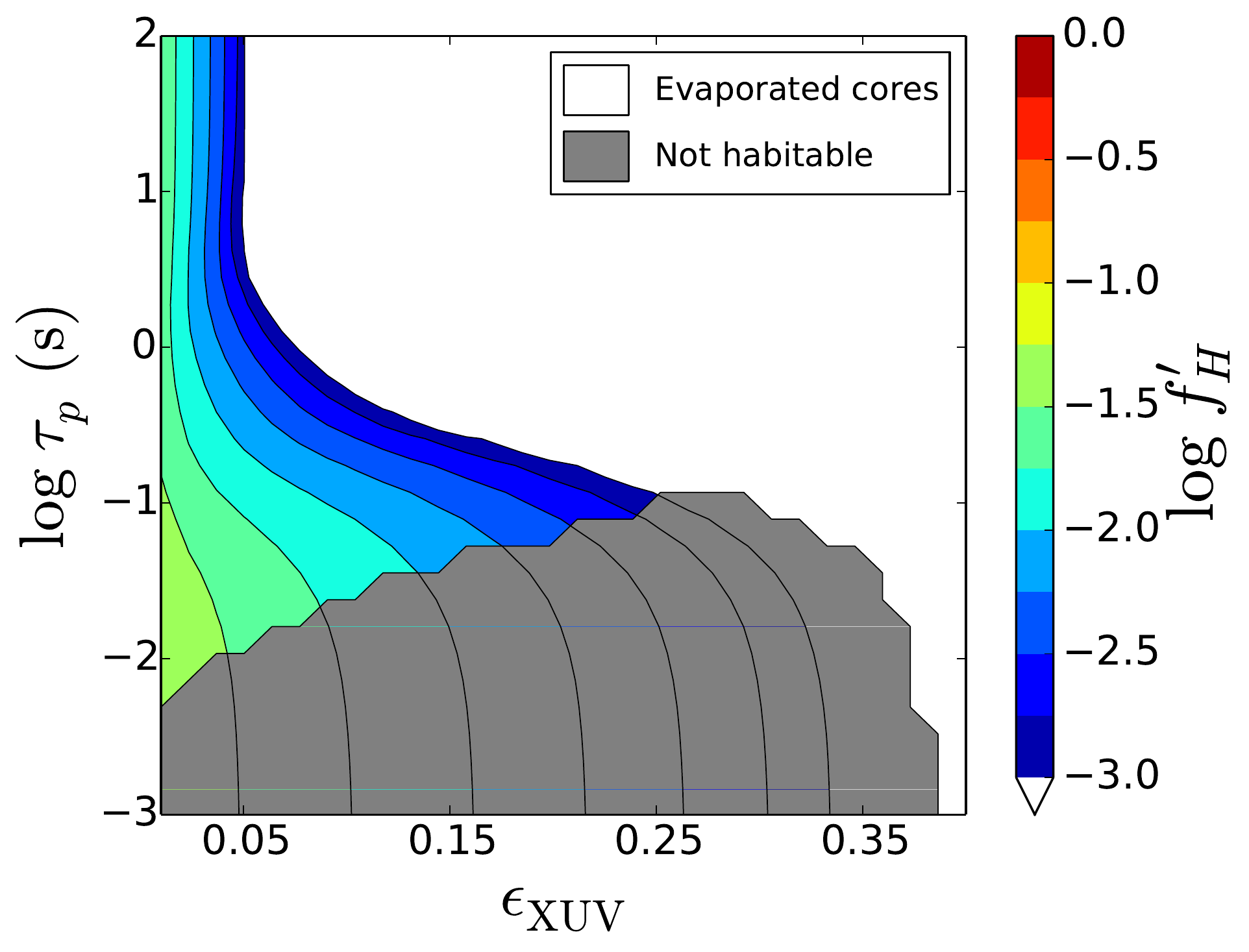}
    }\\
	
	\end{tabular}
	\caption{Contours of the log of the hydrogen mass fraction $f_H'$ at 5 Gyr as a function of the tidal time lag $\tau_p$ and the XUV absorption efficiency $\epsilon_\mathrm{XUV}$ for five integrations of our code. White regions correspond to planets that completely lost their envelopes; dark gray regions indicate planets that are not in the HZ at 5 Gyr. \textbf{(a)} The default run. The planet has a core mass of 1 $\mearth$ with $f_{H_0} = 0.5$, orbiting around a 0.08 $\msun$ star in an initially highly eccentric orbit ($e = 0.8$) at $a = 0.07$ AU. All other parameters are set to their default values (Table~\ref{tab:params}), and the escape mechanism is radiation/recombination-limited at high XUV flux and energy-limited at low XUV flux. Note that the final envelope mass highly depends on \emph{both} $\exuv$ and $\tau_p$. \textbf{(b)} The same as (a), but for a higher core mass $M_c = 1.5\mearth$. No evaporated cores form in this scenario. \textbf{(c)} The same as (a), but for a lower eccentricity $e = 0.7$. Again, no habitable evaporated cores form, and the planet remains outside of the habitable zone for a larger range of $\tau_p$. \textbf{(d)} The same as (a), but for a shorter XUV saturation time of the parent star, $t_\mathrm{sat} = 0.1$ Gyr. No evaporated cores form. Since the XUV flux drops off much more quickly, energy-limited escape is less effective in removing mass and thus $f_H'$ is a weaker function of $\exuv$. \textbf{(e)} The same as (a), but for energy-limited escape only, which would be the case if the flow is X-ray dominated. Note that in this case whether or not the planet becomes an evaporated core is a much stronger function of both $\tau_p$ and $\exuv$.
	\label{fig:contour}}
	\end{figure*}

In Figure~\ref{fig:contour} we show the results of an integration of our code on a grid of tidal time lag $\tau_p$ versus the XUV absorption efficiency $\exuv$. Colors correspond to the final hydrogen mass fraction, $f_H'$; evaporated cores occur in the white regions ($f_H' = 0$). Dark gray indicates planets that either migrated beyond the HZ or remained exterior to it and are therefore not habitable. These plots provide an intuitive sense of the relative importance of tidal evolution ($y$-axis) and energy-limited escape ($x$-axis) in determining whether or not a HEC is formed. We note that once a planet loses its gaseous envelope, we switch the time lag to $\tau_p' = \mathrm{max}(\tau_p,\ 64\mathrm{s})$, where the latter is a typical (gas-free) tidal time lag of a rocky planet \citep[see, for instance,][]{BAR13}.

In the default run (a), we consider a planet with a core mass of $1\mearth$ and an envelope mass of $1\mearth$ ($f_H = 0.5$) orbiting at 0.07 AU in a highly eccentric ($e = 0.8$) orbit around a 0.08$\msun$ star. The mass loss mechanism is radiation/recombination-limited escape at high XUV flux and energy-limited at low XUV flux. At a given value of $\log\ \tau_p$, say 0, the final hydrogen fraction is a strong decreasing function of $\exuv$, as expected: the higher the evaporation efficiency, the smaller the final envelope. Interestingly, the dependence of $f_H'$ on the tidal time lag can be nearly just as strong. At a given value of $\exuv$, say 0.35, $f_H'$ depends strongly on $\tau$, varying from $10^{-1.5} \approx 3\%$ to 0\%--that is, the tidal evolution controls whether or not a HEC forms. This is due to the fact that at large $\tau_p$, tidal migration is fast, bringing the planet into the HZ while its radius is still inflated and the stellar XUV luminosity is higher. Planets with lower values of $\tau_p$ migrate in later and undergo slower mass loss. At very low $\tau_p$, tidal migration is too weak to bring the planets fully into the HZ. The effect is stronger for higher $\exuv$: these are planets whose radii decrease very quickly (due to the fast mass loss), slowing down the rate of tidal evolution and keeping them outside of the HZ for longer.

In plot (b), we increase the core mass slightly to $1.5\mearth$. The effect on the mass loss is significant, and HECs no longer form. At high $\exuv$ and high $\tau_p$, the lowest value of $f_H'$ is about 1\%. This reinforces what we argued above: HECs are most likely for the lowest mass cores.

In plot (c), we instead decrease the eccentricity to 0.7. As in (b), HECs no longer form in these runs due to the decreased strength of the tidal evolution, and again the minimum $f_H'$ is about 1\%. Note that this does not mean that HECs are more likely at higher eccentricity in general---this is only the case here because we fix the initial semi-major axis at 0.07 AU. At fixed \emph{final} semi-major axis (i.e., what we can readily observe in actual systems), planets on initially circular orbits will still lose more mass than planets on initially eccentric orbits (which must have formed farther out).

In plot (d), we decrease the saturation time to $t_{sat} = 0.1$ Gyr, which is typical of early M/late K dwarfs \citep{JDW12}. No HECs form, and the final hydrogen fraction is less sensitive to both $\tau_p$ and $\exuv$: in general, mass loss is significantly suppressed.

Finally, in plot (e), we force the escape mechanism to be energy-limited only. Since the escape is now entirely controlled by $\exuv$, the dependence on this parameter is naturally much stronger, and complete evaporation occurs in this case for $\exuv \gtrsim 0.32$ at any $\tau_p$. Because evaporation occurs more quickly in this case than in the other plots, at any given time the planet radii are smaller, resulting in less efficient migration at a given $\tau_p$. More planets therefore do not make it into the HZ. Interestingly, for very large $\exuv$ ($\gtrsim 0.4$), all planets make it into the HZ. This is due to the fact that these planets transition from gaseous (low $\tau_p$) to gas-free (high $\tau_p$, equal to 64s in these runs) very early on. Despite their lower radii, they benefit from the stronger tidal dissipation of fully rocky bodies and are able to make it into the HZ after 5 Gyr.

In general, the coupling between tides and mass loss is quite complex. For planets with high initial eccentricities, this coupling can ultimately determine whether or not a HEC will form. We discuss this in more detail in \S\ref{sec:res:hecs}.

\subsection{Evaporated Cores in the Habitable Zone}
\label{sec:res:hecs}
\begin{figure*}[t]
  \begin{center}
      {
		    \fontfamily{cmss}\selectfont
		    \huge{\ \ \ \ Default Run, 1 M$_\oplus$}\\
	    }
      \psfig{file=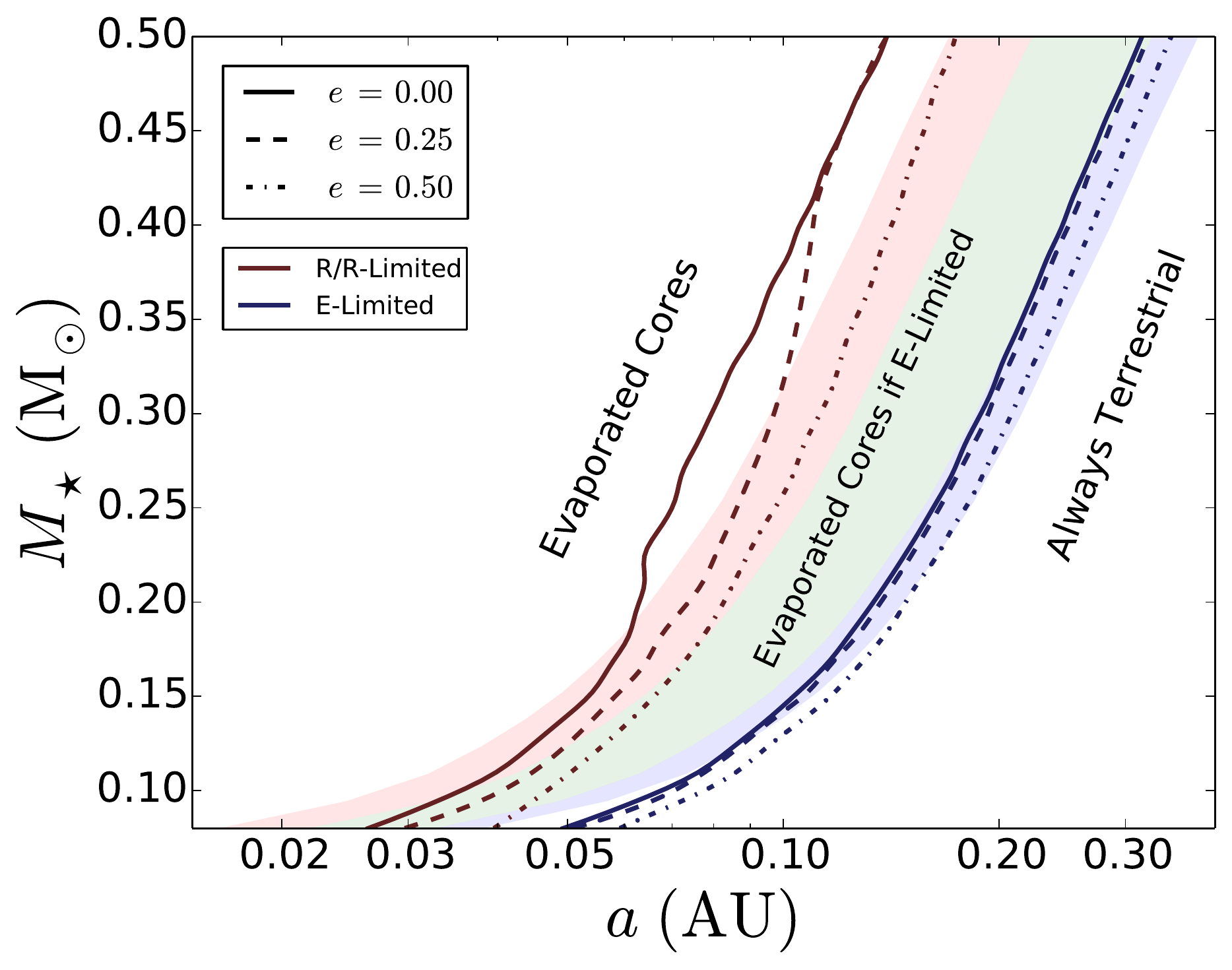,width=4in}
       \caption{Regions of parameter space that may be populated by HECs, for $M_c = 1\mearth$, $M_H \leq 1\mearth$, and default values for all other parameters. Terrestrial planets detected today occupying the space to the left of each contour line could be the evaporated cores of gaseous planets with $f_H \leq 0.5$. Planets detected to the right of the contour lines have always been terrestrial/gaseous. Dark red lines correspond to the conservative mass loss scenario, in which mass loss is radiation/recombination-limited at high XUV flux and energy-limited at low XUV flux. Dark blue lines correspond to mass loss via the energy-limited mechanism only. Planets around stars with significant X-ray emission early on are likely to be in the latter regime. Different line styles correspond to different eccentricities today. Terrestrial planets detected at higher eccentricity (dashed and dash-dotted lines) could be evaporated cores at slightly larger orbital separations than planets detected on circular orbits (solid lines). Note that in the energy-limited regime, \emph{all} $1\mearth$ terrestrial planets in the HZ of low-mass M dwarfs could be habitable evaporated cores. At higher stellar mass, HECs are restricted to planets in the CHZ and IHZ. In the radiation/recombination-limited regime, the accessible region of parameter space is smaller, but around the lowest mass M dwarfs HECs are still possible in the CHZ.}
     \label{fig:money1}
  \end{center}
\end{figure*}

\begin{figure*}[t]
  \begin{center}
      {
		\fontfamily{cmss}\selectfont
		\huge{\ \ \ \ Conservative, 1 M$_\oplus$}\\
	  }
      \psfig{file=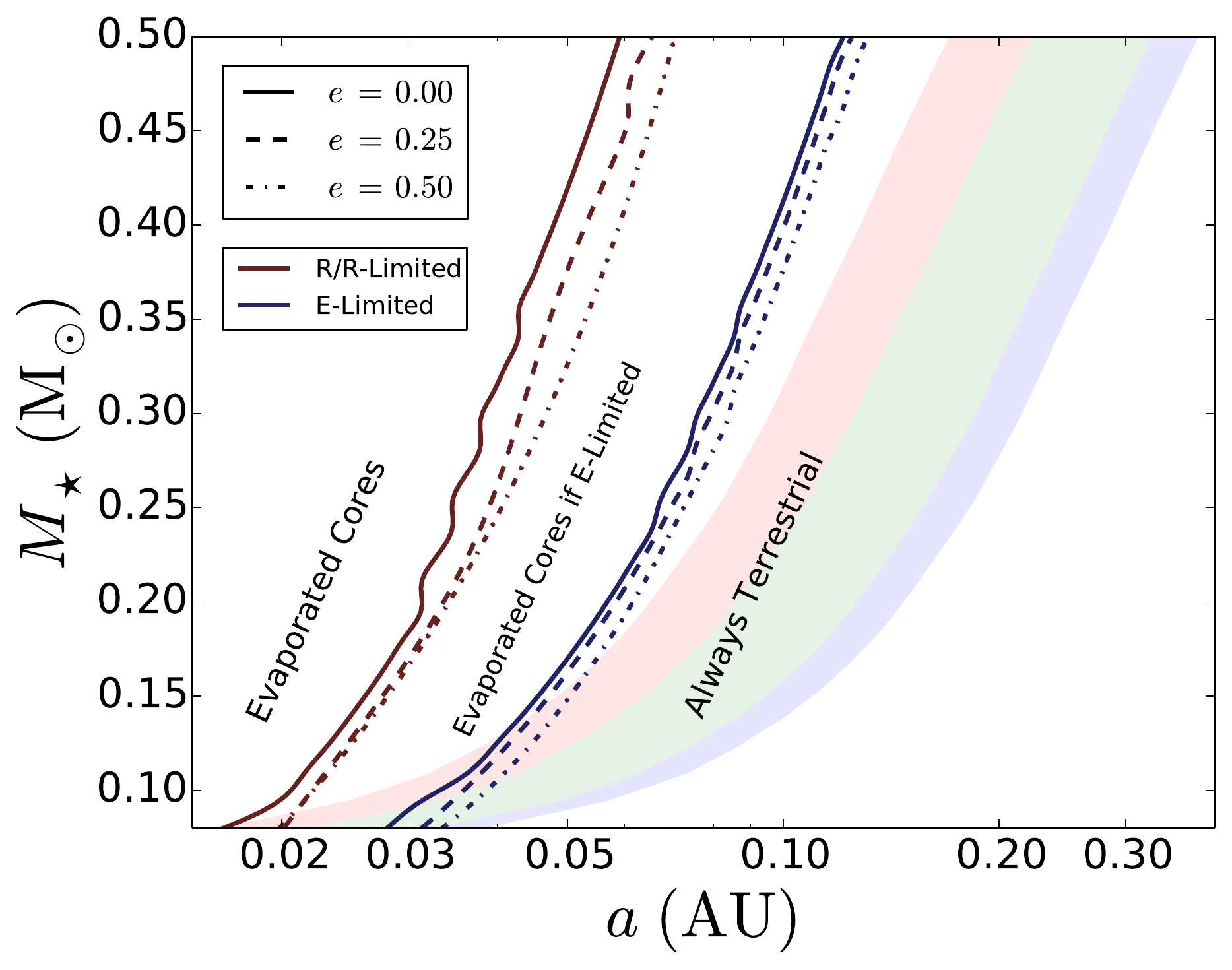,width=4in}
       \caption{The same plot as Figure~\ref{fig:money1}, but for a conservative choice of the parameters governing atmospheric escape: a short XUV saturation time $t_\mathrm{sat} =0.1$ Gyr and a low XUV absorption efficiency $\exuv = 0.15$. In this case, HECs are no longer possible for radiation/recombination-limited escape. For energy-limited escape, HECs are only possible in the IHZ of low-mass M dwarfs and in the CHZ of M dwarfs at the hydrogen-burning limit. At high eccentricity ($e = 0.5$), HECs are only marginally more likely.}
     \label{fig:money3}
  \end{center}
\end{figure*}

\begin{figure*}[t]
  \begin{center}
      {
		\fontfamily{cmss}\selectfont
		\huge{\ \ \ \ Default Run, 2 M$_\oplus$}\\
	  }
      \psfig{file=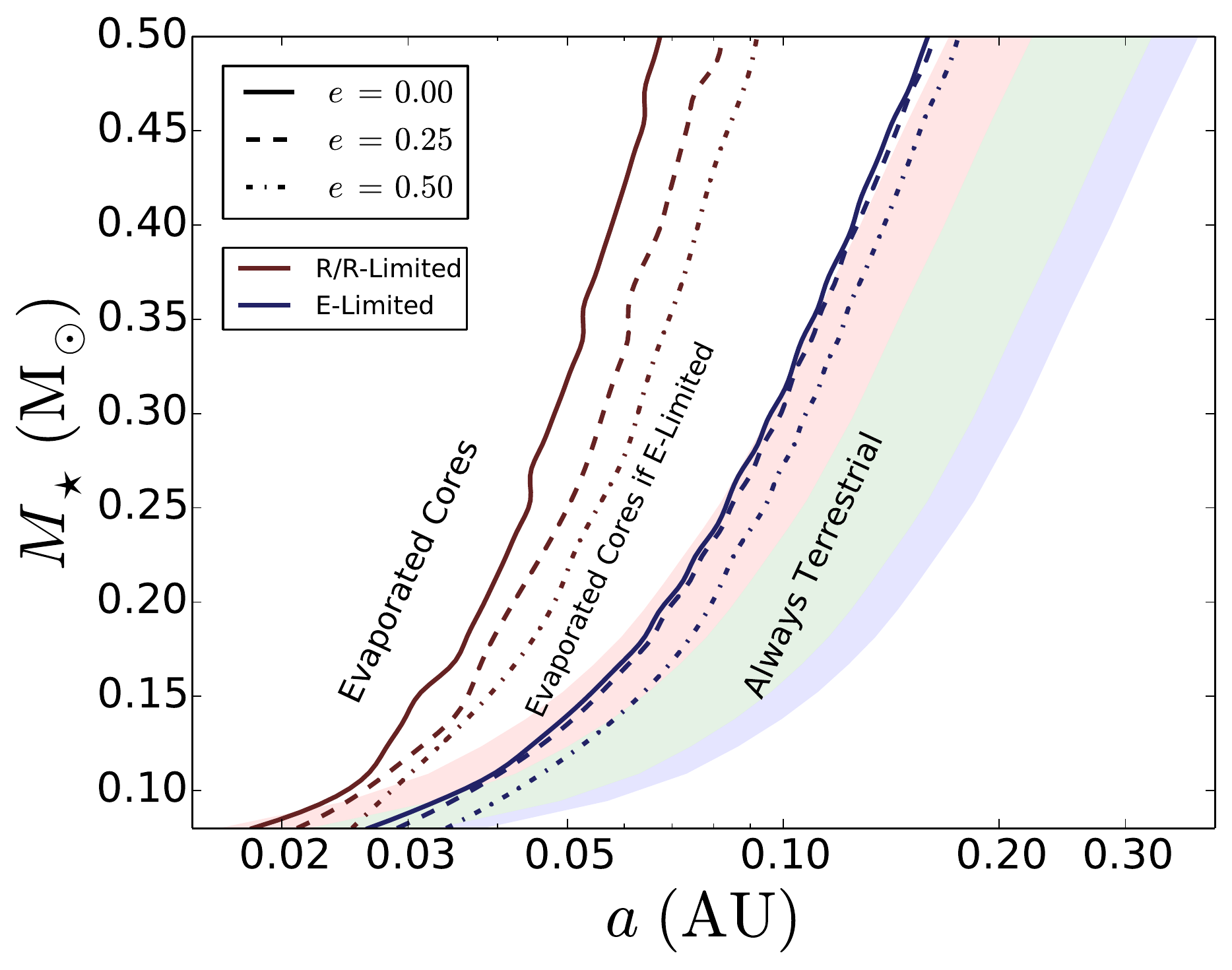,width=4in}
       \caption{The same as Figure~\ref{fig:money1}, but for a higher core mass $M_c = 2\mearth$. HECs are now confined mostly to the IHZ of low-mass M dwarfs in the energy-limited regime. For planets undergoing radiation/recombination-limited escape, super-Earth HECs may not be possible.}
     \label{fig:money5}
  \end{center}
\end{figure*}

Having shown that HECs are possible in certain cases, we now wish to explore where in the habitable zone we may expect to find them. For a ``terrestrial'' (we use this term rather loosely\footnote{As we explain below, it is quite possible that HECs are all but terrestrial, since it is likely that they have large ice mass fractions and are compositionally distinct from Earth. By ``terrestrial,'' in this case, we mean planets that are similar in size and mass to Earth and are not gaseous; these may or may not have a rocky surface like Earth. Our definition of ``terrestrial'' therefore encompasses water worlds.}) planet detected in the HZ, it would be very informative to understand whether or not it may be the evaporated core of a gaseous planet, since its past atmospheric evolution may strongly affect its present ability to host life. To this end, we ran eight grids of $2.7\times 10^6$ integrations each of our evolution code under different choices of parameters in Table~\ref{tab:params}. For initial semi-major axes in the range $0.01 \leq a_0 \leq 0.5$, initial eccentricities in the range $0 \leq e_0 \leq 0.95$, and stellar masses between $0.08\msun \leq M_\star \leq 0.5\msun$, we calculate the final (i.e., at $t_\mathrm{stop} = 5$ Gyr in the default run) values of $a$, $e$, and the envelope mass $M_H$. We then plot contours of $M_H$ as a function of the stellar mass and the \emph{final} semi-major axis and eccentricity (that is, the observable parameters). Since in principle, the mapping $(a_0,e_0,M_{H_0}) \rightarrow (a,e,M_H)$ is not necessarily a bijection (due to the nonlinear coupling between tides and mass loss and the finite resolution of our grid), the function $M_H(a,e)$ may be multi-valued at some points. In these cases, we take $M_H$ at coordinates $(a,e)$ to be the \emph{minimum} of the set of all final values of $M_H$ that are possible at $(a,e)$. Because of this choice, the $M_H = 0$ contour in $a$ versus $M_\star$ plots (Figures~\ref{fig:money1}-\ref{fig:money5}) separates currently terrestrial planets that \emph{could be} the evaporated cores of mini-Neptunes (left) from currently terrestrial planets that have \emph{always} been terrestrial (right). In other words, we are showing where evaporated cores are ruled out.

In Figure~\ref{fig:money1} we perform the calculations described above for planets with $M_c = 1\mearth$ and initial $f_H = 0.5$ (i.e., $M_H = 1\mearth$). The habitable zone is plotted in the background for reference, where again the IHZ, CHZ, and OHZ are indicated by the red, green, and blue shading, respectively. Line styles correspond to different final values of the eccentricity (0, solid; 0.25, dashed; and 0.5, dash-dotted). Dark red lines correspond to the $M_H = 0$ contours in the radiation/recombination-limited escape model; dark blue lines correspond to the energy-limited model. Note that though we designate our two models as ``energy-limited'' and ``radiation/recombination-limited'', at low XUV fluxes the escape is energy-limited in \emph{both} models, since below $F_\mathrm{crit}$ the energy-limited escape rate is always smaller than the radiation/recombination-limited escape rate. In this sense, the ``radiation/recombination-limited model'' is always conservative, as the mass loss rate is set to the minimum of Equations (\ref{eq:dmdt}) and (\ref{eq:dmdtrr}).

As an example of how to interpret this figure, consider a $1\mearth$ rocky planet discovered orbiting a $0.2\msun$ star at 0.1 AU, that is, squarely within the CHZ. Since this planet lies to the left of all blue curves, under the assumption of energy-limited escape it could be a habitable evaporated core. On the other hand, if the atmospheric escape were radiation/recombination-limited, this planet must have always been terrestrial.

Next, consider a putative rocky planet discovered around the same $0.2\msun$ star skirting the inner edge of the HZ (i.e., at 0.07 AU). Under the energy-limited assumption, this planet could be a HEC. For radiation/recombination-limited escape, however, whether or not it could be a HEC depends on its present eccentricity. If the planet is currently on a circular orbit, we infer that it has always been terrestrial (since it lies to the right of the $e = 0$ contour). However, since the planet lies to the left of the higher eccentricity contours, if $e \gtrsim 0.25$, it could be a HEC.

We conclude from Figure~\ref{fig:money1} that if energy-limited escape is the dominant mechanism around M dwarfs, planets with $M_p \sim 1\mearth$ in the CHZ and IHZ of these stars could be habitable evaporated cores. For $M_\star \lesssim 0.15\msun$, HECs may exist throughout the entire HZ. If, on the other hand, these planets are shaped mostly by radiation/recombination-limited escape, HECs are only possible in the IHZ for $M_\star \lesssim 0.2\msun$ and in the CHZ of M dwarfs near the hydrogen-burning limit.

The effect of the eccentricity is significant primarily in the radiation/recombination-limited regime, where a lower mass loss rate keeps the planet's radius large for longer than in the energy-limited regime, allowing it to migrate into the HZ faster, thereby enhancing the mass loss rate. In some cases, particularly near the inner edge of the HZ, the present-day eccentricity yields important information about a planet's evolutionary history: depending on the precise value of $e$, a given terrestrial planet may or may not be a HEC. However, we urge caution in interpreting the results in Figure~\ref{fig:money1} at nonzero eccentricity, given the large uncertainty in the tidal parameters of exoplanets.

On this note, it is important to bear in mind that the curves in Figure~\ref{fig:money1} are a function of our choice of parameters in Table~\ref{tab:params}. In order to assess the impact of our choice of ``default'' parameters on these contours, in Figure~\ref{fig:money3} we repeat the calculation for more conservative values of the two parameters that govern the mass loss mechanism: the XUV saturation time $t_\mathrm{sat}$ and the absorption efficiency $\exuv$. In this figure, we choose $t_\mathrm{sat} = 0.1$ Gyr, the nominal value for earlier K/G dwarf stars (\S\ref{sec:lumevol}), and $\exuv = 0.15$.

In this grid, all curves shift quite dramatically to lower $a$, and HECs are no longer possible under radiation/recombination-limited escape. For energy-limited escape, HECs are confined to the IHZ for $M_\star \lesssim 0.15\msun$ and the CHZ for $M_\star \lesssim 0.1\msun$ at all eccentricities considered here.

Given the large difference between the results of Figures~\ref{fig:money1} and \ref{fig:money3}, care must be taken in assessing whether a planet may be a HEC. Since it is likely that the XUV saturation time is much longer for M dwarfs than for K and G dwarfs, and since our goal at this point is to separate regions of parameter space where HECs can and cannot exist, Figure~\ref{fig:money1} is probably the more relevant of the two. We discuss this point further in \S\ref{sec:discuss}.

In Figure \ref{fig:money5} we raise the core mass to $M_c = 2\mearth$. HECs are now possible only in the IHZ and only in the energy-limited regime. At the lowest $M_\star$, HECs may be possible in the CHZ (energy-limited) and at the very inner edge of the HZ (radiation/recombination-limited). At higher eccentricity, the parameter space accessible to HECs is slightly higher in the energy-limited regime, but it is still a small effect overall. For a conservative choice of escape parameters ($t_\mathrm{sat} = 0.1$ Myr and $\epsilon_\mathrm{XUV} = 0.15$) with $M_c = 2\mearth$, no HECs form anywhere in the HZ.

\section{Discussion}
\label{sec:discuss}
\subsection{Initial Conditions}
\label{sec:discuss:howto}

We have shown that HECs can form from small mini-Neptunes ($M_p \lesssim 2\mearth$) with H/He mass fractions as high as $f_H = 0.5$ around mid- to late M dwarfs. However, instrumental to our conclusions is the assumption that these planets formed beyond the snow line and migrated quickly into the HZ at $t = t_0$. Planets that form \emph{in situ} in the HZ are unlikely to accumulate substantial H/He envelopes due to large disk temperatures and relatively long formation timescales. Significant gas accretion can only occur prior to the dissipation of the gas disk, which occurs on a timescale of a few to $\sim$ 10 Myr; in particular, \cite{LAM14}  showed that a 1 M$_\oplus$ terrestrial planet generally accretes less than $\sim$ 3\% of its mass in H/He in the HZ of a solar-type star. Moreover, planets that form \emph{in situ} do not undergo Roche lobe overflow, since accretion of any gas that would lead to $R_p > R_\mathrm{Roche}$ simply will not take place. 

But can mini-Neptunes easily migrate into the HZ? The large number of recently discovered hot Neptunes and hot Super-Earths \citep[e.g.,][]{HOW12} suggests that inward disk-driven migration is an ubiquitous process in planetary systems, as it is highly unlikely that these systems formed \emph{in situ} \citep{RAY14}. Moreover, systems such as GJ 180, GJ 422, and GJ 667C each have at least one super-Earth/mini-Neptune in the HZ \citep{TUO14,ANG13}, some or all of which may have migrated to their current orbits. On the theoretical front, N-body simulations by \cite{OI09} show that migration of protoplanets into the HZ of M dwarfs is efficient due to the proximity of the ice line and the fact that the inner edge of the disk lies close to the HZ. Mini-Neptunes that assemble early beyond the snow line could in principle also migrate into the HZ, but the migration probability and the distribution of these planets throughout the HZ needs to be investigated further in order to constrain the likelihood of formation of HECs.

A second issue is whether or not Earth-mass cores can accrete large H/He envelopes in the first place. One of our key results is that Earth-mass HECs can form from mini-Neptunes with initial envelope mass fractions of up to 50\%; this would require a 1 M$_\oplus$ planetary embryo to accrete an equivalent amount of gas from the disk. While such a planet would be stable against RLO beyond the snow line, whether or not it could have formed is unclear. Gas accretion takes place in two different regimes: (i) a slow accretion regime, in which the envelope remains in hydrostatic equilibrium and gains mass only as it cools and contracts, evacuating a region that is then filled by nebular gas; and (ii) a runaway accretion regime, in which the rapidly increasing mass of the envelope leads to an increase in the size of the Hill sphere and increasingly faster gas accretion \citep[e.g.,][]{POL96}. Since the critical core mass for runaway accretion is thought to be somewhere in the vicinity of 10-20 M$_\oplus$ \citep{POL96,RAF06}, the progenitors of HECs must accrete their gas slowly. As we mentioned above, \cite{LAM14} find that Earth-mass planets typically form with $f_H \lesssim 0.03$ in the HZ of solar-type stars. This is consistent with recent analyses of Kepler planet data by \cite{ROG14} and \cite{WL14}, who find that most planets with radii less than about 1.5 $\mathrm{R_\oplus}$ (corresponding to masses less than $\sim 5 \mathrm{M_\oplus}$) are rocky, with typical H/He mass fractions of about 1\%. 

Formation beyond the snow line can increase $f_H$: \cite{ROG11} show that a planet with core mass 2.65 M$_\oplus$ accretes only 0.54 M$_\oplus$ of gas (corresponding to $f_H \approx 17\%$). \cite{BH14}, on the other hand, show that planets with cores in the range 2.2-2.5 M$_\oplus$ generally accrete less than 10\% of their mass in H/He. However, these authors terminated gas accretion at 2 Myr. For a longer disk lifetime of 4 Myr, \cite{BH14} show that a planet with core and envelope masses of 2.8 M$_\oplus$ and $\sim 1.2$ M$_\oplus$, respectively, can form ($f_H \approx 30\%$). This is particularly relevant to planet formation around M dwarfs, as these stars may have disk lifetimes in excess of 5 Myr \citep{CAR06,PAS09}, which could allow for larger initial H/He envelope fractions.

Nevertheless, while a 1 M$_\oplus$ core with a 1 M$_\oplus$ envelope is probably an unlikely initial condition, it is important to keep in mind that our results apply to planets that form with smaller H/He fractions as well. Figures \ref{fig:money1}-\ref{fig:money5} show where planets with \emph{up to} 50\% H/He can form HECs; planets with lower H/He mass fractions also form HECs at the same values of $a$ and $M_\star$. See \S\ref{sec:discuss:fh} for a more detailed discussion.

\subsection{Are HECs Habitable?}
\label{sec:discuss:hab}
Under the core accretion mechanism, mini-Neptunes are likely to form close to or beyond the snow line,  where disk densities are higher due to the condensation of various types of ices; these planets should therefore have large quantities of volatiles, including water, ammonia, and CO$_2$ ices. If we assume a disk composition similar to that around the young Sun, the bulk composition of their cores would probably be similar to that of comets: roughly equal parts ice and silicate rock, similar to what studies predict for the composition of water worlds \citep{LEG04,SEL07}. Once these planets have migrated into the HZ and lost their envelopes, it is quite likely that they would be water worlds and therefore not ``terrestrial'' in the strictest sense of the word. Whether or not such planets are habitable may depend on their ability to sustain active geochemical cycles, which are crucial for life on Earth today. 

One concern is a possible interruption of the carbon cycle by a high pressure ice layer at the bottom of the ocean \citep{LEG04}. Recently, \cite{ALI14} calculated the critical ocean mass for high pressure ice formation, finding that it lies between 0.02 and 0.03 M$_\oplus$ for an Earth-mass planet. If HECs are in fact comet-like in composition, a deep ice layer would separate their oceans from their mantles, which could inhibit the recycling of carbon and other bioessential elements between the two reservoirs, a process that is critical to life on Earth. However, whether this is the case is far from settled, as processes involving solid state ice convection could mediate the cycling of these elements. In particular, \cite{LSP13} and \cite{KAL13} presented a mechanism that could recycle CH$_4$ and CO$_2$ between the interior and the atmosphere of water worlds, invoking the ability of these molecules to form clathrates that could be convectively transported through the ice to the surface. Other mechanisms could also exist, and without further modeling, it is unclear whether a high pressure ice layer poses a threat to habitability.

The probable difference in composition between HECs and Earth is likely to have other geophysical implications. Hydrogen-rich compounds such as methane and ammonia could make up a non-negligible fraction of a HEC's mass, leading to extremely reducing conditions at the surface. These could also end up in a secondary atmosphere along with large quantities of CO$_2$, which could lead to strong greenhouse heating, although it is likely that most of the CO$_2$ and NH$_3$ would be sequestered in the ice mantle \citep{LEG04}. The compositional difference of HECs would also likely lead to mantle convection and tectonic activity different from Earth's, as well as differences in magnetic field generation by a possible core dynamo. Since both an active tectonic cycle and a magnetic field may be necessary for habitability, these issues need to be investigated further.

Also critical to the habitability of a HEC is its ability to outgas a secondary atmosphere once its primordial envelope is lost. In particular, the secondary atmosphere must be stable against erosion. While the power-law decay in XUV emission after $t_\mathrm{sat}$ could allow for such an atmosphere to form, low mass M dwarfs may remain active for $t \gg 1$ Gyr. Strong planetary magnetic fields could potentially shield these atmospheres; \cite{SEG10} showed that flares on the extremely active M dwarf AD Leo would have a small effect on the atmosphere of an Earth-like planet in the HZ provided it has an Earth-like magnetic field. However, interactions with the stellar wind could still pose serious problems to these and other planets in the HZs of M dwarfs. In particular, the high XUV/EUV fluxes of active M dwarfs can lead to significant expansion of the upper atmosphere, potentially causing the radius of the exobase to exceed the distance to the stellar magnetopause \citep{LIC10,LAM11}. For a pure N$_2$ atmosphere, \cite{LIC10} showed that nitrogen ionized by EUV radiation above the exobase is subject to ion pickup by the solar wind, leading to the complete erosion of a 1 bar atmosphere in as short a time as 10 Myr. A stronger planetary magnetic field or large quantities of a heavier background gas such as CO$_2$ may be necessary to suppress ion pickup and preserve secondary atmospheres on HECs.

Given the complex processes governing the habitability of HECs, a detailed investigation of these issues is left to future work. As we discuss below, stronger constraints on the details of the X-ray/EUV evolution of M dwarfs of all masses are essential to understanding the fate and ultimately the habitability of planets around these stars.

\subsection{The Need For Better Constraints}
\label{sec:discuss:constraints}
Figure~\ref{fig:money1} shows that the formation of HECs depends strongly on the atmospheric escape mechanism. As we mentioned earlier, whether a flow is closer to radiation/recombination-limited or energy-limited will depend on the ratio of the X-ray to EUV luminosity of the parent star. Low-mass M dwarfs may have XUV luminosities as high as $\sim 4\times 10^{29}$ erg/s early on (\S\ref{sec:lumevol}). If X-rays contribute more than a few percent of this luminosity, low-mass low-density planets in the HZs of these stars may undergo energy-limited X-ray-driven escape \citep[Figure 11 in][]{OJ12}. Unfortunately, knowledge of the exact age-luminosity relation in the X-ray and EUV bands for M dwarfs is still very poor, in great part because of the large uncertainties on these stars' ages. However, recent studies suggest that X-rays contribute a significant fraction of this luminosity \citep[for a review, see][]{SCA07}. In particular, \cite{STE13} report high ($L_X \gtrsim 10^{29}$ erg/s) X-ray luminosities for a sample of early active M dwarfs in the solar neighborhood, with a steeper age dependence than in FUV and NUV bands, which dominate the emission for $t\gtrsim 1$ Gyr. This is consistent with \cite{OJ12}, who argue that close-in mini-Neptunes may undergo a transition from X-ray-driven escape at early times to EUV-driven escape at later times.

Moreover, atmospheric X-ray heating and cooling is primarily done by metals. As \cite{OJ12} point out, atmospheric composition is also likely to play a role in determining whether hydrodynamic flows are EUV- or X-ray-driven. Additionally, the presence of dust in the envelopes of these planets could greatly affect their ability to cool via Lyman $\alpha$ radiation. Absorption of recombination radiation by dust particles lifted high into the envelope by vigorous convection could convert a significant fraction of this energy into heating, which could effectively increase the absorption efficiency $\exuv$ and bring the flow closer to the energy-limited regime. Unfortunately, our present parametric escape model is unable to address the effect of dust and metal abundances on the escape rate---this issue needs to be revisited in the future with more sophisticated hydrodynamic models.

The difference between Figure~\ref{fig:money1} and Figure~\ref{fig:money3} is just as significant. At lower $t_\mathrm{sat}$ and $\exuv$, the HEC parameter space is greatly reduced. As we discuss in \S\ref{sec:lumevol}, the lower value $t_\mathrm{sat} = 0.1$ Gyr is more representative of K and G dwarfs, while M dwarfs may remain saturated for $t_\mathrm{sat} > 1$ Gyr. In fact, the energy-limited contours in Figure~\ref{fig:money1} closely trace the CHZ/OHZ boundary as $M_\star$ increases, predicting that HECs may even be possible around solar type stars (not shown). This is of course not the case, since solar-type stars are known to leave the saturation phase around $t_\mathrm{sat} = 0.1$ Gyr \citep{RIB05}. For some stellar mass near the M/K Dwarf transition ($0.5\msun \lesssim M_{\star,crit} \lesssim 0.7\msun$), Figure~\ref{fig:money3} becomes a better description for HECs, but for low-mass M dwarfs Figure~\ref{fig:money1} is probably more appropriate.

Naturally, the exact value of $\exuv$ will also affect the possible distribution of HECs within the HZ. Recently, \cite{SHE14} performed numerical simulations to solve the kinetic Boltzmann equation for XUV-irradiated hydrogen atmospheres, finding an upper limit to the heating efficiency of $\exuv \approx 0.20$. Our nominal value $\exuv = 0.3$ may therefore be an overestimate for many mini-Neptunes. However, given that our goal in this paper is to explore where in the HZ HECs are possible and to map regions where the transition from gaseous to terrestrial is not possible, our present approach should suffice. Nevertheless, further studies constraining $\exuv$ are crucial to improving our understanding of this parameter space.

\subsection{Eccentricity Effects}
From Figures~\ref{fig:money1}-\ref{fig:money5}, we see that as we consider planets with higher current eccentricity, the region where evaporated cores are possible overlaps increasingly more of the HZ, particularly in the radiation/recombination-limited regime. This does not necessarily mean that HECs are more likely at high eccentricity---but rather that planets in circular orbits at certain $a$ cannot be HECs, while planets at higher eccentricity can. However, that being said, there is an interesting negative feedback between tides and mass loss that may enhance the probability of HECs on eccentric orbits. A planet with initially high eccentricity will undergo fast tidal evolution (due to the strong $\beta$ dependence in Equation~\ref{eq:dedtctl}), particularly if its radius is large (since $de/dt \propto R_p^5$). If a planet loses mass quickly (as is usually the case for gaseous planets with large $R_p$ and at large eccentricity), its radius will shrink and $de/dt$ will decrease. The earlier a planet sheds its envelope, the more likely it is to maintain a nonzero eccentricity in the long run (i.e., after $\sim 5$ Gyr). Since HECs typically form from such quickly evaporating planets, they are more likely to end up in eccentric orbits than their gaseous counterparts that did not lose their envelopes. There is, of course, a trade-off here in the sense that gas-free planets should have higher $\tau_p$ (and therefore faster $de/dt$) than gaseous planets, but the dependence on $\tau_p$ is linear and therefore much weaker than the dependence on the radius. What this means is that there is an interesting link between present eccentricity and mass loss history. Translating a planet's present eccentricity into a probability that it is an evaporated core is no easy task, however, and likely requires a detailed understanding of its initial orbital state and the migration mechanism(s) it underwent. On the other hand, statistical surveys of planets found to the left of the contours in Figure~\ref{fig:money1} could uncover interesting trends.

We note that Figures~\ref{fig:money1}-\ref{fig:money5} show eccentricity contours only up to $e = 0.5$. At final eccentricities higher than 0.5 today, the contour lines should move farther to the right, increasing the parameter space populated by HECs. However, since high eccentricities today in general require extremely high eccentricities in the past, we do not calculate contours above this level. We note, finally, that even though we run simulations with $e_0$ as high as 0.95, Figures~\ref{fig:money1}-\ref{fig:money5} remain unchanged for a lower cutoff $e_0 \lesssim 0.7$.

\subsection{Orbital Effects Due To Roche Lobe Overflow}
\label{sec:discuss:rlo}
In the present work we neglect any orbital effects due to Roche lobe overflow, which could in some cases lead to significant outward migration of the planet; see (\ref{eq:dadtrecoil}). We argued in section \S\ref{sec:model:escape} that since RLO should occur during the initial stage of migration (i.e., from beyond the snow line into the HZ), modeling it was outside the scope of the paper. However, a careful investigation of this initial migration process could uncover interesting couplings. For instance, mini-Neptunes that initially overshoot the HZ would experience strong RLO, which could in principle cause them to migrate \emph{back} into the HZ. These planets could have lost significantly more mass than the ones we considered here, since both RLO and XUV-driven escape would be stronger in their closer-in orbits.

A second important point concerning RLO-induced migration is that for nonzero eccentricity, (\ref{eq:dadtrecoil}) does not apply. In this case, it can be shown that angular momentum exchange between the gas and the planet at pericenter (where overflow is strongest) leads to a net \emph{increase} in the eccentricity. In the limiting case that both bodies may be treated as point masses and the mass transfer rate may be approximated as a delta function at pericenter (which is appropriate for high $e$), \cite{SEP09} show that
\begin{align}
\frac{da}{dt} &= -\frac{a}{\pi}\frac{\dot{M}_p}{M_p}(1-e^2)^{1/2}\left(1 - \frac{M_p}{M_\star}\right) \\
\frac{de}{dt} &= -\frac{1}{\pi}\frac{\dot{M}_p}{M_p}(1-e^2)^{1/2}(1-e)\left(1 - \frac{M_p}{M_\star}\right),
\end{align}
predicting that both $a$ and $e$ will tend to increase with time (recall that $\dot{M}_p$ is negative). Their relative rates of change are
\begin{align}
\frac{de}{dt} &= (1-e)\frac{1}{a}\frac{da}{dt}.
\end{align}
In other words, at intermediate values of $e$, the fractional rates of change in the semi-major axis and the eccentricity are comparable, and the eccentricity will increase proportionally to the semi-major axis.

As $e$ increases, so does the atmospheric mass loss rate (via $K_\mathrm{ecc}$), the Roche lobe overflow rate (modulo the increase in $a$), and the rate of tidal evolution, leading to potentially faster mass loss and rich couplings between these different processes. We will consider all these effects in a future paper.

\subsection{Different $f_H$}
\label{sec:discuss:fh}
Figures~\ref{fig:money1}-\ref{fig:money5} correspond to planets with initial hydrogen fraction $f_H = 0.5$. Planets with lower initial $f_H$ are naturally more unstable to complete loss of their envelopes---this would move all contours to the right, expanding the region where HECs are possible. However, for a different choice of initial $f_H$ in the range $0.1 \lesssim f_H \lesssim 0.5$, the figures change very little: at constant core mass, it is only marginally harder to fully evaporate a $f_H = 0.5$ envelope than it is to evaporate a $f_H = 0.1$ envelope (see discussion below). This is due to both Roche lobe overflow, which dramatically reduces the envelope mass early on, and the fast atmospheric escape for extremely inflated planets. In Figure~\ref{fig:int}(a), for instance, the envelope fraction decreases by a factor of 10 within the first $\sim 10$ Myr due to energy-limited escape. The escape process is in general very fast for large $f_H$, and bottlenecks for $f_H \lesssim 0.1$; thus the HEC boundary is relatively insensitive to the exact choice of the initial $f_H$.

\begin{figure}[h]
  \begin{center}
      \psfig{file=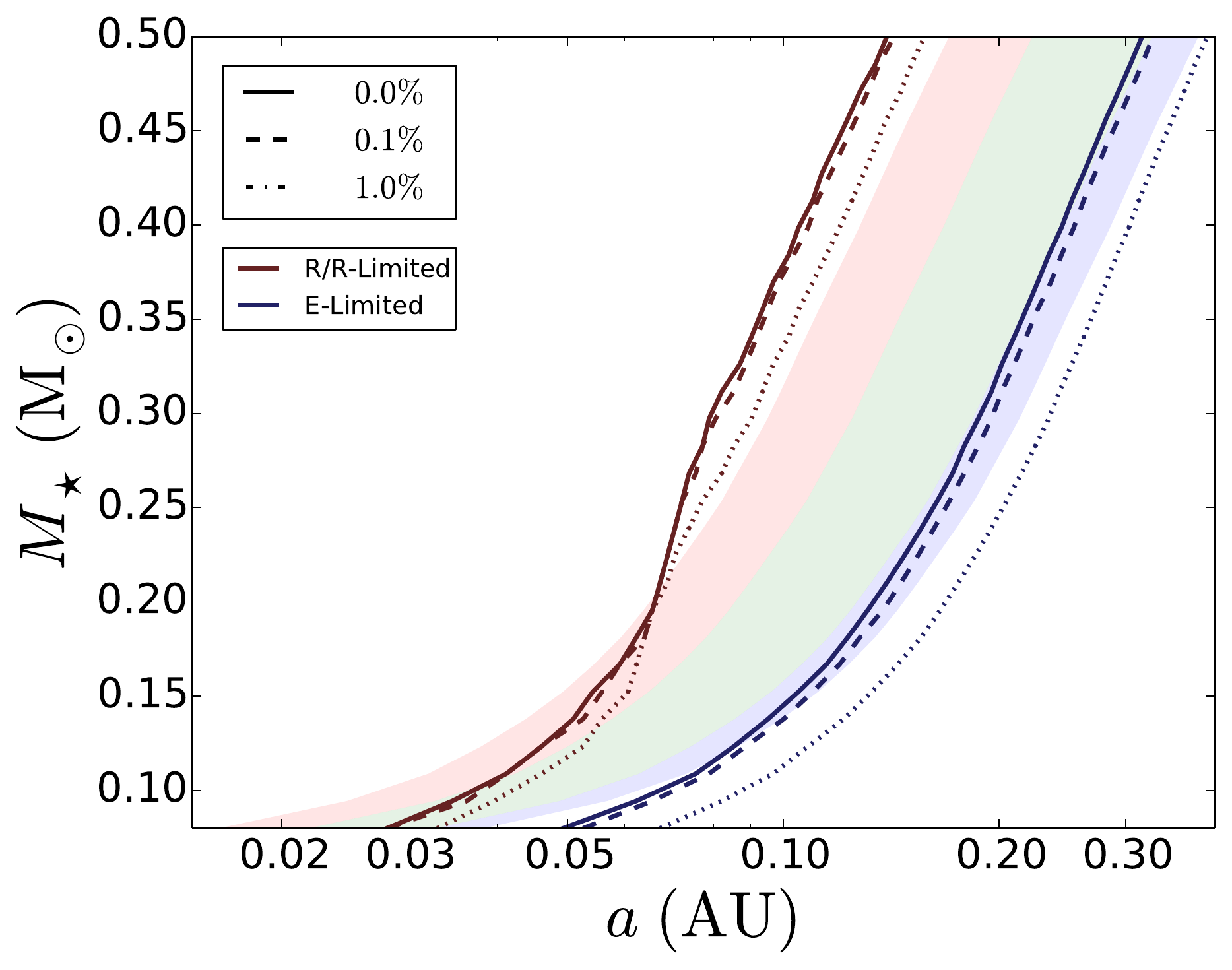,width=\columnwidth}
       \caption{Similar to Figure~\ref{fig:money1}, but here contours correspond to different choices of final $f_H$ for a $1\mearth$ core on a circular orbit. The solid lines correspond to $f_H' = 0$ and are the same as the $e=0$ contours in Figure~\ref{fig:money1}. Dashed lines correspond to the transition between planets that have less than (left) and more than (right) 0.1\% H/He ($f_H' = 0.001$) at 5 Gyr. Dotted lines correspond to the 1\% H/He ($f_H' = 0.01$) transition. While the $f_H' = 0.001$ contours are barely distinguishable from the $f_H' = 0$ contours, at final $f_H' = 0.01$ the HEC parameter space is significantly larger. However, it is unclear whether planets with $f_H' = 0.01$ could be habitable.}
     \label{fig:money1percent}
  \end{center}
\end{figure}

Along the same lines, we can ask how the choice of \emph{final} $f_H$ affects our results. In this study, we define an evaporated core as a planet with $f_H' = 0$ and no atmosphere. However, mini-Neptunes with substantial gaseous envelopes that evaporate down to $f_H' \sim 0.01$ become fundamentally solid planets. At such low $f_H'$, additional non-thermal mass loss processes, including flaring events and interactions with the stellar wind, may significantly erode the remaining envelope. Investigating the habitability of planets with nonzero $f_H'$ is beyond the scope of this paper, but it is worthwhile to consider how an alternative definition of a ``habitable evaporated core'' affects our results. To this end, in Figure~\ref{fig:money1percent} we plot contour lines corresponding to three different choices of the critical $f_H'$ below which we consider a planet to be a HEC. Solid lines correspond to $f_H' = 0$ (the default choice); dashed lines correspond to $f_H' = 0.001$; dotted lines, to $f_H' = 0.01$. Note that in the radiation/recombination-limited regime, this choice does not significantly affect evaporated cores in the HZ. However, for $f_H' = 0.01$, \emph{all} planets undergoing energy-limited escape in the HZ could be HECs.

Earths and super-Earths with up to a few percent of their mass in H/He may still harbor liquid water oceans, and so at this point their habitability can not be ruled out. At higher $f_H'$, however, the surface pressures in excess of $\sim$ 1 GPa \citep[e.g.,][]{CG07} result in the formation of high-pressure ices, at which point the planet will likely no longer be habitable.

\subsection{Other Atmospheric Escape Mechanisms}
We modeled the XUV emission of M dwarfs as a smooth power law, but active M dwarfs are seldom so well behaved. Frequent flares and coronal mass ejections punctuate the background XUV emission and will lead to erosion of planet atmospheres beyond what we model here. During flares, the ratio of the X-ray to bolometric luminosity ($L_\mathrm{X}/L_\mathrm{BOL}$) can increase by up to two orders of magnitude \citep{SCA07}. Moreover, interactions with the stellar wind and nonthermal escape mechanisms such as ion pick-up and charge-exchange can lead to a few Earth-ocean equivalents of hydrogen from these planets \citep{KIS13,KIS14}. However, as \cite{KIS13} demonstrate, the escape rate due to these processes generally amounts to only a few percent of the hydrodynamic escape rate. Including these non-thermal mechanisms would thus have a minor effect on our results.

In this study we also neglect the effects of magnetic fields. The presence of a strong planetary magnetic field may inhibit escape during flares and possibly decelerate the mass loss rate if the planetary wind is ionized. Since a strong magnetic field is likely a requirement of surface habitability around an M dwarf \citep{SCA07}, this point must be addressed in future work.

Finally, stars less massive than about 0.1M$_\odot$ may be in a ``supersaturation'' regime early on, saturating at XUV fluxes one or even two orders of magnitude below higher mass M dwarfs \citep{COO14}. This could reduce escape rates from planets around M dwarfs close to the hydrogen burning limit.

\subsection{Multi-Planet Systems}
In this paper we restricted our calculations to single-planet systems. However, many of the super-Earths and mini-Neptunes recently detected by Kepler reside in multi-planet systems, such as Kepler-32 \citep{FAB12}, Kepler-62 \citep{BOR13}, and Kepler-186 \citep{QUI14}. Moreover, \cite{SWI13} demonstrated how the five-planet system Kepler-32 could be representative of the entire ensemble of Kepler M dwarf systems, arguing that multiplicity could be the rule rather than the exception for these stars.

While tidal processes still operate in systems with multiple planets, the orbital evolution of individual planets will be much more complex. In general, though tidal dissipation may still lead to a decrease in the semi-major axis, the eccentricity evolution will be governed by secular interactions between the planets. The coupling between mass loss and tidal evolution we investigated in \S\ref{sec:res:dep_ecc} would likely be weaker, particularly for closely-packed coplanar systems where the eccentricities are necessarily low. However, planet-planet interactions could allow for even richer couplings to the atmospheric evolution of planets in the HZ. 

Consider, for instance, the case of a terrestrial planet in the HZ and a mini-Neptune interior to the HZ. Both mass loss and tidal dissipation scale inversely with the semi-major axis, so both effects could strongly shape the fate of the inner planet, which could in turn affect both the orbital and atmospheric evolution of the HZ planet in complex ways. The same could happen for a planet overflowing its Roche lobe interior to the HZ, which as we argued in \S\ref{sec:discuss:rlo} could experience large changes in $a$ and $e$.

Similarly, consider a scenario in which a planet in the HZ loses mass at the same time as it is perturbed by a more massive companion exterior to the HZ. As the planet's mass decreases, it will undergo larger swings in eccentricity, which in turn lead to faster mass loss and stronger orbital evolution.

Another interesting scenario involves planets close to mean motion resonances. For a system of two planets in resonance, \cite{LW12} and \cite{BM13} showed that the presence of a dissipative mechanism such as tidal evolution naturally leads to the repulsion of the two planets' orbits: the inner planet's orbit decays, while the outer planet gets pushed outward. In the hypothetical case of two mini-Neptunes in resonant orbits interior to the HZ, this mechanism could result in the migration of the outer planet \emph{into} the HZ, having lost significantly more mass (due to its initially smaller orbit) than if it had originated exterior to the HZ and migrated inwards. A chain of resonant planets, which is a potential outcome of disk-driven migration \citep{TP07,OI09}, could have similarly interesting consequences on planets in the HZ.

All of these scenarios, which involve coupling between atmospheric mass loss, tidal evolution, and secular planet-planet interactions, probably occur even for planets outside the HZ. Modeling this coupling could be critical to understanding systems like Kepler-36 \citep{CAR12}, where two planets with semi-major axes differing by $\sim$ 10\% have a density ratio close to 8, which \cite{LF13} show could be the result of starkly different mass loss histories. Future work should investigate how mass loss modifies the orbital interactions in multiplanet systems.

\subsection{Other Caveats}
\label{sec:discuss:caveats}

\subsubsection{The Habitable Zone}
Recently, \cite{YAN13} argued that cloud feedback on tidally locked planets can greatly increase the planetary albedo and move the IHZ in by a substantial amount. Similarly, \cite{ABE11} showed that planets with limited surface water are stable against runaway greenhouses at greater insolation than Earth, which also decreases the IHZ distance. Since evaporated cores are more likely at smaller $a$, a closer-in inner edge to the HZ could greatly increase the parameter space available to these planets. However, as we argued in \S\ref{sec:discuss:hab}, HECs are likely to have abundant surface water.

\cite{LUG14} showed that terrestrial planets in the HZ of M dwarfs may experience long runaway greenhouses during their host stars' extended pre-main sequence phases, during which time water loss to space can lead to their complete desiccation, rendering them uninhabitable. HECs are naturally more robust against severe water loss, given that their initially dense H/He envelopes can shield the surface and lower atmosphere from XUV radiation. Once the envelope is lost, water loss from the surface could occur, but by that point the stellar XUV flux will be much lower. Future work will address the fate of the water on HECs in detail.

\subsubsection{Thermal Evolution}
As mentioned before, we do not consider how the mass loss rate affects the planet radius, but instead assume the radius instantaneously returns to the ``equilibrium'' value for the given age, mass, and hydrogen fraction. In reality, the radius is likely to remain inflated for some time, particularly for fast mass loss \citep{LOP12}. Because our radii decrease too quickly, the mass loss feedback on the tidal evolution is always negative: mass loss always acts to dampen the tidal evolution. However, accurate modeling of the radius could enable a positive feedback, in which the effect of the decreasing mass (higher $da/dt$) overpowers the effect of the decreasing radius (lower $da/dt$) in (\ref{eq:dadt}) and (\ref{eq:dadtctl}), leading to faster orbital decay.

Currently, our planet radii are also independent of both the insolation and the degree of tidal heating. Under the extremely high fluxes and strong tidal forces at early times, planets in the HZs of M dwarfs are likely to be significantly more inflated than modeled here, resulting in faster atmospheric mass loss and likely a higher probability of complete envelope loss. We also do not model radiogenic heating, which could further add to the inflation of the planet. In this sense, our results are conservative, and HECs may in fact be possible at larger distances from their parent stars. A self-consistent thermal evolution model must be developed to accurately address this point.

We also emphasize that the quasi-static approximation (see the Appendix) may not be valid for the most inflated planets on eccentric orbits above $e \approx 0.4$. Given the large uncertainty concerning the orbital migration due to RLO for planets on eccentric orbits, we urge caution in interpreting our results quantitatively for highly eccentric planets.

\subsubsection{Tidal Evolution}
Given that the CPL tidal model is accurate only to second order in eccentricity, most of our integrations were performed under the CTL framework. At low $e$, however, these models predict qualitatively similar results for both the orbital migration and the coupling to the mass loss processes. At high $e$, on the other hand, tidal models lack observational validation, given the relatively low eccentricities of the major bodies in our solar system. We again urge caution in interpreting quantitative results for $e \gtrsim 0.5$.

Recall that our values of $Q_p$ and $\tau_p$ are fixed throughout the integrations and change discontinuously to the more dissipative rocky values as soon as the envelope is lost; this is  certainly an oversimplification. Though we defined $Q$ via its relationship to the phase lag in (\ref{eq:Q}), it may also be defined as the \emph{specific dissipation function} \citep{GS66}, defined as
\begin{align}
Q^{-1} &\equiv \frac{1}{2\pi E_0}\int_0^{2\pi}\left( -\frac{dE}{dt} \right)dt,
\end{align}
where $E_0$ is the maximum tidal energy stored in the system and the integral is the energy dissipated per cycle. $Q$ is therefore inversely proportional to the dissipation per unit energy stored. Planets with large values of $f_H$, where the
dissipation throughout most of the body is small, should therefore have large values of $Q_p$; rocky planets with small $f_H$, conversely, have large specific dissipation rates and correspondingly low values of $Q_p$, in agreement with the remarks in \S\ref{sec:tides:cpl}. Therefore, we would expect that as planets lose mass, $Q_p$ should decrease, corresponding to an increase in $\tau_p$. Modeling $Q_p(f_H)$ and $\tau_p(f_H)$, however, is outside the scope of this paper, and will be considered in future work. We simply note that as planets lose mass, the decreasing value of $Q_p$/increasing value of $\tau_p$ should offset the decreasing radii, leading to stronger tidal evolution and stronger couplings than reported here. 

One must likewise be careful in choosing the planet radius that goes into calculating the tidal evolution. This radius should be the effective radius of the dissipating material, and is thus dependent on the chosen value of $Q_p$ and $\tau_p$; for high $Q_p$/low $\tau_p$, the radius should probably be that of the envelope. We took this to be $R_p$, the 20 mbar radius shown in the tracks in Figure~\ref{fig:radius}; other choices of the tidal radius will lead to different evolutions.

We only consider planets with zero initial obliquity. For nonzero obliquity, the tidal evolution equations must be modified and will lead to differences in the evolution; see \cite{HLB11} and \cite{BAR13}. We also ignore the spin-up of the planet as it thermally contracts over time, since it should return to the equilibrium spin almost instantaneously. The excess angular momentum would most likely be absorbed into the orbit, but it is a small enough fraction of the orbital angular momentum that it can be safely neglected.

Finally, we note that throughout this paper we set $t_\mathrm{stop} = 5$ Gyr, based on the age of the solar system. For systems older than 5 Gyr, our results should change little, since both the tidal and atmospheric evolution will have strongly tapered off by this time. Younger systems, however, may still be in the throes of tidal decay and mass loss processes, and this must be kept in mind when searching for HECs.

\section{Conclusions}
\label{sec:conclusions}
We have shown that gas-rich mini-Neptunes that migrate early ($t \lesssim 10$ Myr) into the HZs of M dwarfs can naturally shed their gaseous envelopes and form gas-free Earth-mass planets. Together, Roche lobe overflow and hydrodynamic escape can remove up to a few Earth masses of hydrogen and helium from these planets, potentially turning them into ``habitable evaporated cores'' (HECs). This process is most likely for mini-Neptunes with solid cores on the order of 1$\mearth$ and up to about 50\% H/He by mass, and can occur around all M dwarfs, particularly close to the inner edge fo the HZ. HECs are less likely to form around K and G dwarfs because of these stars' shorter super-luminous pre-main sequence phases and shorter XUV saturation timescales. Furthermore, we find that HECs cannot form from mini-Neptunes with core masses greater than about 2$\mearth$ and more than a few percent H/He by mass; thus, massive terrestrial super-Earths currently in the HZs of M dwarfs have probably always been terrestrial. Our results are thus similar to those of \cite{LAM14}, who showed that planets more massive than $\sim$ 1.5 M$_\oplus$ typically cannot lose their accreted nebular gas in the HZs of solar-type stars.

Whether or not a given mini-Neptune forms a HEC is highly dependent on the early XUV evolution of the host star. In particular, a long XUV saturation timescale ($t_\mathrm{sat} \gtrsim 1$ Gyr) is needed to fully evaporate the envelopes of mini-Neptunes in the HZs of early and mid M dwarfs. While a large $t_\mathrm{sat}$ is consistent with the long activity timescales \citep{WES08} and long spin-down times \citep{PIZ00} of M dwarfs, more observations are needed to pin down $t_\mathrm{sat}$ for these stars. Moreover, the relative strength of the X-ray and EUV luminosity early on also affects whether or not HECs can form, as this determines whether the atmospheric escape is energy-limited or radiation/recombination-limited. In the energy-limited regime, which occurs for an X-ray dominated flow, the escape is fast and HECs can form throughout most of the HZ of all M dwarfs. In the radiation/recombination-limited regime, which applies to an EUV dominated flow, HECs only form close to the inner edge of low mass M dwarfs.

We further find that HECs can form from planets on circular as well as eccentric orbits, though they are marginally more likely to have higher $e$ in the long run. While there exist feedbacks between atmospheric mass loss and tidal evolution, we find that these are only significant at $e \gtrsim 0.5$; in these cases, whether or not a HEC forms can depend just as strongly on the tidal properties of the planet as on the efficiency of the atmospheric escape.

Many of the Earth-mass terrestrial planets detected in the HZs of M dwarfs in the coming years could be habitable evaporated cores. These planets should have abundant surface water and are likely to be water worlds, whose potential for habitability should be investigated further.

\vspace*{0.5in}

\small
\noindent We wish to thank Russell Deitrick, Eric Agol, and the rest of the VPL team for their insightful suggestions and many rich discussions. We also wish to thank two anonymous referees for their excellent comments and suggestions. This work was supported by the NASA Astrobiology Institute's Virtual Planet Laboratory under Cooperative Agreement solicitation NNA13AA93A and by a generous fellowship from the ARCS Seattle chapter.

\normalsize

\clearpage
\begin{appendix}
\section{Mass Loss for Eccentric Orbits}
\subsection{The Quasi-Static Approximation}
In \S\ref{sec:massloss:ecc} we argued that for planets on eccentric orbits, the Roche lobe radius may be calculated by replacing $a$ with the instantaneous planet-star distance $r$ in (\ref{eq:roche}), leading to tidally enhanced mass loss close to pericenter. However, this is true only if the orbital period is large compared to the dynamical timescale of the planet, thus allowing sufficient time for the atmosphere to assume the equilibrium shape due to the changing potential. \cite{SEP07} call this limit the \emph{quasi-static approximation}. Below we show that this approximation is valid for most of our planets.

As in \cite{SEP07}, we begin by introducing the parameter
\begin{align}
\alpha(e,f) = \frac{(1+e)^{1/2}}{(1-e)^{3/2}}\times |1-f|
\end{align}
where
\begin{align}
f = \frac{P_\mathrm{orb}}{P_\mathrm{rot}}\frac{(1-e)^{3/2}}{(1+e)^{1/2}}
\end{align}
is the ratio of the rotational angular velocity to the orbital angular velocity at periastron. \cite{SEP07} show that provided the condition
\begin{align}
\frac{P_\mathrm{orb}}{\tau_\mathrm{dyn}} \gg \alpha(e,f)
\end{align}
holds, the system may be treated as quasi-static. The orbital period is given by Kepler's law,
\begin{align}
P_\mathrm{orb} = 2\pi\sqrt{\frac{a^3}{GM_\star}},
\end{align}
while $\tau_\mathrm{dyn}$, the dynamical timescale of the planet, is
\begin{align}
\tau_\mathrm{dyn} \sim \frac{1}{\sqrt{GM_p/R_p^3}}.
\end{align}
The ratio of the two will be smallest for close-in, low-mass planets with large radii. The minimum value in our runs occurs for a super-inflated $2\mearth$, $30\rearth$ planet in the IHZ of a $0.08\msun$ M dwarf, for which $P_\mathrm{orb}/\tau_\mathrm{dyn} \approx 2.4$.

\begin{figure*}[h]
  \begin{center}
      \psfig{file=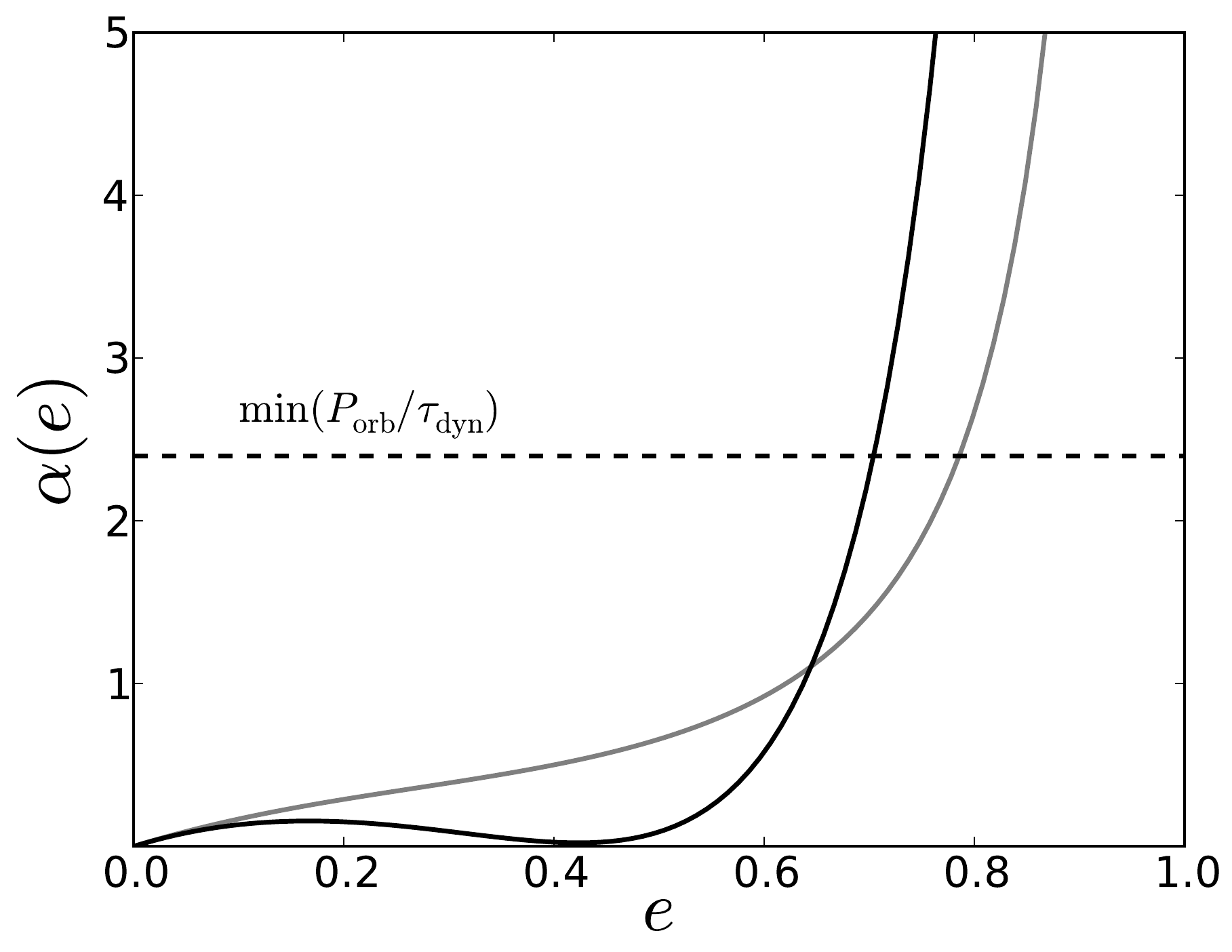,width=3in}
       \caption{The \cite{SEP07} parameter $\alpha$ as a function of $e$ for a synchronously-rotating planet in the CPL model (black) and in the CTL model (gray). The dashed line corresponds to the minimum value of the ratio $P_\mathrm{orb}/\tau_\mathrm{dyn}$
       across all our runs. Note that for $e \lesssim 0.5$, the quasi-static approximation is probably valid.}
     \label{fig:alpha}
  \end{center}
\end{figure*}

We must now compare this to $\alpha(e,f)$. The equilibrium rotational period for a synchronously-rotating planet is obtained from (\ref{eq:omeqcpl}) and (\ref{eq:omeqctl}) in the CPL and CTL models, respectively,
so that we may write $\alpha(e)$ as
\begin{align}
\label{ap:eq:alpha}
\alpha(e,f) = \frac{(1+e)^{1/2}}{(1-e)^{3/2}}\times \bigg|1-\frac{(1-e)^{3/2}}{(1+e)^{1/2}}\frac{\omega_{2,\mathrm{eq}}}{n} \bigg|.
\end{align}
This equation is plotted in Figure~\ref{fig:alpha}. Note that $\alpha$ only begins to approach $P_\mathrm{orb}/\tau_\mathrm{dyn}$ for $e \gtrsim 0.6$ in the CPL model and  $e \gtrsim 0.4$ in the CTL model, and only for the most inflated planets in the IHZ.
We therefore urge caution in interpreting results above $e \gtrsim 0.5$, where the quasi-static approximation may not hold for some planets.

\subsection{The Mass Loss Enhancement Factor}
\label{ap:sec:kecc}
Since the flux $F_\mathrm{XUV}$ is not constant over the course of one orbit, (\ref{eq:dmdt}) must be modified slightly:
\begin{align}
\label{ap:eq:dmdt}
\dot{M}(t) = \frac{R_\mathrm{XUV}^3\epsilon_\mathrm{XUV}L_\mathrm{XUV}}{4GM_p}\left[r(t)^2\left(1-\frac{3}{2\xi(t)} + \frac{1}{2\xi(t)^3}\right)\right]^{-1},
\end{align}
where $r(t)$ is the instantaneous separation between the centers of mass of the star and the planet and where we have plugged-in for $F_\mathrm{XUV}$ in terms of $L_\mathrm{XUV}$ and made use of (\ref{eq:ktide}). The parameter $\xi$ must also be modified,
as the value of the Roche radius will also change as the planet's distance from the star changes during one orbit:
\begin{align}
\label{ap:eq:xi}
\xi(t) 	&\equiv \frac{R_\mathrm{Roche}}{R_\mathrm{XUV}} \nonumber \\
		&= \left[\left(\frac{M_p}{3M_\star}\right)^{1/3}\frac{1}{R_\mathrm{XUV}}\right]r(t) = \frac{r(t)}{a}\xi.
\end{align}
To avoid confusion, henceforth $\xi$ will denote the original expression for circular orbits (Equations \ref{eq:xi} and \ref{eq:roche}), while $\xi(t)$ is the time-dependent parameter given by the expression above.
The time-average of $\dot{M}$ over one orbit is
\begin{align}
\label{ap:eq:avgdmdt}
\avg{\dot{M}}_t = \frac{1}{2\pi/n}\int_0^{2\pi/n}\dot{M}(t)dt
\end{align}
where $n$ is the mean motion. Now, we know that the relationship between $n$ and the eccentric anomaly $E$ is
\begin{align}
nt = E - e\sin E
\end{align}
and that
\begin{align}
\label{ap:eq:ra}
r(E) = a(1-e\cos E),
\end{align}
so
\begin{align}
\label{ap:eq:dtdE}
\frac{dt}{dE} 		&= \frac{1}{n}(1-e\cos E) = \frac{r(E)}{an}.
\end{align}
Substituting into (\ref{ap:eq:avgdmdt}), we have
\begin{align}
\label{ap:eq:avgdmdt2}
\avg{\dot{M}}_t 	&= \frac{1}{2\pi/n}\int_0^{2\pi}\dot{M}(E)\frac{dt}{dE}dE \nonumber \\
			&= \frac{1}{2\pi a}\int_0^{2\pi}\dot{M}(E)r(E)dE.
\end{align}
Now, introducing the mass loss rate for $e = 0$ and $K_\mathrm{tide} = 1$ from (\ref{eq:dmdt}),
\begin{align}
\dot{M}_0 &\equiv \frac{R_\mathrm{XUV}^3\exuv L_\mathrm{XUV}}{4GM_pa^2},
\end{align}
it follows from (\ref{ap:eq:dmdt}) and (\ref{ap:eq:xi}) that
\begin{align}
\dot{M}(E) = \dot{M}_0a^2\left[r(E)^2\left(1-\frac{3}{2\xi}\left(\frac{a}{r(E)}\right)+ \frac{1}{2\xi^3}\left(\frac{a}{r(E)}\right)^{3}\right)\right]^{-1}.
\end{align}
Plugging this into (\ref{ap:eq:avgdmdt2}), and using (\ref{ap:eq:ra}),
\begin{align}
\label{ap:eq:exact}
\avg{\dot{M}}_t	&= \frac{\dot{M}_0}{2\pi}a\int_0^{2\pi} \left(r(E)-\frac{3}{2\xi}a + \frac{1}{2}\left(\frac{a}{\xi}\right)^3r(E)^{-2}\right)^{-1}dE \nonumber\\
				&= \frac{\dot{M}_0}{2\pi}\int_0^{2\pi} \left[(1-e\cos E) - \frac{3}{2\xi} + \frac{1}{2\xi^3(1-e\cos E)^2}\right]^{-1}dE \nonumber \\
				&= \frac{\dot{M}_0}{K_\mathrm{ecc}},
\end{align}
where we define the eccentric mass loss enhancement factor
\begin{align}
\label{ap:eq:kecc}
1/K_\mathrm{ecc} \equiv \frac{1}{2\pi}\int_0^{2\pi} \left[(1-e\cos E) - \frac{3}{2\xi} + \frac{1}{2\xi^3(1-e\cos E)^2}\right]^{-1}dE.
\end{align}
Note that for $e = 0$, $K_\mathrm{ecc}$ reduces to the circular version of $K_\mathrm{tide}$, (\ref{eq:ktide}). Unfortunately, the integral (\ref{ap:eq:exact}) is not analytic. However,
if the last term in the integral is small compared to the first two, an analytic solution is possible. In particular, if we write
\begin{align}
\frac{1}{2\xi^3(1-e\cos E)^2} = \eta\left[(1-e\cos E) - \frac{3}{2\xi}\right],
\end{align}
then we may neglect the last term provided $\eta \ll 1$. Since the term is largest at pericenter ($E = 0$), we may instead require that
\begin{align}
\label{ap:eq:etamax}
0 < \frac{1}{2\xi^3(1-e)^2(1-\frac{3}{2\xi}-e)} \ll 1.
\end{align}
If this holds, (\ref{ap:eq:exact}) simplifies to
\begin{align}
\label{ap:eq:analytic}
\avg{\dot{M}}_t	&\approx \frac{\dot{M}_0}{2\pi}\int_0^{2\pi} \left(1-\frac{3}{2\xi}-e\cos E\right)^{-1}dE \nonumber \\
				&\approx \frac{\dot{M}_0}{2\pi(1-\frac{3}{2\xi})}\int_0^{2\pi} \left[1-\left(\frac{e}{1-\frac{3}{2\xi}}\right)\cos E\right]^{-1}dE \nonumber \\
				&\approx \frac{\dot{M}_0}{1-\frac{3}{2\xi}}\frac{1}{\sqrt{1-\left(\frac{e}{1-\frac{3}{2\xi}}\right)^2}} \nonumber \\
				&\approx \frac{\dot{M}_0}{\left( 1- \frac{3}{\xi} - \frac{9}{4\xi^2} -e^2 \right)^{1/2}}.
\end{align}
Thus, provided condition (\ref{ap:eq:etamax}) holds (typically for $\xi \gtrsim 10$), we may write
\begin{align}
\label{ap:eq:keccapp}
K_\mathrm{ecc} \approx \sqrt{1- \frac{3}{\xi} - \frac{9}{4\xi^2} -e^2}.
\end{align}
We note that for $e = 0$ and keeping only first-order terms in $\xi$, the above expression agrees with that of \cite{ERK07} for $\xi \gg 1$.

\begin{figure*}[t]
  \begin{center}
      \psfig{file=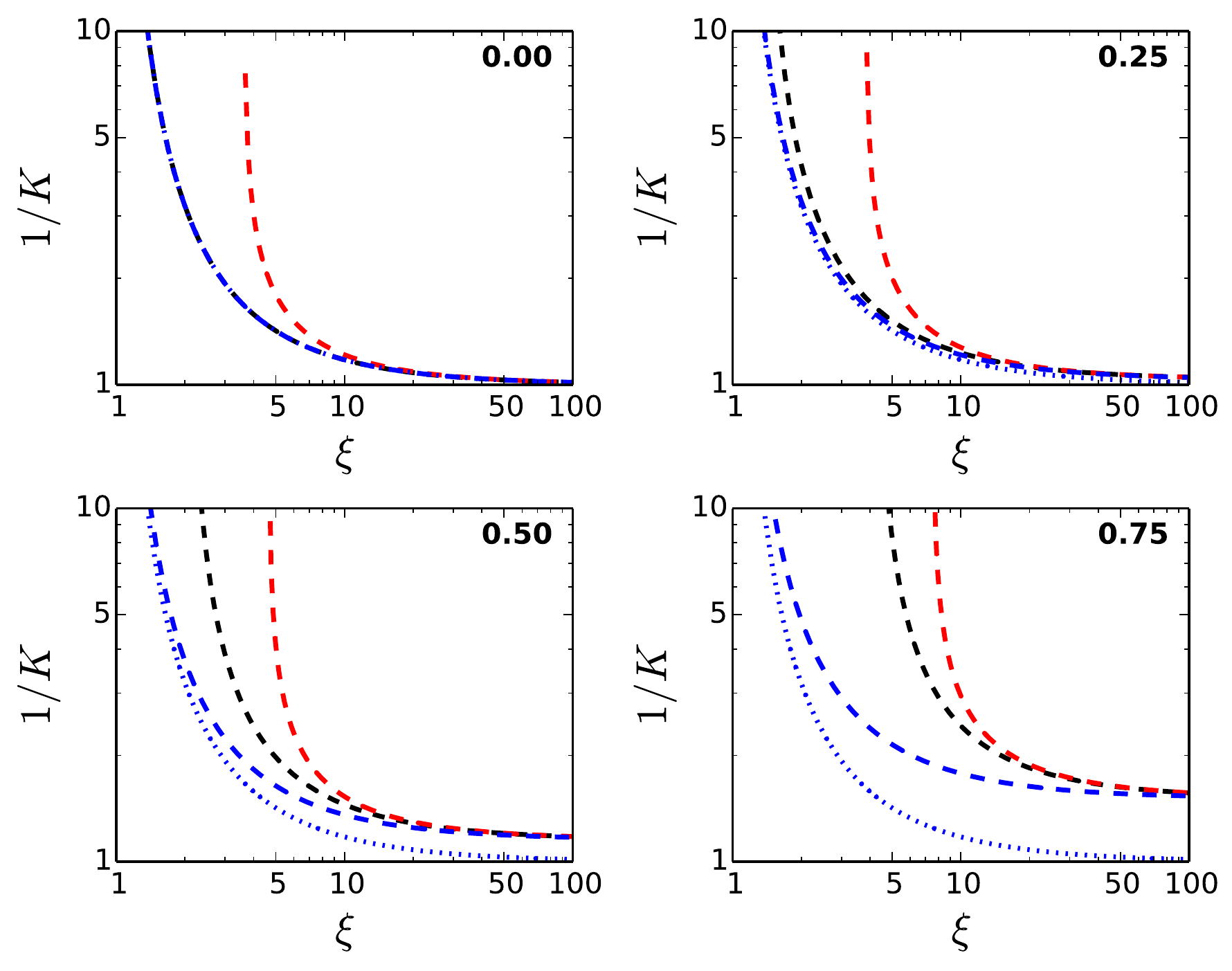,width=4in}
       \caption{Loss rate enhancement factor $1/K$ as a function of the normalized Roche lobe distance $\xi = R_\mathrm{roche}/R_p$ for $e = 0,\ 0.25,\ 0.5$ and $0.75$.
       The \cite{ERK07} model ($K_\mathrm{tide}$) is plotted in blue, with the flux enhancement factor $1/\sqrt{1-e^2}$ (dashed) and without it (dotted). The exact expression $K_\mathrm{ecc}$ derived
       in this work (\ref{ap:eq:exact}) is plotted in black and the analytic approximation (\ref{ap:eq:analytic}) is plotted in red. For low values of $\xi$,
       condition (\ref{ap:eq:etamax}) is not satisfied and the approximation diverges significantly. However, at high eccentricity, as $\xi$ increases, the analytic
       approximation converges to the exact value of $K_\mathrm{ecc}$ sooner than $K_\mathrm{tide}$. Finally, we note that for sufficiently large $\xi$, 
       all curves accounting for the flux enhancement converge to the same value.}
     \label{fig:kecc}
  \end{center}
\end{figure*}

In Figure~\ref{fig:kecc} we plot the loss rate enhancement factor of \cite{ERK13}, $1/K_\mathrm{tide}$ (blue lines), along with the eccentric version, $1/K_\mathrm{ecc}$ (exact\footnote{While we refer to (\ref{ap:eq:kecc}) as the ``exact'' expression for the (inverse of the) mass loss enhancement factor, it is important to remember that
it is the eccentric version of the third-order expression derived by \cite{ERK07} and is, in this sense, still an approximation to the true enhancement.}
expression in black, approximate expression in red). Recall that there are two distinct effects 
contributing to the extra mass loss for eccentric orbits: the $1/\sqrt{1-e^2}$ flux enhancement and the smaller Roche lobe distance during
part of the orbit. In order to better distinguish between these two, we display both $1/K_\mathrm{tide}$ (dotted blue lines) and $1/(K_\mathrm{tide}\sqrt{1-e^2})$ (dashed blue lines).
Thus, any difference between $1/K_\mathrm{tide}$ and $1/K_\mathrm{ecc}$ in the figure is due \emph{solely} to the Roche lobe effect.

Let us first compare the curves corresponding to the exact expressions (blue and black). For $e = 0$, $K_\mathrm{ecc} = K_\mathrm{tide}$, as expected. As the eccentricity increases to 0.25, the flux
enhancement effect remains small (i.e., the dashed and dotted curves are similar), while the Roche lobe effect begins to become significant (the dashed black curve exceeds the blue dashed curve), particularly for low values of $\xi$. 
For $e \gtrsim 0.5$, the effect becomes even more important, leading to an enhancement of a factor of several for $\xi \lesssim 10$.

A particularly important effect is that high eccentricities will cause the planet to undergo Roche lobe overflow at larger values of $\xi$. In the circular \cite{ERK07} model,
Roche lobe overflow occurs when $\xi = 1$ by definition. However, it is straightforward to show that for
\begin{align}
\xi = \xi_\mathrm{crit} = \frac{1}{1-e},
\end{align}
the expression in the integral (\ref{ap:eq:kecc}) diverges, resulting in an infinite mass loss rate. This occurs because at pericenter, where $r = a(1-e)$, $\xi(r) = 1$. In other
words, for $\xi \leq \xi_\mathrm{crit}$, the planet will overflow its Roche lobe during at least part of its orbit, leading to rapid mass loss.

It is important to note that the approximate expression (dashed red curve) converges only for $\xi \gtrsim 10$. For smaller values of $\xi$, it overestimates the mass loss enhancement significantly
and should therefore not be used. However, for high eccentricities (see the last panel, for instance), it converges to $1/K_\mathrm{ecc}$ quicker than $1/K_\mathrm{tide}$. Provided condition (\ref{ap:eq:etamax})
is satisfied, the approximate analytic expression (\ref{ap:eq:analytic}) does a good job at modeling the actual mass loss.

\subsection{Orbits Crossing the Critical Radius}
\label{ap:sec:crossing}
As we described in \S\ref{sec:massloss:rrlim}, the escape regime for an EUV-dominated flow may switch from energy-limited to radiation/recombination-limited above a certain critical value of the flux, $F_\mathrm{crit}$, given by
\begin{align}
F_\mathrm{crit} = \left(\frac{B}{A} \right)^2,
\end{align}
where
\begin{align}
A \equiv \frac{\pi \epsilon_\mathrm{XUV} R_\mathrm{XUV}^3}{G M_p K_\mathrm{tide}}
\end{align}
and
\begin{align}
B \equiv 7.11\times 10^{7}\ \mathrm{g^\frac{1}{2}s^\frac{1}{2}} \left( \frac{R_\mathrm{p}}{R_\oplus} \right)^\frac{3}{2}
\end{align}
are the coefficients multiplying the flux in the energy-limited and radiation/ recombination-limited equations, respectively (Equations \ref{eq:dmdt} and \ref{eq:dmdtrr}). At fixed XUV luminosity, this corresponds to a certain critical orbital radius,
\begin{align}
r_\mathrm{crit} = \sqrt{\frac{L_\mathrm{XUV}}{4\pi F_\mathrm{crit}}}.
\end{align}
Planets on low-eccentricity orbits that do not cross $r_\mathrm{crit}$ are therefore safely within either the energy-limited ($a \gg r_\mathrm{crit}$) or radiation/recombination-limited ($a \ll r_\mathrm{crit}$) regime. In this case, the orbit-averaged mass loss rate is determined simply by replacing $F$ in (\ref{eq:dmdt}) and (\ref{eq:dmdtrr}) by its orbit-averaged value, $\left< F \right>$, given by (\ref{eq:savg}).

However, when an orbit is sufficiently eccentric that the atmospheric escape regime switches from energy-limited to radiation/recombination-limited over the course of a single orbit, the method outlined above is no longer rigorously correct. We must instead integrate the two mass loss rate expressions over the portions of the orbit where they apply. Fortunately, as we demonstrate below, these integrals are analytic. This allows us to calculate the time-averaged value of the mass loss rate, $\avg{\dot{M}}_t$, much as we did in \S\ref{sec:massloss:ecc}:
\begin{align}
\avg{\dot{M}}_t = \frac{A}{P}\int_\mathrm{EL} F(t)dt + \frac{B}{P}\int_\mathrm{RRL} F^\frac{1}{2}(t)dt,
\end{align}
where $P$ is the orbital period, $F(t)$ is the instantaneous XUV flux, and $EL$ and $RRL$ correspond to the energy-limited and radiation/recombination-limited portions of the orbit, respectively. By noting as in (\ref{ap:eq:dtdE}) that
\begin{align}
dt = \frac{r(E)}{an}dE = \frac{r(E)P}{2\pi a}dE,
\end{align}
where $E$ is the eccentric anomaly, we may write
\begin{align}
\avg{\dot{M}}_t &= 
	\frac{B}{2\pi a}\int_{0}^{E_\mathrm{crit}}F^\frac{1}{2}(E)r(E)dE\ + \nonumber \\
	&\phantom{X}\frac{A}{2\pi a}\int_{E_\mathrm{crit}}^{2\pi-E_\mathrm{crit}}F(E)r(E)dE\ + \nonumber \\ 
	&\phantom{X}\frac{B}{2\pi a}\int_{2\pi-E_\mathrm{crit}}^{2\pi}F^\frac{1}{2}(E)r(E)dE,
\end{align}
where 
\begin{align}
E_\mathrm{crit} = \cos^{-1}\left( \frac{1}{e} - \frac{r_\mathrm{crit}}{ae} \right)
\end{align}
is the value of the eccentric anomaly when $r = r_\mathrm{crit}$. The three integrals above follow from the fact that starting from pericenter ($E = 0$), the planet is (by assumption) in the RRL regime up until $E = E_\mathrm{crit}$, switches to EL until $E = 2\pi - E_\mathrm{crit}$, and completes the orbit in the RRL regime. By symmetry of the orbit, the first and last integrals are identical, so we may simplify:
\begin{align}
\avg{\dot{M}}_t &= 
	\frac{A}{2\pi a}\int_{E_\mathrm{crit}}^{2\pi-E_\mathrm{crit}}F(E)r(E)dE\ + \nonumber \\
	&\phantom{X.}\frac{B}{\pi a}\int_{0}^{E_\mathrm{crit}}F^\frac{1}{2}(E)r(E)dE.
\end{align}
Now, noting that $F(E) = L_\mathrm{XUV}/4\pi r^2(E)$ and $r(E) = a(1-e\cos E)$, we have
\begin{align}
\avg{\dot{M}}_t &= 
	\frac{AL_\mathrm{XUV}}{8\pi^2 a^2}\int_{E_\mathrm{crit}}^{2\pi-E_\mathrm{crit}}\frac{dE}{1-e\cos E} + \frac{B}{\pi a}\sqrt{\frac{L_\mathrm{XUV}}{4\pi}}\int_{0}^{E_\mathrm{crit}}dE \nonumber \\[0.25in]
	&= \frac{AL_\mathrm{XUV}}{8\pi^2 a^2}\int_{E_\mathrm{crit}}^{2\pi-E_\mathrm{crit}}\frac{dE}{1-e\cos E} + \frac{BE_\mathrm{crit}}{\pi a}\sqrt{\frac{L_\mathrm{XUV}}{4\pi}}.
\end{align}
By evaluating the integral, we may finally write
\begin{align}
\avg{\dot{M}}_t = A\bar{F}\left[ 1 - \frac{2}{\pi}\tan^{-1}\left( \frac{(1+e)\tan(E_\mathrm{crit}/2)}{\sqrt{1-e^2}} \right) \right] + B\bar{F}^\frac{1}{2} \left[ \frac{(1-e^2)^\frac{1}{4}E_\mathrm{crit}}{\pi} \right].
\end{align}
Note, importantly, that the expressions above are valid only for planets that cross the critical radius during their orbit; that is, planets for which the expression $a(1-e) < r_\mathrm{crit} < a(1+e)$ holds. The mass loss rate for planets outside this region must be calculated via the method described at the beginning of the section.

Finally, note that the formalism derived here makes use of $K_\mathrm{tide}$ rather than its eccentric version $K_\mathrm{ecc}$, which we derived in \S\ref{ap:sec:kecc}. It is in principle possible to account for the tidal enhancement during the energy-limited portion of the orbit, but the resulting integral would not be analytic. Moreover, the tidal enhancement due to the eccentricity is important primarily near pericenter, where the mass loss is radiation/recombination-limited and therefore independent of $K_\mathrm{tide}$. Thus, while the method outlined above may underpredict the mass loss rate in some cases, the effect will be small.

\vspace*{0.1in}
\section{Tidal Evolution Expressions: CPL}
For zero inclination and zero obliquity, the tidal evolution equations for $a$, $e$, and $\omega_i$ in a two-body system are \citep{BAR13}
\begin{align}
\label{eq:dadt}
\frac{da}{dt}&=\frac{a^2}{4GM_\star M_p}\sum_{i\neq j}Z_i'\left(4\epsilon_{0,i}+e^2\left[-20\epsilon_{0,i} + \frac{147}{2}\epsilon_{1,i} + \frac{1}{2}\epsilon_{2,i} - 3\epsilon_{5,i}\right]\right) \\[0.2cm]
\label{eq:dedt}
\frac{de}{dt}&=-\frac{ae}{8GM_\star M_p}\sum_{i\neq j}Z_i'\left(2\epsilon_{0,i}-\frac{49}{2}\epsilon_{1,i} + \frac{1}{2}\epsilon_{2,i}+3\epsilon_{5,i}\right) \\[0.2cm]
\label{eq:domdt}
\frac{d\omega_i}{dt}&=-\frac{Z_i'}{8M_ir_{g,i}^2R_i^2n}\left(4\epsilon_{0,i}+e^2\left[-20\epsilon_{0,i} + 49\epsilon_{1,i} + \epsilon_{2,i}\right]\right),
\end{align}
where the sums are taken over the two bodies. Here, $G$ is the gravitational constant, $R_i$ is the radius of the $i^{th}$ body (planet or star), and $r_{g,i}$ are the radii of gyration. The parameter $Z_i'$ is defined
\begin{align}
Z_i'&\equiv 3G^2k_{2,i}M_j^2(M_i+M_j)\frac{R_i^5}{a^9}\frac{1}{nQ_i},
\end{align}
where $k_{2,i}$ is the Love number of degree 2 of the $i^{th}$ body, which is of order unity and quantifies the contribution of the tidal deformation to the total potential, $n$ is the mean motion of the secondary body (i.e., the planet), and $Q_i$ are the tidal quality factors.  The parameters $\epsilon_{N,i}$ are the signs of the phase lags (assumed equal in magnitude) of the $N^{th}$ wave on the $i^{th}$ body, calculated from
\begin{align}
\label{eq:lags}
\epsilon_{0,i}&=\mathrm{sgn}(2\omega_i-2n)\nonumber\\
\epsilon_{1,i}&=\mathrm{sgn}(2\omega_i-3n)\nonumber\\
\epsilon_{2,i}&=\mathrm{sgn}(2\omega_i-n)\nonumber\\
\epsilon_{5,i}&=\mathrm{sgn}(n)\nonumber\\
\epsilon_{8,i}&=\mathrm{sgn}(\omega_i-2n)\nonumber\\
\epsilon_{9,i}&=\mathrm{sgn}(\omega_i).
\end{align}
Since short-period planets around M dwarfs are likely to be tidally locked, one need not calculate the planetary spin from (\ref{eq:domdt}). Instead, the planet's rotation rate may be calculated from \citep{GOL66}:
\begin{align}
\label{eq:omeqcpl}
\omega_{p,\mathrm{eq}}^\mathrm{CPL} = n(1+9.5e^2),
\end{align}
where $n$ is the mean motion.

\vspace*{0.1in}
\subsection{The Typical Case}
Because of the fast rotation of M dwarfs at early times, planets in the HZ are likely to be far outside the corotation radius of their parent stars. It is useful at this point to consider as an example the specific case of a tidally-locked planet for which $n \ll \omega_\star$. In this case, the stellar phase lags (\ref{eq:lags}) are all positive and $\epsilon_{N,\star} = +1$. For a tidally locked planet, $\epsilon_{2,p} = \epsilon_{5,p} = \epsilon_{9,p} = 1$, and  $\epsilon_{1,p} = \epsilon_{8,p} = -1$. The parameter $\epsilon_{0,p}$, however, is less straightforward to calculate. If tidal locking is to be maintained, the average angular acceleration over one period must be zero. \cite{FER08} show that the only self-consistent way to ensure this is if $\epsilon_{0,2}$ has a different magnitude than the other lags, equal to
\begin{align}
\varepsilon_{0,p}&= 12e^2\varepsilon_{2,p},
\end{align}
which in our notation corresponds to
\begin{align}
\epsilon_{0,p}&=+12e^2.
\end{align}
Note that if $e=0$, $\epsilon_{0,p} = 0$, consistent with $\omega_p = n$ in (\ref{eq:lags}).

In the limit $M_\star \gg M_p$ and keeping only terms up to order $e^2$, equations (\ref{eq:dadt}) - (\ref{eq:domdt}) now reduce to
\begin{align}
\label{eq:dadttyp}
\frac{1}{a}\frac{da}{dt} &= \left[3\sqrt{\frac{G}{M_\star}} \frac{k_{2,\star}R_\star^5M_p}{Q_\star}\left( 1 + \frac{51}{4}e^2 \right) - 21\sqrt{GM_\star^3} \frac{k_{2,p}R_p^5}{Q_pM_p}e^2\right]a^{-13/2} \\[0.2cm]
\label{eq:dedttyp}
\frac{1}{e}\frac{de}{dt} &= \left[\frac{57}{8}\sqrt{\frac{G}{M_\star}} \frac{k_{2,\star}R_\star^5M_p}{Q_\star} - \frac{21}{2}\sqrt{GM_\star^3} \frac{k_{2,p}R_p^5}{Q_pM_p}
\right]a^{-13/2}\\[0.2cm]
\label{eq:domdttyp}
\frac{d\omega_\star}{dt} &= -\left[\frac{3}{32}\frac{G^2M_p^2k_{2,\star}R_\star^3}{r_{g,\star}^2n^2Q_\star} \left(1+\frac{15}{2}e^2\right)\right]a^{-9}.
\end{align}
We need not calculate $d\omega_p/dt$, since we assume the planet's spin is instantaneously set to the equilibrium value. The first term in (\ref{eq:dadttyp}) and (\ref{eq:dedttyp}) is the orbital effect of the tide raised by the planet on the star. In both equations this term is positive, implying that the stellar tide acts to increase both $a$ and $e$. This makes sense under the assumption that the planet is outside of corotation; the stellar bulge therefore leads the planet in the orbit, torquing the planet such that it speeds up and migrates outwards. For an eccentric orbit, the strongest impulse occurs at pericenter; since the planet must return to that point in the orbit, the pericenter distance is preserved, but the faster orbital speed results in a more distant apocenter, and thus higher eccentricity. A similar analysis for the tide raised by the star on the planet (the second term in each of the above equations) yields the result that such a tide acts to \emph{decrease} both $a$ and $e$. 

The evolution of the orbit will therefore depend on the relative magnitudes of the stellar and planetary tides. A simple order of magnitude calculation will show that the ratio of the tide generated by the star on the planet to the tide generated by the planet on the star is
\begin{align}
\bigg|\frac{\dot{a}_{\star\rightarrow p}}{\dot{a}_{p\rightarrow\star}}\bigg| \approx 7\left(\frac{M_\star}{M_p}\right)^2\left(\frac{R_p}{R_\star}\right)^5\left(\frac{Q_\star}{Q_p}\right)\left(\frac{e^2}{1+\frac{51}{4}e^2}\right).
\end{align}
For a 10$\mearth$, 2$\rearth$ mini-Neptune with $Q_p = 10^4$ orbiting a 0.1$\msun$, 0.15$\unit{R_\odot}$ star with $Q_\star = 10^6$, this becomes
\begin{align}
\label{eq:mnex}
\bigg|\frac{\dot{a}_{\star\rightarrow p}}{\dot{a}_{p\rightarrow\star}}\bigg| \approx 3\times 10^4\left(\frac{e^2}{1+\frac{51}{4}e^2}\right),
\end{align}
which is $\gg 1$ for all $e \gtrsim 0.01$. For an Earth-like planet with $Q_p = 10^2$, the ratio is $\gg 1$ for all $e \gtrsim 0.001$. Therefore, for planets starting with eccentricities $\gtrsim 0.1$, we would expect the planetary tide to dominate the evolution, such that the planet's orbit will shrink.
\vspace*{0.1in}
\section{Tidal Evolution Expressions: CTL}
The expressions for the evolutions of the orbital parameters are \citep{BAR13}
\begin{align}
\label{eq:dadtctl}
\frac{da}{dt} &= \frac{2a^2}{GM_\star M_p}\sum_{i\neq j}Z_i\left(\frac{f_2(e)}{\beta^{12}(e)}\frac{\omega_i}{n} - \frac{f_1(e)}{\beta^{15}(e)} \right) \\
\label{eq:dedtctl}
\frac{de}{dt} &= \frac{11ae}{2GM_\star M_p}\sum_{i\neq j}Z_i\left(\frac{f_4(e)}{\beta^{10}(e)}\frac{\omega_i}{n} - \frac{18}{11}\frac{f_3(e)}{\beta^{13}(e)} \right) \\
\label{eq:domdtctl}
\frac{d\omega_i}{dt} &= \frac{Z_i}{M_ir_{g,i}^2R_i^2n}\left( \frac{f_2(e)}{\beta^{12}(e)} - \frac{f_5(e)}{\beta^9(e)}\frac{\omega_i}{n} \right),
\end{align}
where
\begin{align}
\label{eq:zictl}
Z_i&\equiv 3G^2k_{2,i}M_j^2(M_i+M_j)\frac{R_i^5}{a^9}\tau_i
\end{align}
and
\begin{align}
\beta(e) &\equiv \sqrt{1-e^2} \nonumber \\
f_1(e) &\equiv  1 + \frac{31}{2}e^2 + \frac{255}{8}e^4 + \frac{185}{16}e^6 + \frac{25}{64}e^8 \nonumber \\
f_2(e) &\equiv  1 + \frac{15}{2}e^2 + \frac{45}{8}e^4 + \frac{5}{16}e^6 \nonumber\\
f_3(e) &\equiv  1 + \frac{15}{4}e^2 + \frac{15}{8}e^4 + \frac{5}{64}e^6 \nonumber\\
f_4(e) &\equiv  1 + \frac{3}{2}e^2 + \frac{1}{8}e^4 \nonumber\\
f_5(e) &\equiv  1 + 3e^2 + \frac{3}{8}e^4.
\end{align}
As before, if we assume the planet is tidally locked, its rotation rate is given by
\begin{align}
\label{eq:omeqctl}
\omega_{p,\mathrm{eq}}^\mathrm{CTL} &= n\left(\frac{f_2(e)}{\beta^3(e)f_5(e)}\right).
\end{align}
\vspace*{0.1in}
\subsection{The Typical Case}
As in the CPL model, we expect that for the typical case the tides raised on the planet will dominate the evolution. Plugging the result of (\ref{eq:omeqctl}) into (\ref{eq:dadtctl}) and (\ref{eq:dedtctl}), we find that the second (negative) terms dominate and the orbit should therefore shrink and circularize. Proceeding as before, we have
\begin{align}
\bigg|\frac{\dot{a}_{\star\rightarrow p}}{\dot{a}_{p\rightarrow\star}}\bigg| \approx \left(\frac{M_\star}{M_p}\right)^2\left(\frac{R_p}{R_\star}\right)^5\left(\frac{\tau_p}{\tau_\star}\right) \bigg|F\left(e,\frac{\omega_\star}{n}\right)\bigg|,
\end{align}
where
\begin{align}
F\left(e,\frac{\omega_\star}{n}\right) 	&\equiv \frac{\frac{f_2^2(e)}{f_5(e)f_1(e)}-1}{\frac{f_2(e)\beta^3(e)}{f_1(e)}\frac{\omega_\star}{n}-1} \nonumber \\
								&\approx -\frac{241}{16\left(\frac{\omega_\star}{n}-1\right)}e^2
\end{align}
for small $e$. For the same mini-Neptune considered in the CPL case (\ref{eq:mnex}), this becomes
\begin{align}
\bigg|\frac{\dot{a}_{\star\rightarrow p}}{\dot{a}_{p\rightarrow\star}}\bigg| \approx 5\times 10^5\left(\frac{e^2}{\frac{\omega_\star}{n}-1}\right),
\end{align}
which is $\gg 1$ for all $e \gtrsim 0.001\sqrt{\omega_\star/n-1}$, implying that the CTL model also predicts a net inward migration due to tides.
\vspace*{0.1in}
\section{Rate of Change of the Flux}
Conservation of angular momentum requires that for tidal evolution, the rates of change of the eccentricity and the semi-major axis must be related through
\begin{align}
\frac{1}{e}\frac{de}{dt} = \frac{1}{2a}\frac{da}{dt}.
\end{align}

The rate of change of the flux due to orbital changes is
\begin{align}
\label{ap:eq:dfdt}
\frac{dF}{dt} 	&= \frac{L_\mathrm{bol}}{4\pi}\frac{d}{dt}\left( a^2\sqrt{1-e^2} \right)^{-1} \nonumber \\
			&= - \frac{L_\mathrm{bol}}{4\pi}\left( a^2\sqrt{1-e^2} \right)^{-2} \left[ 2a\frac{da}{dt}\sqrt{1-e^2} + \left(\frac{a^2}{2\sqrt{1-e^2}}\right)\left(-2e\frac{de}{dt}\right) \right] \nonumber \\
			&=  \frac{L_\mathrm{bol}}{4\pi}\frac{1}{a^2}\left[-\frac{2}{a}\frac{da}{dt}\frac{1}{\left(1-e^2\right)^{1/2}} + \frac{e}{\left( 1-e^2 \right)^{3/2}}\frac{de}{dt}\right] \nonumber \\
			&= - F \frac{de}{dt}\left[\frac{4}{e\left(1-e^2\right)^{1/2}} - \frac{e}{\left( 1-e^2 \right)^{3/2}}\right] \nonumber \\
			&= - F \frac{de}{dt}\left[ \frac{4-5e^2}{e(1-e^2)^{3/2}} \right].
\end{align}
For $e \lesssim 0.7$, this simplifies to
\begin{align}
\label{ap:eq:dfdtlowe}
\frac{dF}{dt} 	&= - 4F \frac{1}{e}\frac{de}{dt},
\end{align}
which is equivalent to the trivial result
\begin{align}
\frac{1}{F}\frac{dF}{dt} &= - \frac{2}{a}\frac{da}{dt}.
\end{align}
Thus at low eccentricities, the flux will always increase as the orbit shrinks. However, at very high eccentricities ($e \gtrsim 0.8$), (\ref{ap:eq:dfdt}) predicts that $dF/dt$ is \emph{negative} when $da/dt < 0$: the decrease in the flux due to the circularization of the orbit overpowers the increase due to the shrinking orbit.
\end{appendix}

\bibliographystyle{apj}
\bibliography{hecs}
\end{document}